\documentclass[%
 superscriptaddress,
nofootinbib,
 amsmath,amssymb,
 aps,
prd,
twocolumn
]{revtex4-2}

\usepackage{xcolor}
\definecolor{darkpowderblue}{rgb}{0.0, 0.2, 0.6}
\definecolor{powderblue}{rgb}{0.0,0.3,0.5}
\usepackage[unicode,
colorlinks=true,
linkcolor={darkpowderblue},
citecolor={powderblue},
filecolor={darkpowderblue},
urlcolor={darkpowderblue},
pdfusetitle]{hyperref}

\usepackage{bm}
\usepackage{graphicx}
\usepackage{slashed}
\usepackage[normalem]{ulem}

\newcommand{\rmi}{\mathrm{i}}
\newcommand{\rme}{\operatorname{e}}

\newcommand{\xeff}{\chi_{\rm eff}}
\newcommand{\bhat}[1]{\hat{\bm{ #1 }}} 
\newcommand{\sn}{\operatorname{sn\!}{}} 
\newcommand{\order}[1]{\mathcal{O}\!\left( #1 \right)}
\newcommand{\norm}[1]{\left\| #1 \right\|}
\newcommand{\heavi}{\hat{\Theta}} 
\newcommand{\Msun}{\mathrm{M}_\odot} 
\newcommand{\Qlm}[1]{Q^{lm}_{ #1 }}

\newcommand{\inner}[2]{\left( #1 \middle| #2 \right)} 

\begin{document}

\title{Efficient gravitational-wave model for \texorpdfstring{\\}{} fully-precessing and moderately-eccentric, compact binary inspirals}

\author{J. Nijaid Arredondo}
\affiliation{
Illinois Center for Advanced Studies of the Universe, Department of Physics, University of Illinois at Urbana-Champaign, Urbana, Illinois 61801, USA.
}
\author{Antoine Klein}
\affiliation{
Institute for Gravitational Wave Astronomy \& School of Physics and Astronomy, University of Birmingham, Birmingham, B15 2TT, UK.
}
\author{Nicol\'as Yunes}
\affiliation{
Illinois Center for Advanced Studies of the Universe, Department of Physics, University of Illinois at Urbana-Champaign, Urbana, Illinois 61801, USA.
}

\date{\today}

\begin{abstract}
Future gravitational-wave detectors, especially the Laser Interferometer Space Antenna (LISA), will be sensitive to black hole binaries formed in astrophysical environments that promote large eccentricities and spin-induced orbital precession. Approximate models of gravitational waves that include both effects have only recently begun to be developed. The Efficient Fully Precessing Eccentric (EFPE) family is one such model, covering the inspiral stage with small-eccentricity-expanded gravitational-wave amplitudes accurate for initial time eccentricities $e < 0.3$ at 4 years before reaching an orbital frequency of 1 Hz. In this work, we extend this model to cover a larger range of initial eccentricities.  The new EFPE for moderate eccentricities (EFPE\_ME) model is able to accurately represent the leading-order gravitational-wave amplitudes to $e \leq 0.8$. Comparing the EFPE and the EFPE\_ME models in the LISA band, however, reveals that there is no significant difference when the eccentricity at 4 years before merger, $e_0$, is less than or equal to 0.5, as radiation reaction circularizes supermassive black hole binaries too quickly. This suggests that the EFPE model may have a larger regime of validity in eccentricity space than previously thought, making it suitable for some inspiral parameter estimation with LISA data. On the other hand, for systems with $e_0 > 0.5$, the deviations between the models are significant, particularly for binaries with total masses below $10^5\, \Msun$. This suggests that the EFPE\_ME model will be crucial to avoid systematic bias in parameter estimation with LISA in the future, once this model has been hybridized to include the merger and ringdown and the computation of the amplitudes is optimized.

\end{abstract}

\maketitle


\allowdisplaybreaks[4]
\section{Introduction}
\label{sec:intro}
Gravitational waves (GWs) carry a wealth of information about their source to the detectors prepared to observe them.
The latest GW catalog from the ground-based observatories LIGO-Virgo-KAGRA (LVK) detector network lists 90 candidate signals of compact binary coalescences (CBCs), the majority of which have been identified as black hole (BH) binaries, with some likely to contain a neutron star \cite{KAGRA:2021}.
Of the current observations, their spin-angular momenta are poorly constrained \cite{KAGRA:2021}, and while they have been found to be consistent with quasi-circular (non-eccentric) binaries (see, e.g., \cite{Iglesias:2022,Ramos-Buades:2023}), some events contain evidence for eccentricity (see, e.g., \cite{Romero-Shaw:2021,Romero-Shaw:2022xko}).
It is expected that the current LIGO design sensitivity can measure BH binary eccentricities above 0.05 at its low frequency end of 10 Hz \cite{Lower:2018}, allowing for constraints to be placed on binaries formed through dynamical interactions characterized by large eccentricities \cite{Rodriguez:2018,Zevin:2018}.

For future space-based detectors, systems with eccentricity and spin-precession are not only possible, but they are actually unavoidable.
Space-based missions, such as the Laser Interferometer Space Antenna (LISA) observatory~\cite{LISA:2017} and TianQin detector~\cite{TianQin:2015}, will cover the millihertz frequencies when launched within the next couple of decades. One class of system these detectors will observe are binaries far from the merger phase, which the LVK collaboration will later also observe at merger.
Binaries that may be circular in the frequency band of ground-based detectors may be eccentric enough earlier in their inspiral phase to be detected with measurable eccentricity by space-based detectors \cite{Sesana:2016}.
As LISA will cover roughly four decades of frequency space, binaries whose components' spins are misaligned from their orbital angular momentum can undergo numerous spin-precession cycles within that window, also displaying precession-induced modulation in their radiated GWs \cite{Apostolatos:1994}.

Once launched, future space-based detectors will also be sensitive to systems not yet probed by current GW observations, some of which may contain the signs of both eccentricity and spin-precession \cite{LISA:2022}.
Processes such as GW captures, three-body interactions, and hierarchical mergers of compact objects can form significantly eccentric binaries in dense stellar environments, according to simulations of globular clusters \cite{Rodriguez:2018,Antonini:2012,Antonini:2015,Samsing:2017,Samsing:2013,Liu:2020}.
Combined with an isotropic distribution of spins, these dynamical formation channels yield eccentric, spin-precessing GW sources \cite{Rodriguez:2016}.
The ability to detect such systems by accurately measuring their GWs will be key to discriminating between dynamical formation and isolated binary evolution channels \cite{LISA:2023}.

In order to accurately infer the properties of GW-generating systems, signals extracted from the detector noise are compared against GW models through Bayesian parameter estimation.
Until recently, no state-of-the-art inspiral model, however, could accurately represent binaries with \textit{both} high eccentricity and spin-precession.
This is because, up until recently, modeling eccentricity had been a low priority, as most binaries are expected to have circularized by the time they reach the LIGO band \cite{Amaro-Seoane:2015}. 
Therefore, the development of models with eccentricity has largely proceeded separate from the development of models with misaligned spins.

Such siloed development of models applies to many GW template families for CBCs, including the time- and frequency-domain phenomenological (Phenom) models~\cite{Khan:2018,IMRPhenomXHM,IMRPhenomXPHM,IMRPhenomT,IMRPhenomTPHM} and the effective-one-body (EOB) models~\cite{SEOBNRv4EHM,SEOBNRv5HM,SEOBNRv5PHM,TEOBResumS-Dali,TEOBResumS-Giotto}. The former hybridize the waveform across the inspiral, merger, and ringdown (IMR) through phenomenological fitting coefficients. The latter connect the inspiral and the merger-ringdown in the time domain by sewing together a calibrated EOB description of the inspiral with a quasi-normal mode description of the ringdown. In the inspiral stage of CBCs, both of these template families rely on post-Newtonian (PN) methods, i.e.~perturbative expansions in weak fields and slow velocities\footnote{When referring to a term of $n$PN order, we imply that it is of order $(v/c)^{2n}$ relative to the controlling factor in a PN expansion, where $v$ is the orbital velocity of the binary and $c$ is the speed of light.}. 

Since the development of eccentric PN methods has proceeded, historically, and for the most part, separately from the development of spin-precessing PN methods, so has the development of Phenom and EOB GW models, as can be seen in Table \ref{tab:models}, which we discuss later in this section.
When spin has been included with eccentricity, the models usually assume either that the eccentricity is small or that the spins are aligned or anti-aligned with the orbital angular momentum, avoiding precession of the orbital plane.
Only recently was a spin-precessing, eccentric EOB model announced, matching numerical relativity simulations with an initial $e \approx 0.2775$ \cite{SEOBNRPE}.
To capture the structure of the complicated merger-ringdown phase of eccentric orbits, relationships between the GW amplitude peaks at merger and the properties of the final compact objects are currently being analyzed \cite{Wang:2023,Carullo:2023}.
Thus, we see that the development of models that include both eccentricity and spin-precession is still in its infancy.

In this paper, we will focus on the inspiral stage of Phenom models, so let us summarize their historical development in the spin-precessing and eccentric space. The leading PN-order spin effects of a binary were introduced to state-of-the-art quasi-circular waveforms in the IMRPhenomP series \cite{Hannam:2013}, in particular with \texttt{IMRPhenomPv3}, which models binaries with double simple-precession \cite{Khan:2018}.  
In order to avoid expensive computations, eccentricity has been included in Phenom models only approximately. The GW Fourier phase and amplitudes are expanded in powers of small eccentricity in an approach known as the \textit{post-circular approximation}~\cite{Yunes:2009}.
However, models based on this approximation are accurate in a limited range of eccentricity, leaving them vulnerable to biases in parameter estimation if the signal is sufficiently eccentric \cite{Favata:2021,Divyajyoti:2023}.
Only recently have methods been developed to 3PN order for modeling arbitrary eccentricities in the inspiral of non-spinning bodies, achieving faithful waveforms for LIGO up to $e \approx 0.8$ \cite{Moore:2018,Moore:2019}.

Neither the spin-precessing IMRPhenomP family, nor moderately eccentric inspiral models, however, are capable of describing both spin-precession and eccentricity \textit{simultaneously}. This is a problem because, even if we detect GWs from spin-precessing binaries with relatively small eccentricity, the non-circularity of the signal can still lead to systematic biases in the estimation of parameters, and can make it difficult to distinguish one effect from the other \cite{Romero-Shaw:2022}. Therefore, combining these two effects to create a spin-precessing and eccentric GW model is essential to cover the parameter space of generic binaries and enable accurate parameter estimation.

A way to model the inspiral with both eccentricity and spin-precession is by combining the approaches mentioned above.
A first step in this direction was taken through the Efficient Fully-Precessing Eccentric (EFPE) model, which constructs frequency-domain waveforms with excellent agreement relative to numerical inspiral waveforms and at computational speeds comparable to that of the circular \texttt{TaylorF2} model \cite{Klein:2021}.
Like all post-circular models, however, its accuracy is limited to low eccentricities.
Following the EFPE model, we henceforth refer specifically to the \emph{time} eccentricity, which the EFPE was shown to be accurate for  initial $e < 0.3$ at 4 years before the orbits considered by that study reached a frequency of 1 Hz \cite{Klein:2018}.
The range of valid eccentricities of the EFPE model is controlled by the power of $e$ to which its GW amplitudes are expanded; the highest order it currently implements is $\order{e^6}$.
The expansion order of the EFPE model can be systematically increased, but as the expansion must be truncated, it cannot represent binaries with arbitrarily large eccentricity accurately.

In this paper, we develop and analyze an extended EFPE model aimed at describing spin-precessing CBCs with arbitrary eccentricities.
Building upon the EFPE framework, we consider the accuracy of the GW amplitudes, which are represented by sums of infinite series in orbital harmonics.
Instead of Taylor expanding the GW amplitudes in small eccentricity as the post-circular approach does, we study how many harmonics are needed to meet a specified threshold for accuracy within a range of eccentricity.
At leading-order in the PN approximation, these amplitudes are independent of spin, and thus, they allow for a systematic determination of the number of necessary harmonics as a function of eccentricity.
We find that this can be done for $e \leq 0.8$; past this point, the harmonic series converge poorly and require a more careful analysis.

\begin{table*}[th]
    \caption{A sample of recent IMR BH binary models within the Phenom and EOB (of both \texttt{TEOBNR} and \texttt{SEOBNR} flavour) families, which include spins (aligned or misaligned) and/or eccentricity (Ecc), and inspiral-only models at moderate eccentricities. Our new model is the first inspiral-only model to include both spin precession and moderate eccentricities.}
    \label{tab:models}
    \begin{ruledtabular}
    \begin{tabular}{cccc@{\hskip 0.5cm}cc}
    \multicolumn{4}{c}{IMR} & \multicolumn{2}{c}{Inspiral-only  w/\textit{moderate} ecc.} \\
        Aligned spins & Aligned spins + Ecc & Spin precession & Spin precession + Ecc & Ecc & Spin prec. + Ecc \\
        \hline
        \texttt{IMRPhenomXHM} \cite{IMRPhenomXHM} & \texttt{SEOBNRv4EHM} \cite{SEOBNRv4EHM} & \texttt{IMRPhenomXPHM} \cite{IMRPhenomXPHM} & \texttt{SEOBNRPE} \cite{SEOBNRPE} & Moore \& & \textbf{EFPE\_ME} \\
        \texttt{SEOBNRv5HM} \cite{SEOBNRv5HM} & \texttt{TEOBResumS-Dal\'i} \cite{TEOBResumS-Dali} & \texttt{SEOBNRv5PHM} \cite{SEOBNRv5PHM} &  & Yunes \cite{Moore:2019} & \\
         &  & \texttt{TEOBResumS-Giotto} \cite{TEOBResumS-Giotto} & & & \\
    \end{tabular}
    \end{ruledtabular}
\end{table*}

Tabulating the number of harmonics needed for $e \in (0,0.8]$, we then implement these series in the EFPE model, creating a new \textit{EFPE\_ME model} that accurately represents the GW amplitudes for moderate $e$(hence the name)\footnote{We follow Ref.~\cite{Moore:2019} in the choice of calling $e = 0.8$ a \emph{moderate} eccentricity, as truly high eccentricities ($e \gtrsim 0.9$) are not included in this work.}. 
We then explore the parameter space where the EFPE and EFPE\_ME waveforms differ within the sensitivity of LISA for binaries that merge within 4 years, finding that the overlaps between these models decrease for equal-mass binaries with high initial eccentricities ($e_0 > 0.5$) and total masses below $10^5 \,\Msun$.
Thus, for lower eccentricities and higher masses that coalesce in the same amount of time, binaries circularize quickly, making the EFPE sufficient for analyzing such LISA sources.
On the other hand, the EFPE\_ME is needed to more fully cover the parameter space of BH binaries.
Although the EFPE\_ME model takes longer to evaluate compared to the EFPE at large eccentricities, future work can be done to optimize the numerical implementation of this effect and reduce its computational cost, a task we have not endeavored to undertake in this work.

Table \ref{tab:models} presents a summary of the models we have discussed, including our new model, indicating the effects that each model implements.
Although the IMR models are based on similar PN methods as the inspiral-only models, they are validated and calibrated against numerical relativity results in the very late inspiral and merger.
The EOB models use more numerical integration steps than the Phenom family and inspiral-only models, along with the discrete transformation of their waveform into the frequency domain, requiring methods to reduce their computational costs (see, i.e., \cite{Purrer:2015,Smith:2016,Thomas:2022}).
The EFPE\_ME model described in this paper is the first model to combine spin precession with moderate eccentricities directly within the frequency domain, and can thus be adapted for the Phenom family to efficiently generate waveforms for data analysis.

In the remainder of this paper, we detail the derivation of the EFPE\_ME model and the results summarized above.
Starting with Sec.~\ref{sec:waveform}, we review the binary dynamics and waveform of the EFPE model that we extend in this work.
We then introduce the leading-order amplitudes that enter the waveform in Sec.~\ref{sec:amplitudes}, expressing them in a form valid for general eccentricities, as Fourier series in orbital harmonics.
Setting a threshold for accuracy, we then establish a procedure for truncating these series to meet our requirements and tabulate the truncation order for a set of maximum eccentricities we wish to model.
This completes our new waveform for moderate eccentricities, the EFPE\_ME.
We compare our new amplitudes against the previous EFPE amplitudes in Sec.~\ref{sec:comparisons}, and comparing them to the previous amplitudes, and then comparing the two models in the frequency domain for inspiraling binaries.
We conclude in Sec.~\ref{sec:conclusion}, re-emphasizing the need for models with arbitrary eccentricities.

Henceforth, we use the following conventions. The units are geometric, with $G=1=c$.
Spatial vectors are denoted by bold letters $\bm{A}$, their magnitudes with $A = \sqrt{\bm{A}\cdot\bm{A}}$, and their unit vectors with a hat $\bhat{A} = \bm{A}/A$.
The complex conjugate of a number $z$ will carry an asterisk as $z^*$, and derivatives with respect to time will be denoted with dots, i.e. $\dot{a} \equiv da/dt$.
All angular momenta are scaled by the total mass squared $M^2$, such that, for instance, $\bm{L} \equiv \bm{L}_{\text{physical}}/M^2$.
We distinguish the eccentricity symbol $e$ from the exponential function symbol $\rme^x$ by typesetting the latter differently and as shown.


\section{Waveform for a spin-precessing eccentric binary}
\label{sec:waveform}
\begin{figure}[t]
    \includegraphics[width=0.4\textwidth]{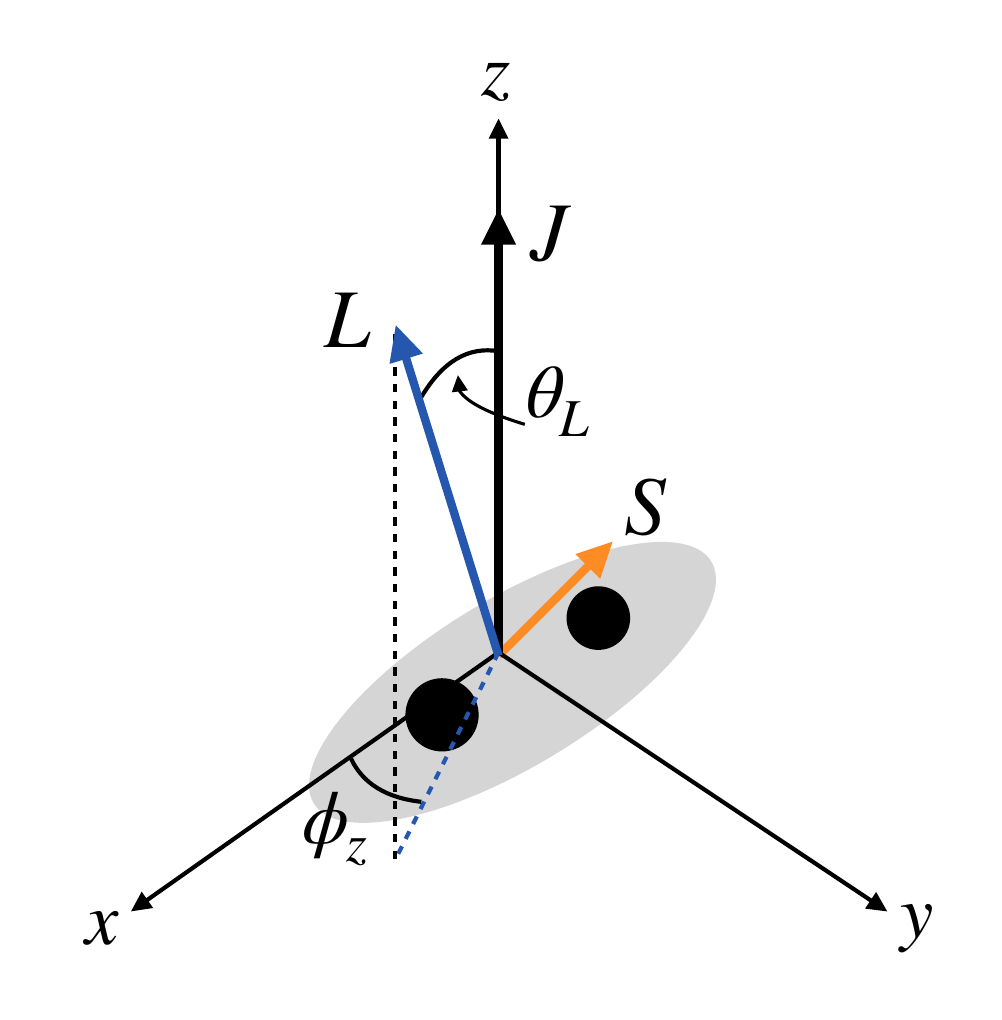}
    \caption{A spin-precessing black hole binary (black circles) with its angular momenta vectors (thick arrows).
    The $x$-$y$-$z$ coordinate axes are fixed in the inertial frame, while the orbital angular momentum $\bm{L}$, being normal to the orbital frame (grey ellipse),
    is inclined from the total angular momentum $\bm{J}$ at an angle $\theta_L$ and precesses around it by an angle $\phi_z$.
    The spin angular momentum $\bm{S}$ is the sum of each of the black hole's spins.}
    \label{fig:frame}
\end{figure}

We consider a binary BH orbit of total mass $M = m_1+m_2$,
where each BH has a mass $m_i$ and a spin angular momentum $\bm{S}_i$ vector with $i=1,2$.
To leading order in the PN approximation and spin-spin couplings,
the Newtonian orbital angular momentum vector $\bm{L}$ and the total spin angular momentum vector $\bm{S} = \bm{S}_1 + \bm{S}_2$ precess around the conserved total angular momentum vector $\bm{J} = \bm{L} + \bm{S}$ \cite{Apostolatos:1994}.
Ignoring the special case where $\bm{J}$ undergoes transitional precession due to precisely anti-aligned $\bm{L}$ and $\bm{S}$,
we take the direction of the total angular momentum $\bhat{J}$ to be fixed, as it does not vary significantly to leading order in radiation-reaction \cite{Chatziioannou:2017}.
Defining our fixed inertial frame with $\bhat{J}$ along its $z$-axis,
we can construct the waveform for this orbit by tracking the three angles that represent the orbit's precession:
\begin{itemize}
    \item $\theta_L$, the angle between $\bhat{L}$ and $\bhat{J}$;
    \item $\phi_z$, the angle of $\bm{L}$'s projection onto the plane perpendicular to $\bhat{J}$; and
    \item $\zeta$, the phase difference between the orbital frequency measured in the inertial frame and the non-inertial frame.
\end{itemize}
We illustrate these angles in Fig.~\ref{fig:frame}.
The angle $\zeta$ is not drawn as it is not a geometric quantity,
but it nonetheless informs the observed precession \cite{Schmidt:2012}.
The evolution of these Euler angles is described in Sec.~\ref{sub:dynamics},
where we review the dynamics of our binary,
and use them to construct our gravitational waveform in Sec.~\ref{sub:waveform}.
We follow the same derivations as in Ref.~\cite{Klein:2021}, but make clear any distinctions between conventions.

\subsection{Spin-precession and radiation-reaction of the orbital plane}
\label{sub:dynamics}

The dynamics for our BH binary are split into two sectors: the \textit{conservative} dynamics and the \textit{dissipative} dynamics.
The conservative dynamics are integrable,
while the dissipative dynamics radiate energy and angular momentum away from the orbit, causing it to inspiral.
These two sectors have a distinct separation of timescales,
since the dissipation occurs on the radiation-reaction timescale
\begin{equation}
  T_{\text{rr}} = \order{\frac{M}{v^8}},
\end{equation}
where $v$ is the orbital velocity,
while the conservative dynamics involves three timescales:
that of the orbit,
\begin{equation}
    T_{\text{orb}} = \order{\frac{M}{v^3}},
\end{equation}
that of spin precession,
\begin{equation}
    T_{\text{pr}} = \order{\frac{M}{v^5}},
\end{equation}
and that of periastron precession,
\begin{equation}
    T_{\text{pp}} = \order{\frac{M}{v^5}}.
\end{equation}
In the early inspiral, where the orbit is widely separated and the orbital velocity is small, we see that the timescales follow a hierarchy $T_{\text{orb}} \ll T_{\text{pr}} \sim T_{\text{pp}} \ll T_{\text{rr}}$.
This allows one to perform the following approximations.
First, we can average the equations of motion over the orbital period, neglecting the fast oscillations that occur on that timescale.
Second, we can average the radiation-reaction over the precession timescales.
This constitutes a multiple-scale analysis of the equations of motion to leading order,
leaving only the dominant timescale, $T_{\text{rr}}$, to be integrated.
Such approximations are common and have been shown to be accurate, at least when the hierarchy from above is obeyed \cite{Chatziioannou:2017}.
The advantage of this analysis is that it allows one to write the solution for the conservative dynamics in closed form, and then evolve the constants of motion of that sector on the radiation-reaction timescale,
which is computationally more efficient than tracking all the dynamics on the shorter timescales.

We first review the conservative dynamics.
The spinning orbit itself can be expressed to 2PN order using the quasi-Keplerian (QK) parameterization \cite{Klein:2010},
which defines the orbital elements in the absence of radiation.
The elements of interest to us are the eccentricity $e$ (specifically, the \emph{time} eccentricity), the orbital phase $\phi$, and the PN parameter
\begin{equation}
    y = \frac{(M\omega)^{1/3}}{\sqrt{1-e^2}} = \order{v},
\end{equation}
where $\omega$ is the mean orbital frequency.
The phase itself can be decomposed as $\phi = \lambda + W$,
where $\lambda$ grows linearly in time as $\dot{\lambda} = \omega$, while $W$ is a more complicated periodic function of the orbit.
Their full expressions can be found in Appendix B of Ref.~\cite{Klein:2018}.

The 2PN spin dynamics allow $\bm{L}$ and $\bm{S}$ to precess around the total angular momentum vector $\bm{J}$,
which is conserved along with its magnitude $J$ and the magnitudes of the individual angular momenta $S_i$ and $L$.
Using the orbit-averaged precession equations for $\bm{L}$ and  $\bm{S}_i$ (see Eqs.~(1)-(3) of \cite{Klein:2021}),
the system can then be closed by defining the total reduced spin $\bm{s} = \bm{s}_1 + \bm{s}_2$, where we introduce the reduced individual spin vectors
\begin{equation}
    \bm{s}_i = \frac{M}{m_i} \bm{S}_i
\end{equation}
to write down a conserved quantity, known as the mass-weighted effective spin\footnote{We note that this is only conserved by the orbit-averaged equations of motion \cite{Racine:2008}.}
\begin{equation}
    \xeff = \bhat{L} \cdot \bm{s}.
\end{equation}
The dynamical variable that determines the evolution of the angular momenta is then commonly chosen to be the total spin magnitude $S$ (see Ref.~\cite{Chatziioannou:2017} and references therein).
However, the evolution of this quantity becomes singular for equal mass ($m_1 = m_2$) systems.
Instead, we use the reduced effective spin difference,
\begin{equation}
    \delta\chi = \bhat{L} \cdot (\bm{s}_1 - \bm{s}_2),
\end{equation}
as the dynamical variable, which does not exhibit this singularity.
The geometry of the system then allows one to write the evolution of the Euler angles as
\begin{align}
    \label{eq:thetaL}
    \cos\theta_L &= \bhat{L}\cdot\bhat{J} = \frac{1}{2J} \left( 2L+\chi_{\text{eff}}+\delta\mu \delta\chi \right), \\
    \label{eq:dphizdt}
    \frac{d\phi_z}{dt} &= \frac{1}{\sin^2\theta_L} \left(\frac{d\bhat{L}}{dt}\right) \cdot \left(\bhat{J}\times\bhat{L}\right), \\
    \label{eq:dzetadt}
    \frac{d\zeta}{dt} &= -\cos\theta_L \frac{d\phi_z}{dt},
\end{align}
where $\delta\mu = (m_1 - m_2)/M$.
All of the variables above can be written in terms of $\delta\chi$, of which we've shown it explicitly for the simplest expression (Eq.~\eqref{eq:thetaL}) as the rest are more complex (see Sec.~II of \cite{Klein:2021}).
Thus, $\delta\chi$ contains the time dependence of the conserved angular momenta.
This dependence is determined through the differential equation
\begin{align}\label{eq:Dchi}
  \left(M\frac{d\delta\chi}{dt}\right)^2 &= -\frac{1}{y}\left(\frac{3}{2}\kappa y^6\right)^2 \nonumber \\
  &\times (\delta\chi-\delta\chi_+)(\delta\chi-\delta\chi_-)(\delta\chi_3-\delta\mu \delta\chi),
\end{align}
where $\kappa \equiv (1-e^2)^{3/2} (1-y\xeff)$ and the roots $\delta\chi_+$, $\delta\chi_-$, and $\delta\chi_3$ are derived in Ref.~\cite{Klein:2021}.
Equation~\eqref{eq:Dchi} is closed for $\delta \chi$, so it can be solved independently of any other; its solution can be written in terms of the Jacobi elliptic function $\sn(\psi_{\rm p},m)$ as
\begin{equation}
\label{eq:deltax}
  \delta\chi = \delta\chi_- + (\delta\chi_+ - \delta\chi_-) \sn^2(\psi_{\rm p},m),
\end{equation}
where $\psi_{\rm p}$ is a new angle that satisfies the differential equation
\begin{equation}\label{eq:Dpsip}
  M\frac{d\psi_{\rm p}}{dt} = \frac{3}{2}\kappa y^6 \sqrt{\frac{1}{y}(\delta\chi_3 - \delta\mu\delta\chi_-)}
\end{equation}
that determines the function's \emph{phase}, while
\begin{equation}
  m = \frac{\delta\chi_+ - \delta\chi_-}{\delta\chi_3/\delta\mu - \delta\chi_-}
\end{equation}
is its \emph{modulus} (see Ref.~\cite{Brizard:2007} for a brief overview of the function and its appearance in physics).
In the absence of gravitational radiation, the right-hand side of Eq.~(\ref{eq:Dpsip}) is constant and thus the phase is linear in time.
A particular property of the Jacobi sine function is that $\sn(\psi_{\rm p},0) = \sin(\psi_{\rm p})$,
so when $m = 0$ in the equal-mass case,
$\delta\chi$ reduces to a trigonometric function.
As Eqs.~(\ref{eq:dphizdt})-(\ref{eq:dzetadt}) can be written in terms of $\delta\chi$,
integrating them yields our Euler angles in terms of elliptic functions \cite{Klein:2021}.
Thus, we have the conservative dynamics of our binary in closed analytic form and can proceed to add radiation-reaction to them.

The emission of gravitational radiation from a binary dissipates the system's energy and angular momentum,
driving the orbiting bodies to merger.
This occurs on the longest timescale discussed, $T_{\rm rr}$,
and is determined by the following orbit-averaged differential equations for the orbital elements up to 3PN,
\begin{subequations}
\label{eq:rr}
\begin{align}
    M\frac{dy}{dt} &= \left(1-e^2\right)^{3/2} \eta\; y^9 \sum_{n=0}^{6} a_n y^n, \label{eq:dydt}\\
    M\frac{de^2}{dt} &= -\left(1-e^2\right)^{3/2} \eta\; y^8 \sum_{n=0}^{6} b_n y^n, \\
    M\frac{d\lambda}{dt} &= \left(1-e^2\right)^{3/2} y^3,
\end{align}
\end{subequations}
where $\eta \equiv m_1 m_2 / M^2$ is the symmetric mass ratio and the $a_n$ and $b_n$ can be found in Appendix C of \cite{Klein:2018}.
These equations depend on the so-called enhancement factors (see, e.g., Eq.~(C3) of \cite{Klein:2018}), which we have expanded to $\order{e^{20}}$.
The $a_n, b_n$ coefficients contain spin couplings at 2PN order that are averaged over the precession timescale.
Note that the evolution of $\lambda$ is of the same form as in the conservative sector, but now with $y$ growing secularly in time.
The dissipation of the Newtonian orbital angular momentum can be seen from the definition of its magnitude, $L = \eta/y$.
We also see that the orbit tends to circularize as $e$ is driven to small values, although contributions from the spins can keep it from reaching arbitrarily small values \cite{Klein:2010}.

With the evolution of $y(t)$ and $e(t)$ determined by Eq.~(\ref{eq:rr}),
the evolution of $\delta\chi$ (Eq.~(\ref{eq:deltax})) with radiation-reaction is also determined,
and thus our Euler angles (Eqs.~(\ref{eq:thetaL})-(\ref{eq:dzetadt})).
One can then proceed to build the gravitational waveform for a spin-precessing, eccentric binary.

\subsection{Waveform for a spin-precessing binary}
\label{sub:waveform}

The methods reviewed in this section follow the same works referenced in the previous section.
These are procedures that have been performed in the past,
but as done by Ref.~\cite{Moore:2019} for non-spinning binaries,
we now extend them to higher eccentricities.
This extension requires calculating the waveform amplitudes necessary to represent the waveform at a given eccentricity.
We present these new amplitudes in Sec.~\ref{sec:amplitudes},
but first describe the construction of the waveform in this section.

We proceed in two steps.
In Sec.~\ref{sub:time-domain}, we first compute the waveform in the time domain by decomposing it into orbital harmonics, which gives us a form amenable for Fourier transformation.
This transformation is desirable as GW detection and data analysis is most efficiently performed in the frequency domain.
However, the waveform is a complicated function of time and cannot be explicitly Fourier-transformed.
Instead, we use an approximation scheme in Sec.~\ref{sub:freq-domain} that makes use of the harmonic decomposition to produce an accurate representation of the waveform in the frequency domain.

\subsubsection{Harmonic decomposition in time}
\label{sub:time-domain}

Our starting point for the waveform  is the spherical harmonic decomposition of the plus, $h_+$, and cross, $h_\times$, polarizations \cite{Kidder:2007},
\begin{equation}\label{eq:harmonic decomp}
  h_+ - \rmi h_\times = \sum_{l=2}^\infty \sum_{m=-l}^l h^{lm} ~{}_{-2} Y^{lm}(\Theta, \Phi),
\end{equation}
where $(\Theta, \Phi)$ are the spherical angles from the binary to the detector as measured in the inertial frame,
and ${}_{-2} Y^{lm}$ are the spin-weighted spherical harmonics of spin weight $-2$.
The conventions for these functions can be found in Appendix A of \cite{Klein:2021}.

The decomposition of the waveform in $h^{lm}$ modes is useful for comparisons with numerical simulations of the radiation emitted by binaries \cite{Kidder:2007,Blanchet:2008}, but suffer from the drawback that they cannot be readily compared between binary configurations whose orbital planes are precessing, as precession mixes modes together \cite{Schmidt:2012}.
A \emph{twisting-up} procedure has thus been developed to compare precessing numerical simulations and to compute analytic waveforms by dynamically rotating the modes from the frame aligned with the orbital plane (where precession effects are minimized, but not completely absent) to the inertial frame that measures the precession in $h^{lm}$ \cite{Schmidt:2010,OShaughnessy:2011,Boyle:2011}.
This method ``twists-up'' the modes via
\begin{equation}\label{eq:twist-up}
  h^{lm} = \sum_{m'=-l}^l D^l_{m'm}(\phi_z,\theta_L,\zeta) H^{lm'},
\end{equation}
where the $H^{lm}$ modes can be found up to 3PN in Ref.~\cite{Mishra:2015}\footnote{Here we designate the non-twisted modes as capitalized $H^{lm}$, which are not to be confused with the $H^{lm}$ of \cite{Mishra:2015}; that is, our $H^{lm}$ are what they call $h^{lm}$.},
and the Wigner $D$-matrices follow the conventions of Ref.~\cite{Klein:2021}.

Now, what are the modes $H^{lm}$?
These are functions of the orbit as measured in the spherical harmonic basis. Specifically, they can be expressed as functions of the eccentric anomaly $u$, because they are of the form
\begin{align}\label{eq:Hlm}
    H^{lm}(t) &= h_0 \rme^{-\rmi m\phi(t)} K^{lm}[u(t)] \nonumber\\
    &\equiv h_0 \hat{H}^{lm}(t),
\end{align}
where 
\begin{equation}
    h_0 \equiv 4\sqrt{\frac{\pi}{5}} \frac{M\eta x}{R},
\end{equation}
$x = (M\omega)^{2/3}$, and $R$ is the luminosity distance to the binary.
We have defined a reduced mode $\hat{H}^{lm}$ in the last line of Eq.~(\ref{eq:Hlm}).
However, in order to be Fourier transformed,
these modes require explicit knowledge of the time-dependence of the orbit.
In the QK parameterization,
the time-dependence can be determined by the mean anomaly $\ell$,
whose relation to $u$ is
\begin{equation}\label{eq:lofu}
    \ell = u - e \sin u + f_t(u,\vartheta),
\end{equation}
where $\vartheta$ is the true anomaly and the function $f_t$ is defined in Appendix B of Ref.~\cite{Klein:2018}.
The mean anomaly is useful as it is an explicit function of time through its definition, $\dot{\ell} = \omega/(1+k)$ (where $k$ is the periastron advance), which can be integrated linearly in time in the absence of radiation reaction,
or can be solved with Eq.~(\ref{eq:rr}) in the presence of radiation.
As the anomalies are periodic,
the relationships between them (such as Eq.~(\ref{eq:lofu})) can give one the reduced modes $\hat{H}^{lm}$ in terms of a harmonic Fourier series in $\ell$ and $\lambda$,
\begin{equation}
\label{eq:Hlm decomp}
    \hat{H}^{lm}(t) = \sum_{j=-\infty}^{\infty} N^{lm}_j \rme^{-\rmi (j\ell + m \lambda)}.
\end{equation}
Here, we have expanded the phase $\phi(t)$ and absorbed its oscillatory piece $W(t)$ into $N^{lm}_j \rme^{-\rmi j\ell}$.
We detail the steps between Eq.~(\ref{eq:Hlm}) and Eq.~(\ref{eq:Hlm decomp}) in Sec.~\ref{sec:amplitudes}.
We have also chosen a particular phase for the exponential, which will ease the Fourier transformation in Sec.~\ref{sub:freq-domain}.
In the conservative sector, all the time-dependence is held by the complex exponential.
When taking radiation-reaction into account, the amplitudes $N^{lm}_j$ also become time-dependent, as they are functions of the orbital elements.
Note, however, that the amplitudes then vary on the radiation-reaction timescale, while the oscillating exponential varies on the orbital and precession timescales,
a fact that will be useful when Fourier transforming the modes.

Before taking our waveform into the frequency domain,
we separate it into its two polarizations.
Defining
\begin{equation}
    A_{l,m,m'}(t) \equiv h_0\, {}_{-2}Y^{lm}(\Theta, \Phi)  D^l_{m'm}(\phi_z,\theta_L,\zeta),
\end{equation}
we can write Eq.~\eqref{eq:harmonic decomp} as
\begin{equation}
    h_+ - \rmi h_\times = \sum_{l=2}^\infty \sum_{m=-l}^l \sum_{m'=-l}^l A_{l,m,m'} \hat{H}^{lm'}.
\end{equation}
Using this equation and its complex conjugate,
along with the property $H^{l\,-m} = (-1)^l (H^{lm})^*$ and the symmetry of the $m'$ sum,
we find the polarizations
\begin{equation}\label{eq:h+x}
  h_{+,\times}(t) = \sum_{\bm{l}} \mathsf{A}^{+,\times}_{\bm{l}} \hat{H}^{lm'},
\end{equation}
where $\bm{l}$ is the hyperindex $\{l,m,m'\}$ such that
\begin{equation}
    \sum_{\bm{l}} \equiv \sum_{l=2}^\infty \sum_{m=-l}^l \sum_{m'=-l}^l
\end{equation}
and
\begin{subequations}
\begin{align}
  \mathsf{A}^+_{\bm{l}} &= \frac{1}{2} \left[ A_{l,m,m'} + (-1)^l A_{l,m,-m'}^* \right],\\
  \mathsf{A}^\times_{\bm{l}} &= \frac{\rmi}{2} \left[ A_{l,m,m'} - (-1)^l A_{l,m,-m'}^* \right].
\end{align}
\end{subequations}
Equation (\ref{eq:h+x}) thus gives the time domain polarizations for our binary,
decomposed in harmonics of the orbit that are then twisted-up to account for spin precession,
and that can now be Fourier transformed for analysis.

\subsubsection{Frequency-domain waveform}
\label{sub:freq-domain}

Having obtained our time-domain waveform in the previous section,
we now find the frequency-domain waveform for the GW polarizations through their Fourier transformation,
\begin{equation}\label{eq:FTh}
  \Tilde{h}_{+,\times}(f) \equiv \mathcal{F}\{h_{+,\times}(t)\}(f) = \int_{-\infty}^\infty h_{+,\times}(t) \rme^{2\pi\rmi f t}\, dt.
\end{equation}
As we have seen in previous sections,
$h_{+,\times}(t)$ are complicated functions of time and do not lend themselves to a fully analytic Fourier transform.
Numerical (discrete) transforms are possible,
but can incur a heavy computational cost considering the disparate orbital, precession, and radiation-reaction timescales involved;
the waveforms have to be sampled at the shortest timescales, while still covering the longest timescales, if they are to model the inspiral of a binary.
Fortunately, it is this separation of scales that allows one to accurately approximate Eq.~(\ref{eq:FTh}) through other mathematical methods.

First, we separate the terms in the polarizations by timescales.
Equation~(\ref{eq:h+x}) is already a step towards that;
its amplitudes vary on the radiation-reaction timescale, while $\ell$ and $\lambda$ vary on both orbital and precession timescales.
We separate these last two by defining $\delta\lambda = \lambda - \ell$, where
\begin{equation}
  \delta\dot{\lambda} = \frac{k}{1+k} \dot{\lambda}.
\end{equation}
This defines the precession timescale as $\delta\dot{\lambda} / \dot{\lambda} = \order{y^2}$.
Writing the modes in their harmonic decomposition (Eq.~(\ref{eq:Hlm decomp})) and substituting $\ell$ with $\lambda - \delta\lambda$,
we rewrite the polarizations as
\begin{equation}\label{eq:h+x decomp}
  h_{+,\times}(t) = \sum_{n=-\infty}^\infty \sum_{\bm{l}} \mathcal{A}^{+,\times}_{\bm{l},n} \rme^{-\rmi(n\lambda + (m'-n)\delta\lambda)},
\end{equation}
where
\begin{equation}\label{eq:Aln}
  \mathcal{A}^{+,\times}_{\bm{l},n}(t) = \mathsf{A}^{+,\times}_{\bm{l}} N^{lm'}_{n-m'}.
\end{equation}
We now have separated the amplitudes $\mathcal{A}^{+,\times}_{\bm{l},n}$,
which vary on the spin-precession timescale,
from the phase $n\lambda + (m'-n)\delta\lambda$,
which is dominated by the orbital timescale.
Thus, $n$ denotes the $n$th harmonic of the orbit.

At this point, it is common to employ the stationary phase approximation (SPA) to evaluate the Fourier transform of Eq.~(\ref{eq:h+x decomp}), since its amplitudes vary on a slower timescale than its phase.
However, the effects of precession cause ``catastrophes'' that lead to divergences with this method.
We therefore instead use the method of shifted uniform asymptotics (SUA) that was introduced in~\cite{Klein:2014},
which alleviates these issues by expanding the SPA and resumming it \cite{Klein:2014,Klein:2021}.
Applying the SUA to the polarizations yields
\begin{subequations}
\begin{align}\label{eq:SUA wf}
  \begin{split}
  \Tilde{h}_{+,\times}(f) &\overset{\mathrm{SUA}}{=} \sum_{n=1}^\infty \sum_{\bm{l}} \sqrt{2\pi} T_{n,q} \rme^{\rmi\Psi_{n,q}} \\
  &\overset{\hphantom{SUA}}{\times} \sum_{k=-k_\mathrm{max}}^{k_\mathrm{max}} a_{k,k_\mathrm{max}} \mathcal{A}^{+,\times}_{\bm{l},n}(t_{n,q} + kT_{n,q}),
  \end{split}\\
  \label{eq:SUA phase}
  \Psi_{n,q} &= 2\pi f t_{n,q} - n\lambda(t_{n,q})-q\delta\lambda(t_{n,q}) - \pi/4,\\
  T_{n,q} &= \left[n\ddot{\lambda}(t_{n,q})+q\delta\ddot{\lambda}(t_{n,q})\right]^{-1/2},
\end{align}
\end{subequations}
where $q\equiv m'-n$,
the stationary points $t_{n,q}$ are determined by\footnote{As we are interested in positive frequencies $f$,
the stationary points also restrict $n\dot{\lambda}+q\delta\dot{\lambda} > 0$.
We have confirmed that this inequality is satisfied until close to merger for the relevant values of $n$ and $q$.}
\begin{equation}\label{eq:stationarypts}
  2\pi f = n\dot{\lambda}(t_{n,q})+q\delta\dot{\lambda}(t_{n,q}),
\end{equation}
and the constants $a_{k,k_\mathrm{max}}$ satisfy the system of equations
\begin{subequations}
\begin{align}
  &\frac{1}{2}a_{0,k_\mathrm{max}} + \sum_{k=1}^{k_\mathrm{max}} a_{k,k_\mathrm{max}} = 1,\\
  &\sum_{k=1}^{k_\mathrm{max}} a_{k,k_\mathrm{max}} \frac{k^{2p}}{(2p)!} = \frac{(-\rmi)^p}{2^p p!}, &p \in \{1,...,k_\mathrm{max}\},\\
  &a_{-k,k_\mathrm{max}} = a_{k,k_\mathrm{max}},
\end{align}
\end{subequations}
for an arbitrary $k_\mathrm{max}$.

The waveform in Eq.~(\ref{eq:SUA wf}) contains sums over two indices, $n$ and $q$,
which determine the stationary points defined by Eq.~(\ref{eq:stationarypts}).
Solving this equation requires numerical inversion and can become computationally expensive if $n$ is truncated at large values,
as it is for large eccentricities.
To make this more efficient,
we take advantage of the fact that $\delta\lambda$ is due to precession of the binary.
We therefore expand $t_{n,q}$ as a correction to $t_{n} \equiv t_{n,0}$ with
\begin{equation}\label{eq:Delta tnm}
    \Delta t_{n,q} \equiv t_{n,q} - t_n = \sum_{p=1}^P \epsilon^p \Delta t^{(p)}_{n,q},
\end{equation}
where $\epsilon = \order{\delta\dot{\lambda} / \dot{\lambda}}$ is an order-keeping parameter and $P$ is the order at which this expansion is truncated.
In this manner, we can avoid solving for $(n,q)$ pairs of stationary points and instead only calculate $t_n$.
To determine the $\Delta t^{(p)}_{n,q}$,
we Taylor expand Eq.~(\ref{eq:stationarypts}),
\begin{equation}
    2\pi f = \sum_{p=0}^P \frac{1}{p!} \frac{d^p}{dt^p}\left[ n\dot{\lambda}(t) + \epsilon q\delta\dot{\lambda}(t) \right]_{t=t_n} (\Delta t_{n,q})^p.
\end{equation}
Inserting Eq.~(\ref{eq:Delta tnm}) and keeping terms only up to $\order{\epsilon^P}$ in the expansion above,
we can solve order by order for the $\Delta t^{(p)}_{n,q}$.
For the first three orders, we find
\begin{subequations}
\begin{align}
    \label{eq:tn}
    \epsilon^0 &: 2\pi f = n\dot{\lambda}, \\
    \epsilon^1 &: \Delta t^{(1)}_{n,q} = -\frac{q \delta\dot{\lambda}}{n \ddot{\lambda}}, \\
    \epsilon^2 &: \Delta t^{(2)}_{n,q} = -\frac{q^2 \delta\dot{\lambda}}{2n^2 \ddot{\lambda}^3} \left( \dddot{\lambda} \delta\dot{\lambda} - 2 \ddot{\lambda} \delta\ddot{\lambda} \right),
\end{align}
\end{subequations}
where all functions of time are evaluated at $t = t_n$,
and so Eq.~(\ref{eq:tn}) defines $t_n$.
Reference~\cite{Klein:2018} found that Eq.~\eqref{eq:Delta tnm} can be written as
\begin{equation}
    \Delta t_{n,q} = \sum_{p=1}^P \frac{1}{p!} \left( -\frac{q}{n} \right)^p \slashed{D}^{p-1} \left[ \frac{\delta\dot{\lambda}^p}{\ddot{\lambda}} \right]_{t=t_n},
\end{equation}
where we have the differential operator
\begin{equation}
    \slashed{D} (\cdot) \equiv \frac{1}{\ddot{\lambda}} \frac{d}{dt} (\cdot).
\end{equation}
Taylor expanding Eq.~(\ref{eq:SUA phase}) and collecting terms,
one can then, in the same manner, write
\begin{subequations}
\begin{align}
    \Psi_{n,q} &= 2\pi f t_n -n \lambda(t_n) - \pi/4 + \Delta\Psi_{n,q}, \\
    \Delta\Psi_{n,q} &= -q\delta\lambda(t_n) + n \sum_{p=2}^{P+1} \frac{1}{p!} \left(-\frac{q}{n}\right)^p \slashed{D}^{p-2} \left[\frac{\delta\dot{\lambda}}{\ddot{\lambda}}\right]_{t=t_n}.
\end{align}
\end{subequations}
We have now expanded our waveform into two pieces,
\begin{equation}
\label{eq:h+xf}
    \Tilde{h}_{+,\times}(f) \approx \sum_{n=1}^\infty \Tilde{h}_n^{(0)} \Tilde{h}_{+,\times,n}^{\rm PP},
\end{equation}
where first we have the leading terms
\begin{subequations}
\begin{align}
    \Tilde{h}_n^{(0)}(f) &= \sqrt{2\pi} T_n \rme^{\rmi(2\pi f t_n - n\lambda(t_n) - \pi/4)}, \\
    T_n &= \left[ n\ddot{\lambda}(t_n) \right]^{-1/2},
\end{align}
\end{subequations}
and lastly, upon replacing $m'$ with $q+n$,
the precession corrections
\begin{align}
\begin{split}
    \Tilde{h}_{+,\times,n}^{\rm PP}(f) &=
    \sum_{\bm{l}} \rme^{\rmi\Delta\Psi_{n,q}} \\ &\times \sum_{k=-k_\mathrm{max}}^{k_\mathrm{max}} a_{k,k_\mathrm{max}} \mathcal{A}^{+,\times}_{n,q}(t_n + \Delta t_{n,q} + kT_n),
\end{split}
\end{align}
where we have the amplitudes
\begin{equation}
\label{eq:A+x}
    \mathcal{A}^{+,\times}_{n,q} = \mathsf{A}^{+,\times}_{\bm{l}} N^{l,q+n}_{-q}.
\end{equation}

The derivation of the frequency-domain waveform in Eq.~(\ref{eq:h+xf}) mirrors that of Ref.~\cite{Klein:2021},
but here we have been careful to make our definitions explicit to make any differences clear.
In particular, we denote our harmonic amplitudes as $N^{l,q+n}_{-q}$ while (in our notation) theirs are  $G^{l, -q-n}_q$ to highlight (i) our choice of phase for the Fourier decomposition, and (ii) that our amplitudes are not expanded in small eccentricity.
We calculate these amplitudes in the next section,
pushing the waveform presented here to model higher eccentricities.


\section{Amplitudes for eccentric binaries}
\label{sec:amplitudes}
In the previous section,
we introduced the twisted-up waveform for spin-precessing eccentric binaries, following Ref.~\cite{Klein:2021}.
In that work, however, the Fourier amplitudes of the modes $H^{lm}$ were expanded around small eccentricity and were validated only up to $e \approx 0.3$.
In this paper,
we demonstrate how to derive the Fourier amplitudes $N^{lm}_j$ that lead to the mode amplitudes $H^{l m}$ through Eqs.~\eqref{eq:Hlm} and (\ref{eq:Hlm decomp}) for \emph{arbitrary} eccentricities,
where one chooses the accuracy of the Fourier series and truncates it once that accuracy is met.
We validate the accuracy of these amplitudes in Sec.~\ref{sec:comparisons}.

We decompose the (reduced) mode amplitudes by first factorizing them as follows:
\begin{equation}
    \hat{H}^{lm} = \rme^{-\rmi m\lambda} \rme^{-\rmi mW} K^{lm},
    \label{eq:hatHlm}
\end{equation}
where $K^{lm}$ is identified as the piece left over after removing $\rme^{-\rmi m\phi}$ from $\hat{H}^{lm}$.
Each of the terms in the above equation are functions of time, and we, therefore, decompose them into Fourier series in the mean anomaly $\ell$.
However, we do not do this with $\rme^{-\rmi m\lambda}$ because it has to be treated separately when calculating the Fourier transform, as done in Sec.~\ref{sub:freq-domain}.
For the other terms, we must find their harmonic series representation,
\begin{subequations}
\label{eq:P and K}
\begin{align}
\label{eq:Wdecomp}
    \rme^{-\rmi mW} &= \sum_{s=-\infty}^\infty \mathcal{P}^{mW}_s \rme^{-\rmi s\ell},\\
    K^{lm} &= \sum_{s=-\infty}^\infty \mathcal{K}^{lm}_s \rme^{+\rmi s\ell}.
    \label{eq:Kdecomp}
\end{align}
\end{subequations}
Once the Fourier coefficients $\mathcal{P}^{mW}_s$ and $\mathcal{K}^{lm}_s$ are determined,
we can identify the amplitudes in Eq.~(\ref{eq:Hlm decomp}) as the product series
\begin{equation}
\label{eq:Nlmj}
    N^{lm}_j = \sum_{s=-\infty}^\infty \mathcal{P}^{mW}_s \mathcal{K}^{lm}_{s-j}.
\end{equation}

Formally, the Fourier series presented above contain an infinite number of terms, but in practice only a finite number of terms are needed to obtain a representation of the mode to a chosen accuracy.
In general, the $N^{lm}_j$ scale as some positive power of $e$ that increases with $j$, so the series is expected to converge for small enough values of $e < 1$.
Numerically, then, only amplitudes ranging from $j = j_{\min}$ to $j_{\max}$ need to be found to accurately represent a given mode.
As $j = n-m'$ (see Eq.~\eqref{eq:Aln}),
the bounds on $j$ also determine the bound on the number of harmonics $n$ in the waveform.

In Sec.~\ref{sub:0PN amps},
we show the explicit construction of the Newtonian-order amplitudes as an example.
We take the components of the series in Eq.~\eqref{eq:P and K} and describe how to truncate them in Sec.~\ref{sub:truncating 1},
which we then use to obtain the Fourier amplitudes $N^{lm}_j$ and truncate the harmonic series expansions of $H^{lm}$ in Sec.~\ref{sub:truncating 2}.
These sections demonstrate how to properly truncate the infinite series to accurately represent the modes to a chosen numerical precision.
With these series in hand,
one can then generate the waveforms of Sec.~\ref{sec:waveform} for an eccentric and spin-precessing binary.

\subsection{The Newtonian amplitudes for an eccentric binary}
\label{sub:0PN amps}

The ingredients for calculating the amplitudes $N^{lm}_j$ are laid out in Ref.~\cite{Boetzel:2017}.
In that study,
the coefficients are found using PN-accurate representations of the relations between the QK parameters.
While our definition of the $\mathcal{P}^{mW}_s$ coefficients matches theirs,
our $\mathcal{K}^{lm}_s$ are defined as the Fourier coefficients of everything else that remains,
which can include products of other series as well.
Because of this, in order to explain how to construct these $N^{lm}_j$ amplitudes, we will first illustrate the procedure by carrying out the calculation explicitly at Newtonian (0PN) order below.

In the QK parameterization,
the Newtonian modes are~\cite{Mishra:2015}
\begin{subequations}
\label{eq:H2m_0}
\begin{align}
  \hat{H}^{20}_{(0)} &= \sqrt{\frac{2}{3}} \frac{e\cos u}{1-e\cos u},\label{eq:H20_0}\\
  \hat{H}^{22}_{(0)} &= \frac{\rme^{-2\rmi\phi}}{(1-e\cos u)^2} \left(2-2e^2 - e\cos u \,(1-e\cos u)\right. \nonumber\\
  &\left. +2\rmi e\sqrt{1-e^2} \sin u\right), \label{eq:H22_0}
\end{align}
\end{subequations}
where the $(0)$ subscript denotes that they are of 0PN order.
The $m=1$ mode vanishes at this order,
and the $m<0$ modes can be found by conjugation (see Sec.~\ref{sec:waveform}).

Let us now Fourier decompose these two Newtonian modes. In general, 
the base series expansions we will use here can be found in Ref.~\cite{Boetzel:2017}.
As a first example, consider 
the Fourier decomposition of Eq.~(\ref{eq:H20_0}), namely
\begin{align}\label{eq:H20_0 decomp}
  \hat{H}^{20}_{(0)} &= 2\sqrt{\frac{2}{3}} \sum_{j=1}^\infty J_j(je) \cos(j\ell),
\end{align}
where $J_n(x)$ is the $n$th-order Bessel function of the first kind. 
As there is no $W$ function for the $m=0$ modes,
we only need to find the decomposition of $K^{l0}$ in general.
To rewrite the decomposition into complex exponentials,
we use Euler's equation and extend the sum to $j=-\infty$ to find
\begin{equation}
\label{eq:H20_0 c decomp}
  \hat{H}^{20}_{(0)} = K^{20}_{(0)} = \sqrt{\frac{2}{3}} \sum_{j=-\infty}^\infty J_j(je) \rme^{-\rmi j\ell}, \quad (j\neq0).
\end{equation}
We thus identify the Fourier amplitudes $N^{20}_{(0) j} = \mathcal{K}^{20}_{(0) j} = \sqrt{2/3} \; J_j(je)$.

Let us now turn to the Fourier decomposition of Eq.~\eqref{eq:H22_0}, which we begin by writing as
\begin{equation}\label{eq:H22_0 predecomp}
  \hat{H}^{22}_{(0)} = K^{22}_{(0)} \sum_{j=-\infty}^\infty \mathcal{P}^{2W}_{(0)\,j} \rme^{-\rmi(j\ell + 2\lambda)}.
\end{equation}
The first step is to determine the $\mathcal{P}^{mW}_j$ coefficients that enter the $W$ decomposition through Eq.~\eqref{eq:Wdecomp} and the $\mathcal{K}^{lm}_s$ coefficients that enter the $K^{l m}$ decomposition through Eq.~\eqref{eq:Kdecomp}. We first focus on the $\mathcal{P}^{mW}_j$ coefficients.
At 0PN order, $W = \vartheta-\ell$, which in Eq.~(34) of Ref.~\cite{Boetzel:2017} is shown to yield the $\mathcal{P}^{2W}_{(0)\,j}$ coefficients,
\begin{equation}
\label{eq:Pmwj}
  \mathcal{P}^{mW}_{(0)\,j} = \sum_{k=0}^\infty \mathcal{P}^{mW}_{j,k} = \sum_{k=0}^\infty \mathcal{E}^{m}_k \epsilon^{ku}_{j+m},
\end{equation}
where
\begin{align}
  \label{eq:Emk}
  \mathcal{E}^m_k &=
  \begin{cases}
    (-\beta)^m, & k=0 \\
    \binom{k-1}{k-m}\, {}_2F_1(-m,k,k-m+1;\beta^2) \beta^{k-m}, & k>0
  \end{cases}
  \\
  \label{eq:epsks}
  \epsilon^{ku}_{n} &=
  \begin{cases}
    -\frac{1}{2} e \,\delta_{k1}, & n=0 \\
    \frac{k}{n} J_{n-k}(ne), & n>0
  \end{cases}
\end{align}
$\beta = \left( 1-\sqrt{1-e^2} \right)/e$,
$\binom{k}{m}$ is the binomial coefficient,
and ${}_2F_1(a,b,c;z)$ is the generalized hypergeometric function.
The hypergeometric function in Eq.~(\ref{eq:Emk}) can diverge when $m>k$ for any $\beta$,
but we find that its product with the binomial coefficient simplifies to a polynomial,
\begin{multline}
\label{eq:Emk polynomial}
    \binom{k-1}{k-m}\, {}_2F_1(-m,k,k-m+1;\beta^2) = \\
    m \sum_{n=0}^m \frac{(-1)^n (k+n-1)!}{(k-m+n)!\, (m-n)!\, n!} \beta^{2n},
\end{multline}
which is regular for all $m$ and $k$, as we show in Appendix \ref{app:Emk}.

Let us now focus on the $\mathcal{K}^{22}_s$ coefficients by specifically defining
\begin{equation}\label{eq:K220}
  K^{22}_{(0)} = \frac{2(1-e^2)}{(1-e\cos u)^2} - \frac{e\cos u}{1-e\cos u} + 2\rmi e \frac{\sqrt{1-e^2}\sin u}{(1-e\cos u)^2}.
\end{equation}
The first term can be decomposed as $2(1-e^2) \sum_m^\infty A^2_m$,
where we identify $A^p_m$ as the coefficients of the following Fourier decomposition,
\begin{equation}
    \frac{1}{(1 - e\cos u)^p} = \sum_{m=0}^\infty A^p_m \cos(m\ell).
\end{equation}
Expressions for these coefficients can be found up to 3PN order in Ref.~\cite{Boetzel:2017};
to 0PN order, we take the known decomposition for $p=1$,
\begin{equation}
    \frac{1}{1 - e\cos u} = 1+2\sum_{m=1}^\infty J_m(me) \cos(m\ell),
\end{equation}
and square it to get the $p=2$ coefficients.
The resulting products of series can be simplified using Eq.~(B16) of \cite{Pierro:2000} to yield the coefficients
\begin{subequations}\label{eq:A^2m}
\begin{align}
  A^2_{0} &= (1-e^2)^{-1/2},\\
  \begin{split}
  A^2_{m} &= 4 J_m + 2\left[\heavi(m-1) \sum_{k=1}^{m-1} J_k J_{m-k} \right.\\
  &\left.+ \sum_{k=m+1}^\infty J_k J_{k-m} + \sum_{k=1}^\infty J_k J_{k+m}\right],
  \end{split}
\end{align}
\end{subequations}
where $\heavi(\cdot)$ is the Heaviside step function,
and for convenience, we have suppressed the argument of the Bessel functions that match their index (e.g., $J_{m+n} \equiv J_{m+n}\left[(m+n)e\right]$).
While the decomposition of the middle term in Eq.~\eqref{eq:K220} is similar to that of Eq.~(\ref{eq:H20_0}),
the last term is found as another product of Fourier series,
\begin{align}
  \frac{\sin u}{(1-e\cos u)^2} &=
  \left(\frac{1}{1-e\cos u}\right) \left(\frac{\sin u}{1-e\cos u}\right) \nonumber\\
  &= \left(1+2\sum_{m=1}^\infty J_m(me) \cos(m\ell)\right) \nonumber\\
  &\times\left(2\sum_{n=1}^\infty J'_n(ne) \sin(n\ell)\right), \label{eq:K220products}
\end{align}
where $J'_n(x) \equiv \frac{d}{dx} J_n (x) = (J_{n-1}(x) - J_{n+1}(x))/2$.
Following Appendix D of Ref.~\cite{Tessmer:2010},
we can combine these series as
\begin{align}
  \frac{\sin u}{(1-e\cos u)^2} &= 2\sum_{n=1}^\infty (J'_n (ne) + I_n) \sin(n\ell),
\end{align}
where we have introduced the coefficients
\begin{align}
  \label{eq:In}
  \begin{split}
  I_n &= \heavi(n-1) \sum_{m=1}^{n-1} J_m J'_{n-m} \\ 
  &- \sum_{m=n+1}^\infty J_m J'_{m-n} + \sum_{m=1}^\infty J_m J'_{m+n},
  \end{split}
\end{align}
Combining all the series representations of Eq.~\eqref{eq:K220}, we can rewrite $K^{22}_{(0)}$ as a series of both cosines and sines, namely
\begin{align}\label{eq:K220series}
  K^{22}_{(0)} &= \sum_{m=0}^\infty \mathcal{K}^{\text{C}}_m \cos(m\ell) + \rmi \sum_{n=1}^\infty \mathcal{K}^{\text{S}}_n \sin(n\ell),
\end{align}
where
\begin{subequations}
\label{eq:Kcal}
\begin{align}
  \label{eq:KcalCm}
  \mathcal{K}^{\text{C}}_m &= 
  \begin{cases}
  2(1-e^2) A_0^2, & m=0\\
  2(1-e^2) A_m^2 - 2J_m(me), & m>0
  \end{cases}\\
  \label{eq:KcalSn}
  \mathcal{K}^{\text{S}}_n &= 4e\sqrt{1-e^2}(J'_n(ne) + I_n).
\end{align}
\end{subequations}
Note that the coefficients $A^2_m$ and $I_n$ are series themselves.

Ultimately, we wish to express $K^{22}_{(0)}$ as a series in complex exponentials.
Expanding the trigonometric functions in Eq.~(\ref{eq:K220series}) \`a la Euler,
we then write
\begin{equation}
  K^{22}_{(0)} = \sum_{j=-\infty}^\infty \mathcal{K}^{22}_{(0)\,j} \rme^{\rmi j\ell},
\end{equation}
where
\begin{align}
\label{eq:K22j}
  \mathcal{K}^{22}_{(0)\,j} =
  \begin{cases}
  \mathcal{K}^{\text{C}}_0, & j=0 \\
  \frac{1}{2} \left( \mathcal{K}^{\text{C}}_{|j|} + \operatorname{sign}(j) \mathcal{K}^{\text{S}}_{|j|} \right). & j\neq0
  \end{cases}
\end{align}
With this in hand, and using our results for the Fourier expansion of $\mathcal{P}^{mW}_j$ in Eq.~\eqref{eq:Pmwj}, we can rewrite Eq.~\eqref{eq:H22_0 predecomp} to finally have the Fourier expansion of $\hat{H}^{22}_{(0)}$:
\begin{equation}
  \hat{H}^{22}_{(0)} = \sum_{j=-\infty}^\infty N^{22}_{(0)\,j} \rme^{-\rmi(j\ell + 2\lambda)},
\end{equation}
where
\begin{equation}
\label{eq:finalN22j}
  N^{22}_{(0)\,j} = \sum_{s=-\infty}^\infty \mathcal{P}^{2W}_{(0)\,s} ~ \mathcal{K}^{22}_{(0)\,s-j}.
\end{equation}
The $\mathcal{P}^{2W}_{(0)\,s}$ and $\mathcal{K}^{22}_{(0)\,s-j}$ coefficients are found in Eq.~\eqref{eq:Pmwj} and \eqref{eq:K22j}, respectively.
A similar analysis applies for higher PN orders:
using the PN-accurate expressions in \cite{Boetzel:2017}, one finds the coefficients of the series in Eq.~(\ref{eq:P and K}) and combines them into the amplitudes $N^{lm}_j$ as in Eq.~(\ref{eq:Nlmj}) for each mode. We will not carry out this higher PN-order extension here, and instead, leave this for future work. 

\subsection{Truncating the coefficients inside the Fourier amplitudes}
\label{sub:truncating 1}

How then does one determine the number of terms to keep in the sum over $s$ when calculating each $N^{lm}_j$?
The answer may depend on the exact details of the mode under consideration,
but the general structure can be seen from our 0PN example above.
Within $N^{lm}_j$, we must determine the series expansions of $\mathcal{P}^{mW}_s$ and $\mathcal{K}^{lm}_{s-j}$.
Where we truncate these series depends on the size of their coefficients relative to each other's,
a size that we expect to decrease as more terms are added to the series if the total sums are to be finite.
To describe our procedure for truncating all of these series, we use the Newtonian modes defined in the section above.

To organize all these levels of series of series,
we will define their coefficients generically with the variable $\Qlm{j_1, j_2, ..., j_d}$, where $d$ is the depth of the series.
For example, at the top level we have only one variable, $\Qlm{j} = N^{lm}_j$, which depends on $\mathcal{P}^{mW}_s$ and $\mathcal{K}^{lm}_{s-j}$ through Eq.~\eqref{eq:finalN22j}.
Therefore, at the next level, we have $\Qlm{j,s} = \{ \mathcal{P}^{mW}_s, \mathcal{K}^{lm}_{s-j} \}$, but these coefficients themselves are also defined in terms of their own series expansions.
The coefficient $\mathcal{P}^{mW}_s$ is a series with coefficients $ \mathcal{P}^{mW}_{s,k}$ through Eq.~\eqref{eq:Pmwj}, while the coefficient $\mathcal{K}^{lm}_{s-j}$ is a series 
\begin{equation}
    \mathcal{K}^{lm}_{s-j} = \sum_{k=0}^\infty \mathcal{K}^{lm}_{s-j,k},
\end{equation}
where the coefficients $\mathcal{K}^{lm}_{s,k}$ are determined from Eqs.~\eqref{eq:K22j},~\eqref{eq:Kcal} and~\eqref{eq:A^2m};
we will explicitly define them in Eqs.~\eqref{eq:K0}-\eqref{eq:Kk} once we have identified the relative magntiudes of the coefficients.
Thus, at the third level we have $\Qlm{j,s,k} = \{ \mathcal{P}^{mW}_{s,k}, \mathcal{K}^{lm}_{s-j,k} \}$.

The number of terms that need to be kept in these series of series is difficult to determine analytically in general, so to gain some insight, we will consider a leading-order expansion in small eccentricity. For the $m=0$ case at Newtonian order, this expansion is not actually needed because the $N^{20}_j$ amplitudes are a single series that only depends on Bessel functions. Moreover, since $J_j(e) = \order{e^j}$ for $x \ll 1$,
each $j$th term decreases in size with respect to the previous term for small eccentricities.
On the other hand, when $m = 2$, the situation is more complicated. To evaluate the magnitude of the $\mathcal{K}^{22}_{s-j,k}$ coefficients, we first consider its components,
which are also composed of Bessel functions.
Again using the asymptotic behavior of $J_n(x)$,
we can identify the leading-order terms in small eccentricity of $A^2_m$ and $I_n$.
We then define the $k$th coefficient of the decomposition of $\mathcal{K}^{22}_s = \sum_k \mathcal{K}^{22}_{s, k}$ as
\begin{equation}\label{eq:K22sk}
  \mathcal{K}^{22}_{s, k} = \frac{1}{2} \left( \mathcal{K}^{\rm C}_{s,k} + \operatorname{sign}(j) \mathcal{K}^{\rm S}_{s,k} \right).
\end{equation}
To identify the $k=0$ terms,
we combine the leading-order terms with the terms that are not expanded in series; for example, in Eq.~\eqref{eq:KcalSn}, besides the series $I_n$ we also have the single term $J'_n$, which is independent of $k$.
Thus, the $k=0$ terms are
\begin{subequations}\label{eq:K0}
\begin{align}
  \mathcal{K}^{\rm C}_{s\neq0,0} &= 2(1-e^2) \left[4 J_s + 2\heavi(s-1)\sum_{n=1}^{s-1} J_n J_{s-n}\right] - 2 J_s,\\
  \mathcal{K}^{\rm S}_{s,0} &= 4e\sqrt{1-e^2} \left[J'_s + \heavi(s-1)\sum_{n=1}^{s-1} J_n J'_{s-n}\right],
\end{align}
\end{subequations}
and the $k>0$ terms are
\begin{subequations}\label{eq:Kk}
\begin{align}
  \mathcal{K}^{\rm C}_{s,k} &= 8(1-e^2) J_{s+k} J_k, \\
  \mathcal{K}^{\rm S}_{s,k} &= 4e\sqrt{1-e^2} \left[- J_{s+k} J'_k + J_k J'_{s+k}\right].
\end{align}
\end{subequations}
For $s=0$, there is no need to identify the coefficients $\mathcal{K}^{22}_{0,k}$ as we can see from Eq.~\eqref{eq:KcalCm} that $\mathcal{K}^{22}_0 = 2\sqrt{1-e^2}$.
This can be thought of as an instance of a series $\Qlm{j,s} = \sum_k \Qlm{j,s,k}$ where only one of its terms is nonzero.
This also occurs for $\mathcal{P}^{mW}_j$ when $j+m=0$,
as can be seen from the Kronecker-delta involved in Eq.~\eqref{eq:epsks}.
Such expressions can then be immediately evaluated and do not require truncation.

\begin{figure*}
    \centering
    \includegraphics[width=0.45\textwidth]{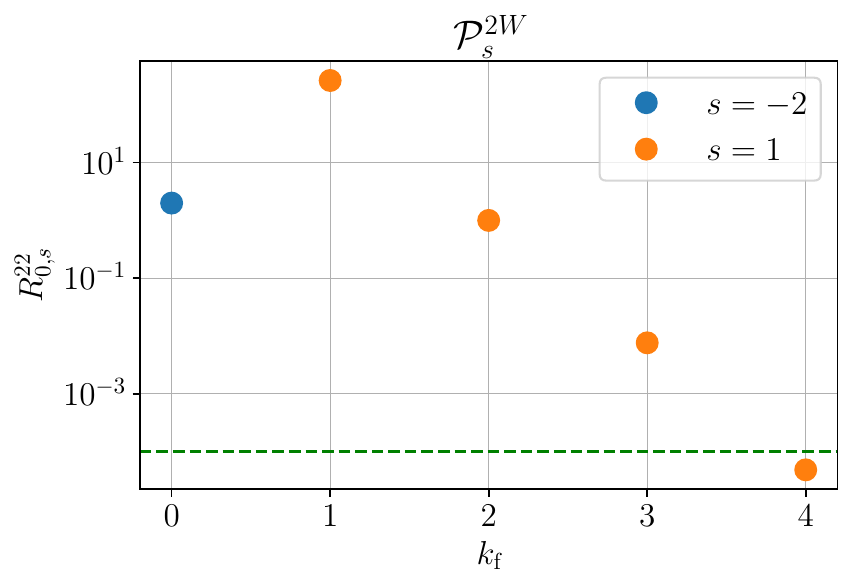}
    \includegraphics[width=0.45\textwidth]{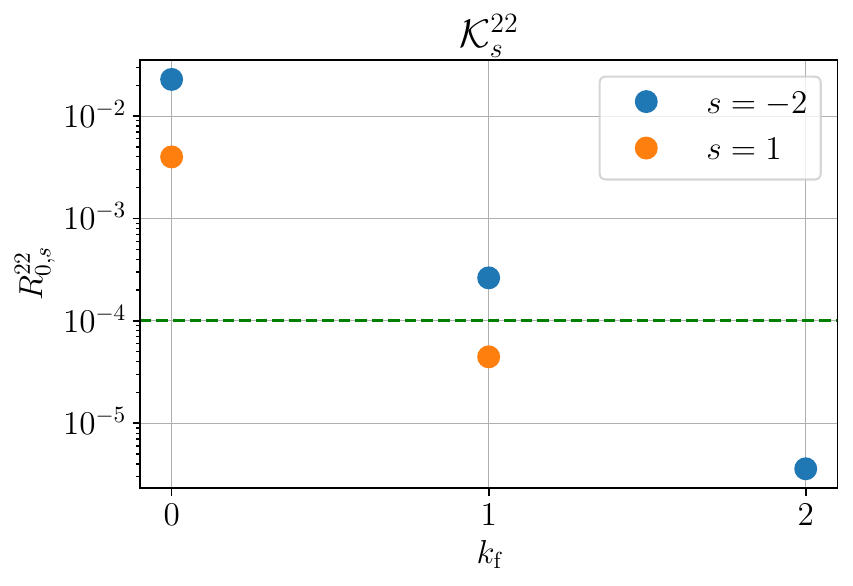}

    \caption{The ratios $R^{22}_{0,s}$ for the $\mathcal{P}^{2W}_s$ (left) and $\mathcal{K}^{22}_s$ (right) coefficients at $s=-2$ (blue circles) and $1$ (orange circles) as $k_{\rm f}$ is increased.
    The eccentricity for these coefficients is $e=0.1$.
    Once the ratio reaches the tolerance $\overline{R}^{lm}_{j,s} = 10^{-4}$ (green dashed line)
    we truncate the series at that $k_{\max} = k_{\rm f} + 1$ term.
    For $\mathcal{P}^{2W}_{-2}$, only the first two terms in its series expansion are non-zero ($R^{22}_{0,-2}(1) = 0)$, and so its truncation is immediately determined without needing to reach $\overline{R}^{lm}_{j,s}$.
    For $\mathcal{P}^{2W}_{1}$, there is no ratio calculated at $k_{\rm f}=0$ as the $k=0$ term is zero, so $R^{22}_{0,1}$ isn't calculated until $k_{\rm f} = 1$.}
    \label{fig:PKConv}
\end{figure*}

Ultimately, we seek to accurately represent a mode $H^{lm}$ as needed by our waveform.
To simplify matters,
we only need to consider $\overline{H}^{lm} \equiv \hat{H}^{lm} / \rme^{-\rmi m\lambda}$ as $\rme^{-\rmi m\lambda}$ is only a prefactor in the series expansion of $\hat{H}^{lm}$.
The first question we might ask in truncating the relevant series is,
how many $N^{lm}_j$ are required to accurately represent $\overline{H}^{lm}$?
However, in answering that question, we find that for each $j$ being explored  we must first answer how many $\mathcal{P}^{mW}_{s,k}$ and $\mathcal{K}^{lm}_{s-j,k}$ are required to calculate $N^{lm}_j$ precisely.
That is, for a specific $l$, $m$, $j$, and $s$,
we must first truncate the series $\Qlm{j,s} = \sum_{k} \Qlm{j,s,k}$.
We use the fact that, for small eccentricities, the magnitudes of $\Qlm{j,s,k}$ decrease as $k$ increases.
Starting with $k = k_{\rm f} = 0$,
we compare the effect of adding the $k = k_{\rm f}+1$ coefficient to its series by computing the ratio
\begin{equation}
\label{eq:Rlm_js}
  R^{lm}_{j,s}(k_{\rm f}) = \left| \frac{ \Qlm{j,s,k_{\rm f} + 1} }{ \sum_{k=0}^{k_{\rm f}} \Qlm{j,s,k} } \right|.
\end{equation}
Specifying a tolerance $ \overline{R}^{lm}_{j,s} $,
we increase $k_{\rm f}$ until the precision criterion $R^{lm}_{j,s} < \overline{R}^{lm}_{j,s}$ is satisfied,
therefore truncating the series at $k_{\max} = k_{\rm f} + 1$.
For $\overline{R}^{lm}_{j,s} = 10^{-4}$ and a sample eccentricity $e=0.1$, we illustrate this procedure in Fig.~\ref{fig:PKConv} for $j = 0$, $s = -2$ and $1$.
The series expansion of $\mathcal{P}^{2W}_{-2}$ has only two non-zero terms, so it is immediately determined,
while the expansion of $\mathcal{P}^{2W}_{1}$ requires $k_{\max} = 5$ to meet our chosen tolerance $\overline{R}^{lm}_{j,s} = 10^{-4}$.
Similarly for $\mathcal{K}^{22}_s$, we find $k_{\max} = 3$ for $s=-2$ and $k_{\max} = 2$ for $s=1$.
This analysis is done for all $s$ of interest; we now proceed to find which  $s$ terms need to be included in the series expansion of a specific $N^{lm}_j$.

\begin{figure}
    \centering
    \includegraphics[width=0.48\textwidth]{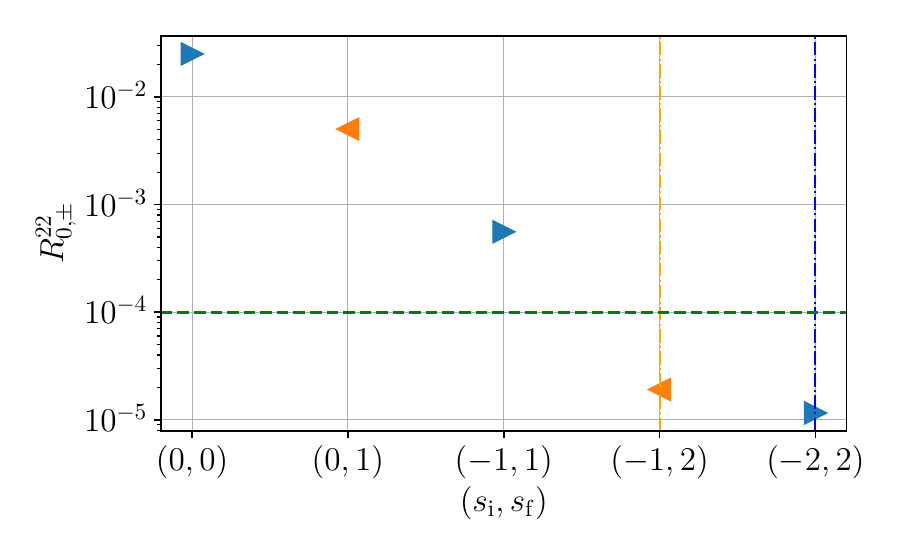}
    \caption{The ratio $R^{22}_0$ for the series expansion of $N^{22}_{(0)\, 0}$ as its bounds $(s_{\rm i}, s_{\rm f})$ are expanded.
    The bounds are identified for each point in the sequence along the horizontal axis.
    The eccentricity is $e=0.1$, as in Fig.~\ref{fig:PKConv}.
    The direction in which the bounds are being expanded at each step is denoted by the color and direction of the triangular markers in the plot;
    blue and towards the right for $s_{\rm f}+1$, orange and towards the left for $s_{\rm i}-1$.
    Once the ratio in a certain direction falls below $\overline{R}^{lm}_j = 10^{-4}$ (green dashed line) the bound is fixed.
    The vertical dot-dashed lines mark where this truncation occurs in the direction denoted by its color.
    The expansion of $N^{22}_{(0)\, 0}$ is thus truncated at $(s_{\min}, s_{\max}) = (-2, 3)$.}
    \label{fig:N220Conv}
\end{figure}

Above, we described the procedure for finding the necessary terms in the expansion of the $\Qlm{j,s}$ coefficients for $l=2=m$, $j=0$, and $s = -2$ and $1$.
The reason we calculate the coefficients for these specific values of $s$ is that they arise when analyzing the expansion of $N^{22}_{(0) 0}$,
which is the first value of $j$ for which we start analyzing the expansion of $\overline{H}^{22}_{(0)}$.
How we determine the expansion of the Fourier amplitudes,
for any $j$,
follows the same reasoning as for the $\Qlm{j,s}$ from above.
We begin at $s=j$,
where the product $\mathcal{P}^{mW}_s \mathcal{K}^{lm}_{s-j}$ generally peaks in magnitude,
and analyze how adding terms towards $s\to \infty$ and $s\to -\infty$ adds to the series expansion of $N^{lm}_j$.
Thus, the ratio test we performed above must now be done in two directions.
In expanding the range of $s$ we generate a sequence of lower and upper bounds, $\{(s_{\rm i}, s_{\rm f})\}$.
Beginning at $s_{\rm i} = j = s_{\rm f}$,
we first step once in the positive direction and calculate the ratio
\begin{equation}
    R^{lm}_{j,+}(s_{\rm i}, s_{\rm f}) = \left| \frac{ \Qlm{j,s_{\rm f}+1}}{ \sum_{s=s_{\rm i}}^{s_{\rm f}} \Qlm{j,s} } \right|.
\end{equation}
We then compare this to a chosen tolerance $\overline{R}^{lm}_j$.
If the criterion $R^{lm}_{j,+} < \overline{R}^{lm}_j$ is satisfied,
we stop adding terms in the positive $s$ direction,
truncating the series at $s_{\max} = s_{\rm f} + 1$.
Whether or not the criterion is satisfied at this step,
we set the new upper bound $s_{\rm f} \to s_{\rm f}+1$ for the next step and then step in the negative direction to calculate 
\begin{equation}
    R^{lm}_{j,-}(s_{\rm i}, s_{\rm f}) = \left| \frac{ \Qlm{j,s_{\rm i}-1}}{ \sum_{s=s_{\rm i}}^{s_{\rm f}} \Qlm{j,s} } \right|.
\end{equation}
If the criterion $R^{lm}_{j,-} < \overline{R}^{lm}_j$ is satisfied,
we stop adding terms in the negative $s$ direction,
truncating the series at $s_{\min} = s_{\rm i} - 1$.
Whether or not the criterion is satisfied, we then set the new lower bound $s_{\rm i} \to s_{\rm i}-1$.
Alternating between expanding in positive and negative directions,
we will eventually satisfy the criteria for both,
converging upon a value for $N^{lm}_j = \sum_{s=s_{\min}}^{s_{\max}} \mathcal{P}^{mW}_s \mathcal{K}^{lm}_{s-j}$.

For the same sample eccentricity as above, Fig.~\ref{fig:N220Conv} shows this procedure for $N^{22}_{(0) 0}$,
where we determine the bounds $(s_{\min}, s_{\max}) = (-2, 3)$.
At each new value of $s$, we must perform the truncation of the expansion of the $Q^{lm}_{j,s}$ coefficients as described previously.
In fact, we showed how the series expansion of the $s = -2$ coefficients was truncated in Fig.~\ref{fig:PKConv}.
The analysis performed here is to be done for every $j$ explored in the series expansion of $\overline{H}^{lm}$,
whose truncation we determine in the following section.

\subsection{Truncating the Fourier amplitudes}
\label{sub:truncating 2}

Now, we must determine the range $j_{\min} \leq j \leq j_{\max}$ for each mode.
This differs from the previous analyses as the $\overline{H}^{lm}$ are Fourier series in time;
while we truncated $\Qlm{j,s}$ and $\Qlm{j}$ based on their \emph{convergence},
the expansion of $\overline{H}^{lm}$ is to be truncated based on its \emph{accuracy},
as one can directly compare the mode to its Fourier expansion.
Parameterizing time through the mean anomaly $u$,
we perform this comparison over an entire orbit $u \in [0, 2\pi)$.
We define the following L2-norm to carry out this comparison:
\begin{equation}
    \norm{f} \equiv \sqrt{\frac{1}{2\pi} \int_0^{2\pi} \left[f(u)\right]^2\, du}
\end{equation}
for some function $f(u)$, and thus define the error between the mode and its series expansion as
\begin{equation}
\label{eq:Deltalm}
    \Delta^{lm}(j_{\min},j_{\max}) = \norm{\overline{H}^{lm} - \sum_{j=j_{\min}}^{j_{\max}} N^{lm}_j \rme^{-\rmi j\ell}},
\end{equation}
where the series is bounded by $(j_{\min},j_{\max})$, and $\ell$ is related to $u$ through Eq.~(\ref{eq:lofu}) (which to Newtonian order is simply $\ell = u - e\sin u$).
The \emph{relative error} is then
\begin{equation}
\label{eq:Rlm}
    R^{lm}(j_{\min},j_{\max}) = \frac{ \Delta^{lm}(j_{\min},j_{\max}) }{ \norm{\overline{H}^{lm}} }.
\end{equation}

Unlike the previous analysis where the ratio test was performed in both directions, however,
we must be efficient in calculating Eq.~\eqref{eq:Deltalm} as the series becomes increasingly oscillatory as more terms are added to it---a situation that arises for high eccentricities.
The integral in the L2-norm is performed numerically and thus can become computationally expensive.
Fortunately, we have confirmed by inspecting the behavior of the Fourier amplitudes at small eccentricities that the magnitude of $N^{22}_j$ for $j>2$ is usually larger by at least an order of magnitude than that of $N^{22}_{-j}$.
Therefore, as we expand $j_{\min}$ towards negative integers and $j_{\max}$ towards positive integers, if at some point $\order{N^{lm}_{j_{\min}}} < \order{N^{lm}_{j_{\max}}}-1$,
we can ignore amplitudes from the negative direction until they become comparable to those from the positive direction.

\begin{figure}
    \centering
    \includegraphics[width=0.48\textwidth]{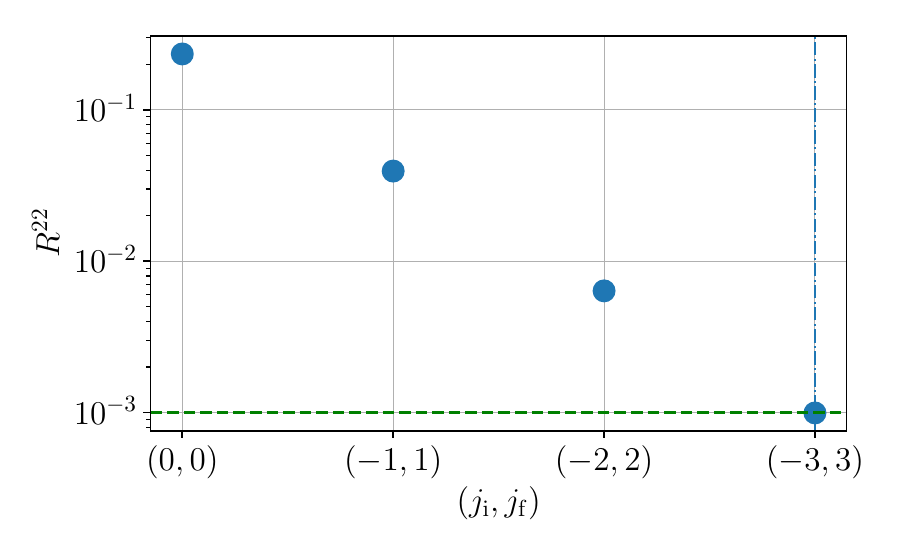}
    \caption{The relative error $R^{22}$ for $\overline{H}^{22}_{(0)}$ as the bounds $(j_{\min},j_{\max})$ of its series are expanded,
    again for $e=0.1$.
    The bounds $(-3,3)$ are determined once the error falls below the tolerance $\overline{R}^{lm} = 10^{-3}$ (green dashed line),
    as indicated by the vertical dot-dashed line.}
    \label{fig:H220Conv}
\end{figure}

Thus, we perform the following procedure to efficiently truncate the series:
\begin{enumerate}
    \item We start by setting $j_{\min} = 0 = j_{\max}$ and setting a tolerance $\overline{R}^{lm}$.
    \item If $R^{lm}(j_{\min},j_{\max})$ is not less than $\overline{R}^{lm}$, we expand the bounds in both directions by adding the $j_{\max} + 1$ and $j_{\min} - 1$ terms to the series and then computing the new $R^{lm}$.
    \item Step 2 is repeated until either the error falls below the tolerance or $j>3$.
    \item In case of the latter,
    the upper bound is set to $j_{\max} \to j_{\max} + 1$ and the new term is added to the series.
    \item We compare the order of magnitude of this new $N^{lm}_{j_{\max}}$ to that of the previous $N^{lm}_{j_{\min}}$.
    If $\order{N^{lm}_{j_{\min}}} < \order{N^{lm}_{j_{\max}}}-1$, we do not compute $N^{lm}_{j_{\min}-1}$, leaving the lower bound as is.
    Otherwise, the lower bound is set $j_{\min} \to j_{\min} - 1$ and its term is added to the series.
    \item The new $R^{lm}$ is computed.
    \item Steps 4-6 are repeated until the error falls below the tolerance.
    During this time, it may be that the order of magnitude of the amplitude from the positive direction decreases to almost that of the last amplitude from the negative direction.
    In that case, step 5 would resume adding terms in that latter direction.
\end{enumerate}
Figure \ref{fig:H220Conv} shows the error found at each of these steps for the Newtonian $(l,m) = (2,2)$ mode for $e=0.1$,
yielding the truncation $(j_{\min},j_{\max}) = (-3,3)$ for $\overline{R}^{lm} = 10^{-3}$.
Thus, a total of 7 harmonics are sufficient to represent this mode to 0.1\% accuracy at this eccentricity.

\subsection{Summary of the truncation procedure}

\begin{figure*}[!htb]
    \centering
    \includegraphics[width=0.9\textwidth]{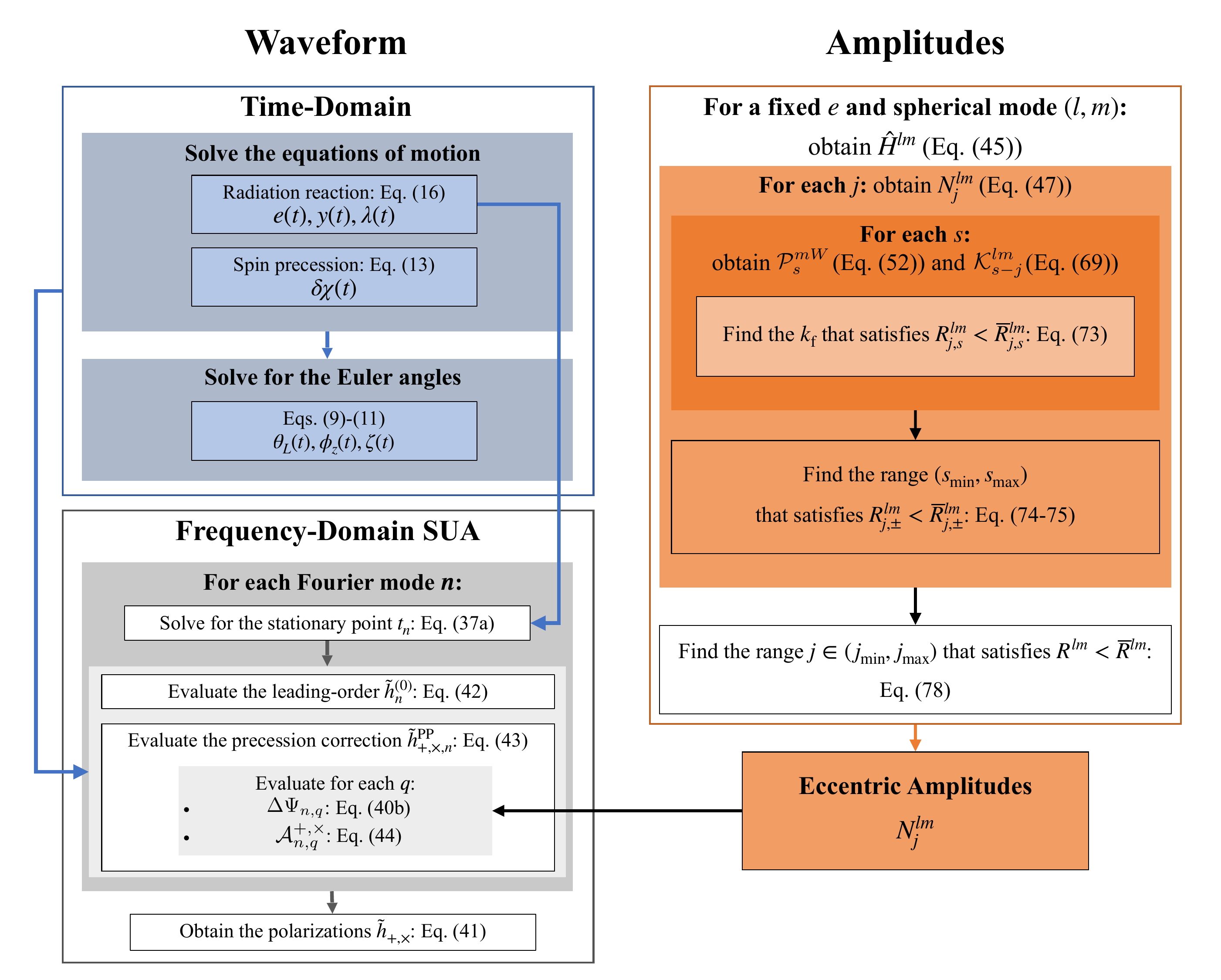}
    \caption{Schematic diagram of the construction of the waveform (left) and Fourier amplitudes (right).
    Each arrow points from one step to the one it informs;
    for example, the amplitudes $N^{lm}_j$ are part of the waveform amplitudes $\mathcal{A}^{+,\times}_{n,q}$ (see Eq.~(\ref{eq:A+x})).
    The limits of the series representation of the amplitudes can be tabulated beforehand following the procedure on the right (see Sec.~\ref{sec:amplitudes}) and then used to generate the waveform described in Sec.~\ref{sec:waveform}.
    }
    \label{fig:construction}
\end{figure*}

We summarize the procedure we have detailed above:
starting with bounds $j_{\min} = 0 = j_{\max}$, we calculate the relative error $R^{lm}(j_{\min},j_{\max})$ for a mode $(l,m)$ and expand the bounds until the error falls below a tolerance $\overline{R}^{lm}(j_{\min},j_{\max})$.
For each $j$ explored, the series expansion of $N^{lm}_j$ has to be truncated,
which means expanding the bounds $(s_{\rm i},s_{\rm f})_j$ until the expansion's precision $R^{lm}_{j,\pm}(s_{\rm i}, s_{\rm f})$ falls below a tolerance $\overline{R}^{lm}_{j}$.
This, in turn, requires the calculation of its components $\mathcal{P}^{mW}_s \mathcal{K}^{lm}_{s-j}$, which are also defined as series with terms $\mathcal{P}^{mW}_{s,k}$ and $\mathcal{K}^{lm}_{s-j,k}$,
whose respective truncations $k_{\max}$ are then found using the precision criterion $R^{lm}_{j,s} < \overline{R}^{lm}_{j,s}$.
For each $j \in [j_{\min}, j_{\max}]$, then, there is a list of ranges $\{ (s_{\min},s_{\max})_j \}$,
and for each $s \in [s_{\min}, s_{\max}]$, there are two sets of $\{(k_{\max})_{j,s}\}$ each for the truncations of $\mathcal{P}^{mW}_s$ and $\mathcal{K}^{lm}_{s-j}$.

The construction described above is illustrated schematically in Fig.~\ref{fig:construction},
where we also sketch at which point the amplitudes enter the waveform.
As one can discern from the multiple steps taken in building these amplitudes,
i.e., in finding the ranges of the multiple series involved for each mode for a specific eccentricity,
the process is computationally intense.
We thus perform this construction \textit{before} generating waveforms,
tabulating the ranges needed for sample values of eccentricity $e_i$ in \texttt{Mathematica}\footnote{The numbers $\Qlm{j,s,k}$ and $\Qlm{j,s}$ were evaluated with a numerical precision of 16 digits in our procedure. The L2-norms were evaluated with the function \texttt{NIntegrate}. For the $m=2$ mode L2-norm, we used an \texttt{AccuracyGoal} and \texttt{PrecisionGoal} of 4 digits with a \texttt{WorkingPrecision} of 8 digits for $e \leq 1/2$, while for $e > 1/2$, the \texttt{PrecisionGoal} was set to 10. For the $m=0$ mode, we set the \texttt{AccuracyGoal} $=\infty$ with a \texttt{PrecisionGoal} of 7 digits and a \texttt{WorkingPrecision} of 16 across all sampled eccentricities. These choices were made to make the integration efficient without losing the precision needed to meet our tolerances.}.
When the waveform calls the amplitudes for an arbitrary $e$,
we use the set of ranges found for the tabulated value of $e_i$ right above it to return the accurate amplitude.
For example, if one calls for the amplitudes for an eccentricity that falls between two neighboring tabulated eccentricities $e_{i-1} < e_i$,
the ranges for $e_i$ are returned,
as the $e_{i-1}$ ranges are guaranteed to be accurate only for $e \leq e_{i-1}$.
We set our accuracy goal $R^{lm} = 10^{-3}$ such that the series expansions are thus 0.1\% accurate,
with a precision goal $R^{lm}_j = R^{lm}_{j,s} = 10^{-4}$.
In Table \ref{tab:limits}, we list the limits of $j$ found for the Newtonian modes for our sampled eccentricities.
This includes $m=0$, which from our previous discussion has a simple series expansion that from Eq.~\eqref{eq:H20_0} can be seen to have symmetric bounds $j_{\max} = -j_{\min}$.
We observe for $m=2$ that while at low eccentricities the bounds are symmetric around $j=0$,
this is not the case for higher eccentricities,
where more positive $j$ terms are kept than negative $j$.
The $j$, $s$, and $k$ bounds for the Newtonian modes can be found in our supplementary material \cite{EccentricAmplitudes}, where we provide a \verb|Mathematica| notebook to retrieve them for any eccentricity up to our maximum tabulated eccentricity, $e=0.8$, as beyond this the convergence behaves poorly for $m=2$.

\begin{table}[b]
\caption{The top-level series truncations found for the expansions of the $(l=2, m=0,2)$ Newtonian modes at our tabulated eccentricities $e_i$.
For $m=0$, we list the bounds $j_{\max} = -j_{\min}$,
while for $m=2$, we list $(j_{\min}, j_{\max})$.
Our relative error tolerance is $\overline{R}^{lm} = 10^{-3}$ while the precision tolerances for the underlying series are $\overline{R}^{lm}_{j,s} = \overline{R}^{lm}_j = 10^{-4}$.
The limits for the lower-level series can be found in the supplementary material.}
\label{tab:limits}
\begin{ruledtabular}
\begin{tabular}{ccc}
    $e_i$ & $m=0$ & $m=2$ \\
    \hline
    0.01 &  2 & $(-1,1)$ \\
    0.10 &  4 & $(-3,3)$ \\
    0.15 &  4 & $(-4,5)$ \\
    0.20 &  5 & $(-5,6)$ \\
    0.25 &  6 & $(-6,7)$ \\
    0.30 &  7 & $(-6,9)$ \\
    0.35 &  8 & $(-7,10)$ \\
    0.40 & 10 & $(-8,12)$ \\
    0.45 & 11 & $(-10,15)$ \\
    0.50 & 13 & $(-11,18)$ \\
    0.55 & 16 & $(-13,22)$ \\
    0.60 & 20 & $(-16,27)$ \\
    0.65 & 24 & $(-20,35)$ \\
    0.70 & 31 & $(-25,44)$ \\
    0.75 & 42 & $(-32,59)$ \\
    0.80 & 59 & $(-45,90)$
\end{tabular}
\end{ruledtabular}
\end{table}

With the amplitudes thus determined,
the GW modes are accurately represented for moderate eccentricities in the time domain.
We now describe briefly the implementation of the eccentric amplitudes into the framework of the EFPE model for their use in the frequency domain.

\subsection{Implementation of the model}

With the number of Fourier amplitudes tabulated above, we can now correctly truncate the sum over $n$ in the frequency-domain waveform, Eq.~\eqref{eq:h+xf}.
Considering the indices in Eq.~\eqref{eq:A+x}, we see that the limits on $j$ found in the previous section depend on the limits of $q$ and the mode number $m'$.
To match the phase convention of Ref.~\cite{Klein:2021}, we first must make the substitutions $q\to-q$ and $n\to-n$ in the indices of the Fourier amplitudes such that
\begin{equation}
    N^{l,q+n}_{-q} \to N^{l,-q-n}_q.
\end{equation}
The relation between the mode number and these indices is now $n = -q-m'$.
As $q=j$ in this notation, we can then identify the limit on $n$ in the waveform:
\begin{equation}\label{eq:nmax}
    n_{\max} = \max(-q-m') = l_{\max} - j_{\min},
\end{equation}
as $j_{\min} \leq 0$ and $l_{\max}$ is the highest mode included in the model.
Incorporating these changes into Eq.~\eqref{eq:h+xf}, the form of the waveform is thus
\begin{equation}
    \tilde{h}_{+,\times}(f) = \sum_{n=1}^{n_{\max}} \Tilde{h}_n^{(0)} \Tilde{h}_{+,\times,n}^{\rm PP},
\end{equation}
where
\begin{align}
\begin{split}
    \Tilde{h}_{+,\times,n}^{\rm PP}(f) &=
    \sum_{k=-k_\mathrm{max}}^{k_\mathrm{max}} \sum_{q=q_{\min}}^{q_{\max}} a_{k,k_\mathrm{max}} \rme^{\rmi\Delta\Psi_{n,q}} \\
    &\times \overline{\mathcal{A}}^{+,\times}_{n,q}(t_n + \Delta t_{n,q} + kT_n),
\end{split}\\
    \overline{\mathcal{A}}^{+,\times}_{n,q} &= \hat{W} \sum_{l=2}^{l_{\max}} \sum_{m=-l}^{l} \mathsf{A}^{+,\times}_{\bm{l}} N^{l,-q-n}_{q},
\end{align}
and $q_{\min} = -(n+l_{\max})$, $q_{\max} = -(n-l_{\max})$.
Here we have used a one-sided Tukey window 
\begin{equation}
    \hat{W}(t) =
    \begin{cases}
        0, & t<t_1 \\
        1, & t_1 \leq t < t_2 \\
        \sin^2 \left( \frac{\pi}{2} \frac{t_3-t}{t_3 - t_2} \right), & t_2 \leq t<t_3 \\
        0, & t_3 \leq t
    \end{cases}
\end{equation}
which was also used in Ref.~\cite{Klein:2021} to reduce spectral leakage from the abrupt ending of the waveform.
The points in time are chosen such that $t_3$ is the end of the integration time for the binary or the point in its evolution where it reaches the innermost stable circular orbit ($y=1/\sqrt{6}$), whichever occurs first, and $t_2$ is 10 orbital cycles before $t_3$.
We thus obtain our GW polarizations with our new amplitudes.

The question remains of when $n_{\max}$, as defined in Eq.~\eqref{eq:nmax}, should be determined, as it depends on $j_{\min} = j_{\min}(e)$.
One may suggest that at each frequency $f$ being sampled, a relation between frequency and eccentricity be used such that $j_{\min}(e) = j_{\min}[f(e)]$.
However, this relation is multi-valued by the stationary point condition, which even to leading order (Eq.~\eqref{eq:tn}) is harmonic-dependent.
Within the EFPE framework, we opt to use the largest eccentricity for a system of interest: the binary's initial eccentricity $e_0$, so $n_{\max} = n_{\max}(e_0)$ is set from the beginning and guarantees the amount of harmonics needed even as the binary circularizes.
Unfortunately, this increases the evaluation time of a waveform considerably for large $e_0$.
We leave finding a more efficient choice for future work, as this implementation suffices for the purpose of validating the eccentric amplitudes.
We emphasize that the work here only changes the Fourier amplitudes of the EFPE model, leaving the binary dynamics and Fourier transform itself intact.
To distinguish this implementation from the EFPE model, we denote the model with our amplitudes as the \textit{EFPE\_ME} model, as it covers moderate eccentricities $e \leq 0.8$.
In the next section, we test the accuracy of these amplitudes throughout an orbit and compare the frequency-domain EFPE and EFPE\_ME models.


\section{Comparisons between amplitudes}
\label{sec:comparisons}
We will now explore the improvement of our eccentric amplitudes over the small-eccentricity-expanded amplitudes  $G^{lm}_j$ of Ref.~\cite{Klein:2018}.
In that study, the series expansions are available up to $\order{e^6}$, yielding an expansion with bounds $j_{\max} = 6 = -j_{\min}$ for all eccentricities.
In the sections that follow, we will compare our eccentric amplitudes against these,
first for fixed eccentricities using the relative error defined in Sec.~\ref{sub:truncating 2},
and second for inspirals of sample binaries in the frequency band of LISA.
The former will demonstrate the accuracy of our amplitudes across the space of eccentricity, while the latter will highlight the regions of the parameter space that benefit from including these eccentric amplitudes.

\subsection{Time-domain comparisons}
\label{sub:TD}

\begin{figure*}[!htb]
    \centering
    \includegraphics[width=0.75\textwidth]{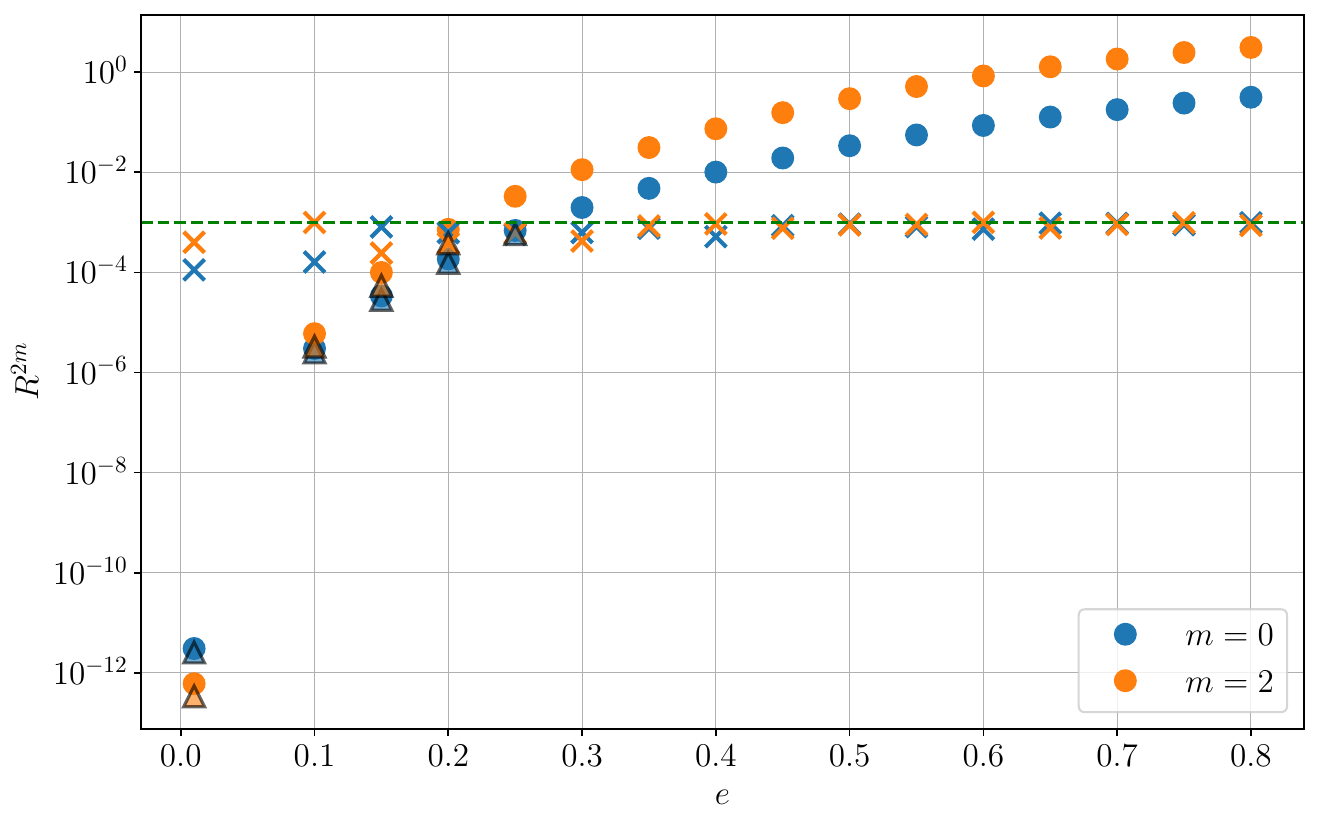}
    \caption{The relative error of the small-eccentricity-expanded amplitudes of Ref.~\cite{Klein:2018} (circles) and the amplitudes presented in Sec.~\ref{sec:amplitudes} (crosses).
    In blue and orange are the errors for $m=0$ and $2$, respectively.
    Our amplitudes maintain an accuracy of 0.1\% as marked by the green dashed line while the $\order{e^6}$ amplitudes do so only for $e \leq 0.25$ ($0.2$) $\equiv e_{\rm match}$ for $m=0$ ($2$).
    To match the high accuracy at low eccentricities, we truncate our amplitudes for $e \leq e_{\rm match}$ at $j_{\max} = 6 = -j_{\min}$ (see Sec.~\ref{sub:TD}).
    The colored triangles outlined in black show how this truncation matches the accuracy of the $\order{e^6}$ amplitudes.}
    \label{fig:L2}
\end{figure*}

As a first test of our amplitudes,
we can compare the accuracy of the eccentricity-expanded modes of Ref.~\cite{Klein:2018} to our modes for the eccentricities listed in Table \ref{tab:limits}.
We calculate the relative error of the $\order{e^6}$ series by replacing $N^{lm}_j$ in Eq.~\eqref{eq:Deltalm} with the $G^{lm}_j$ and calculate their $R^{lm}$ as in Eq.~\eqref{eq:Rlm}.
We compare those errors to those of our amplitudes in Fig.~\ref{fig:L2},
where the former are plotted in circles while the latter are in crosses.
One can see that the $\order{e^6}$ amplitudes accurately represent the modes at low eccentricities before exceeding the threshold of 0.1\% for $e \gtrsim 0.2$,
while our amplitudes maintain this accuracy across our space of eccentricity.
At the highest tabulated eccentricity,
our modes are two orders of magnitude more accurate than the eccentricity-expanded modes.
For high eccentricities then, our amplitudes allow for a more accurate representation of the modes than an expansion in small eccentricity does.
To maintain the low-eccentricity accuracy of the $\order{e^6}$ amplitudes, we set the $N^{lm}_j$ amplitudes to return the $e_{\rm match}=0.25$ ($e_{\rm match}=0.2$) limits for the $m=0$ ($m=2$) modes for $e \leq e_{\rm match}$.
The colored triangles outlined in black in Fig.~\ref{fig:L2} demonstrate how the accuracy of this truncation matches well with the $\order{e^6}$ amplitudes.
We adopt this truncation for the remainder of our analysis beginning in the next section, but we first make one more comparison between the amplitudes within the time-domain.

\begin{figure*}[!htb]
    \centering
    \includegraphics[width=\textwidth]{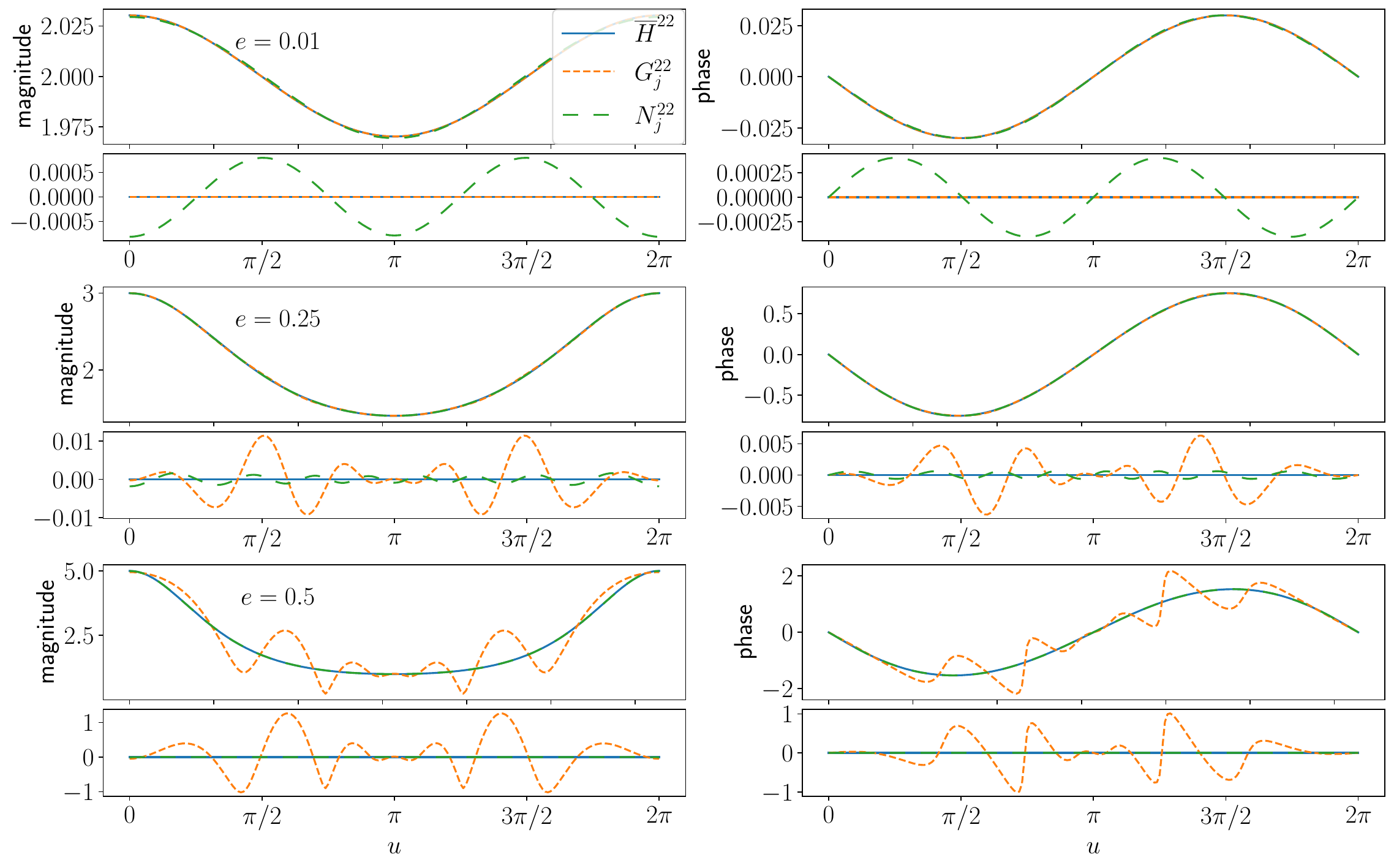}
    \caption{The reduced mode $\overline{H}^{22}_{(0)}$ (blue line) and its series expansions with Fourier amplitudes $G^{22}_j$ (orange dashes) and $N^{22}_j$ (green dashes).
    The left column plots the mode's magnitude while the right plots its phase.
    We show this mode for three different eccentricities in each row, with $e=0.01$ at the top, $e=0.25$ in the middle, and $e=0.5$ at the bottom.
    Underneath each magnitude and phase plot we show the difference between the series expansions and $\overline{H}^{22}_{(0)}$.}
    \label{fig:TD22}
\end{figure*}

We can analyze the behavior of these different representations of the modes throughout a Newtonian orbit.
Figure \ref{fig:TD22} shows the reduced $m=2$ mode, $\overline{H}^{22}_{(0)}$, for a low eccentricity ($e=0.01$), a moderate eccentricity ($e=0.25$), and a higher eccentricity ($e=0.5$) Newtonian orbit.
Plotted with it are its series expansions using the $\order{e^6}$ amplitudes and our amplitudes,
along with the difference of these two against the mode.
Being a complex number, we show both its magnitude and its phase.
As expected from their relative errors,
at low eccentricities the $\order{e^6}$ amplitudes represent the mode more accurately than ours,
but that accuracy is quickly lost at the moderate eccentricity,
where there aren't enough harmonics in the series to approximate the mode accurately.
Instead, our expansion contains the necessary amount of harmonics needed to accurately capture not only the magnitude of the mode, but also its phase, especially at high eccentricities.
Although we do not plot it here, similar behavior is found for $m=0$.

In this section, we have validated our eccentric amplitudes for fixed eccentricities in the time domain, demonstrating their superior accuracy to the small eccentricity expansion adopted in Ref.~\cite{Klein:2021}.
As we have not modified the dynamics of the binary introduced in Sec.~\ref{sub:dynamics}, the addition of radiation and spin-precession will not affect this accuracy as the binary inspirals and its eccentricity evolves.
The remaining question then is how well the effects of our eccentric amplitudes may be captured by a GW detector, an analysis that is performed in the frequency domain.
We now turn to answering this question in the following section.

\subsection{Overlap comparisons}
\label{sub:overlaps}

In the previous section, we validated our eccentric amplitudes in the time domain against both the non-decomposed modes (Eq.~\eqref{eq:H2m_0}) and the decomposed modes of Ref.~\cite{Klein:2018} expanded in powers of eccentricity.
While this comparison demonstrates the controlled accuracy of our eccentric amplitudes with the modes, we must now go to the frequency domain to quantify their effect on the detection of eccentric and spin-precessing binaries.
In this section, we characterize this effect through the \emph{overlap} between the small-eccentricity-expanded amplitudes and our eccentric amplitudes, which will tell us how well these two representations can be distinguished.
The smaller the overlap, the more significant the effect of eccentricity on the amplitudes may be for characterizing these binaries.

Let us now define the two frequency-domain models to be used in this comparison.
The EFPE model of Ref.~\cite{Klein:2021} is the basis of both models, which solves the evolution of a binary in the time-domain to approximate the frequency-domain waveform with the SUA method as described in Sec.~\ref{sec:waveform}.
The EFPE\_ME model, however, contains enough terms in the harmonic decomposition of the modes to accurately represent them across a range of eccentricities, pushing the boundary of the inspiral model to higher eccentricities than before without expanding in small eccentricity.
The inclusion of a different amount of harmonics than the EFPE's maximum of $n_{\max} = 8$ at 0PN (at least 2 harmonics as there are $|m'|=0,2$ modes, 6 harmonics added as the amplitudes were expanded as far as $\order{e^6}$) also changes the truncation of the $n$ harmonics.
In Sec.~\ref{sub:TD}, we described how we set the limits on $j$ in order to match the EFPE's accuracy at eccentricities below $e_{\rm match}$;
at larger eccentricities, we use the limits listed in Table \ref{tab:limits}, which then inform our $n_{\max}$ as defined in Eq.~\eqref{eq:nmax}.

To compare waveforms,
we compute their \emph{overlap} in the LISA frequency band,
which we take to have a lower limit of $f_{\min} = 2 \times 10^{-5}$ Hz and an upper limit $f_{\max} = 1$ Hz.
We compute LISA's response through time-delay interferometry (TDI), which projects the polarizations presented in Sec.~\ref{sec:waveform} onto the responses of the observatory's three arms and takes combinations of them to mitigate fluctuations in the laser frequency \cite{Tinto:2004}.
In particular, we compute the A, E, and T modes in the rigid-arm approximation, yielding three responses in the frequency domain, $\tilde{h}_{i,N}$ ($N =$ A, E, T), from the $i$th gravitational-wave signal.
The overlap between two signals $h_1$ and $h_2$ is then
\begin{equation}
    \mathrm{O} = \frac{\inner{h_1}{h_2}}{\sqrt{\inner{h_1}{h_1} \inner{h_2}{h_2}}},
\end{equation}
where we use the inner product
\begin{equation}
    \inner{h_1}{h_2} =
    4\Re \left(
        \int_{f_{\min}}^{f_{\max}} df
        \sum_{N}
        \frac{\tilde{h}_{1,N}(f)\, \tilde{h}_{2,N}^*(f)}{S_{N}(f)}
    \right),
\end{equation}
with the noise spectral density in each channel $S_{N}$.
The overlap is equal to 1 when the waveforms are completely the same and are less than 1 when they are different; how much the overlap differs from 1 tells us how well the signals are distinguished by the detector.
We calculate the inner products using a Clenshaw-Curtis quadrature scheme after changing the integration variable to $\log f$.
Having defined our overlap, we now use it to quantify the difference between the EFPE and EFPE\_ME waveforms.

We generate waveforms for equal-mass binaries with component masses $m_1 = m_2 = 100, 10^3, 10^4,$ and $10^5$ M$_{\odot}$.
For each of these binaries, we sample across the range of eccentricities covered by the EFPE\_ME, $e \leq 0.8$, and set them as the initial eccentricity $e_0$, i.e.~the eccentricity at the beginning of our simulations, defined at 4 years before merger.
Across all of these systems, we set the remaining initial conditions as follows:
\begin{itemize}
    \item In a fixed $\hat{x}$-$\hat{y}$-$\hat{z}$ frame, the orbital angular momentum is inclined from the vertical $\hat{z}$-axis by $\pi/3$ radians and rotated from the $\hat{x}$-axis by $\pi/6$.
    \item In the same frame, both BH spins point vertically upwards.
    With spin magnitudes $\chi_1 = 0.5$ and $\chi_2 = 0.3$, we obtain systems with $\chi_{\rm eff} \approx 0.28$.
    This also yields an initial effective precession spin parameter $\chi_{\rm p} \approx 0.35$, which approximately measures the largest BH spin component lying within the orbital plane \cite{Schmidt:2014}.
    \item The initial mean orbital phase $\lambda_0$ and the argument of periastron $\delta\lambda_0$ are both set to 0.
    \item The polar angles of the fixed frame to the detector $\theta_{\rm N}$ and $\phi_{\rm N}$ are both set to 0.
    \item Setting the time to the end of the inspiral $t_{\rm f}$ at 4 years (the duration of the LISA mission) determines the remaining parameter, the initial PN parameter $y_0$, which thus also determines the initial orbital frequency and the initial frequency point for evaluation.
    Its value will depend from system to system as it is a function of the masses and $e_0$.
    We describe how we determine $y_0$ from $t_{\rm f}$ in App.~\ref{app:y0}.
\end{itemize}
The evolution of our systems is stopped when $t = t_{\rm f}$, $y = y_{\rm ISCO} = 1/\sqrt{6}$ (the innermost stable circular orbit), or when the orbital frequency reaches $f_{\max}$, whichever occurs first.
We have also set a luminosity distance of $100$ Mpc for our binaries, but as this only appears in $h_0$, it is an overall factor that cancels out in O.
In summary, our model is characterized by 17 parameters: the two component masses, the initial orbital phase, the initial argument of periastron, the initial eccentricity, the two sky angles, the luminosity distance, the initial PN parameter $y_0$, the direction of the orbital angular momentum (two angles), and the two spin vectors (six degrees of freedom).

With these fiducial initial conditions, we aim to characterize how our eccentric amplitudes will be distinguished within the LISA band.
As we are only modifying the effects of eccentricity in the amplitudes, we expect our conclusions to be qualitatively similar across values of spin.
The correlations between spins and eccentricity can instead be found through parameter estimation studies (e.g., \cite{Favata:2021,GilChoi:2022}).
For simplicity and speed, we use the EFPE-(0,0) base model, which uses the $m=0$ approximation in the solution of the spin precession (Eq.~\eqref{eq:deltax}), accurate in the early inspiral, and uses the adiabatic solutions for $\phi_z$ and $\zeta$; see Sec.~V of Ref.~\cite{Klein:2021} for details.

\begin{figure*}
    \centering
    \includegraphics[width=\textwidth]{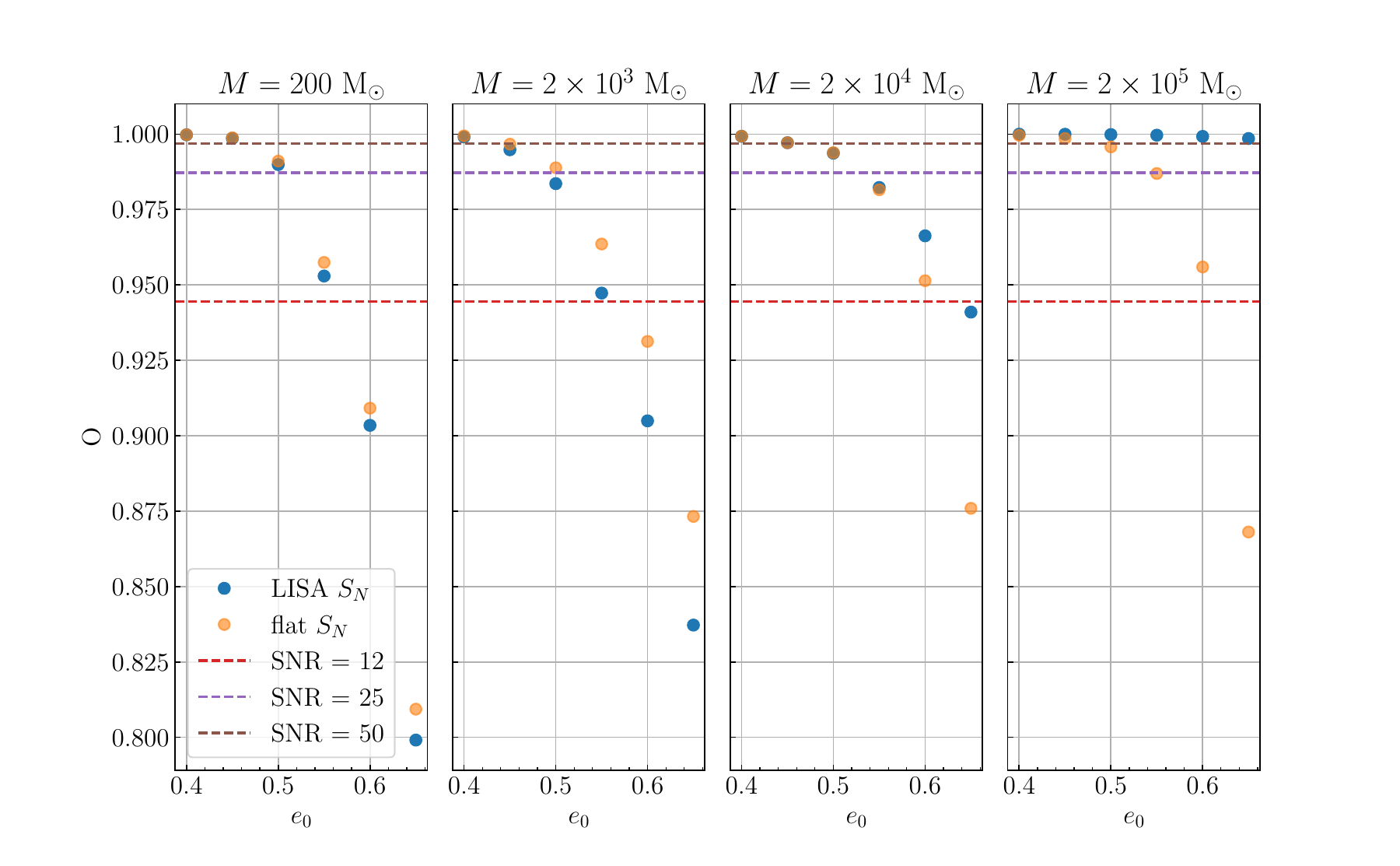}
    \caption{
    Overlaps for five different equal-mass binaries as a function of initial eccentricity $e_0$.
    We compute the overlaps in blue dots using the LISA noise, and in the orange dots with a flat noise to illustrate the effects of LISA's sensitivity, which yields higher overlaps for larger masses at higher $e_0$.
    For SNRs of 12, 25, and 50, we show the expected overlap thresholds in red, purple, and brown lines (O $\approx$ 0.944, 0.987, and 0.997, respectively) if a model is to have smaller systematic errors than the statistical errors of the detector.
    Systems with overlaps below these thresholds will require eccentric amplitudes to be included.
    }
    \label{fig:matchecc}
\end{figure*}

\begin{figure*}
    \centering
    \includegraphics[width=0.8\textwidth]{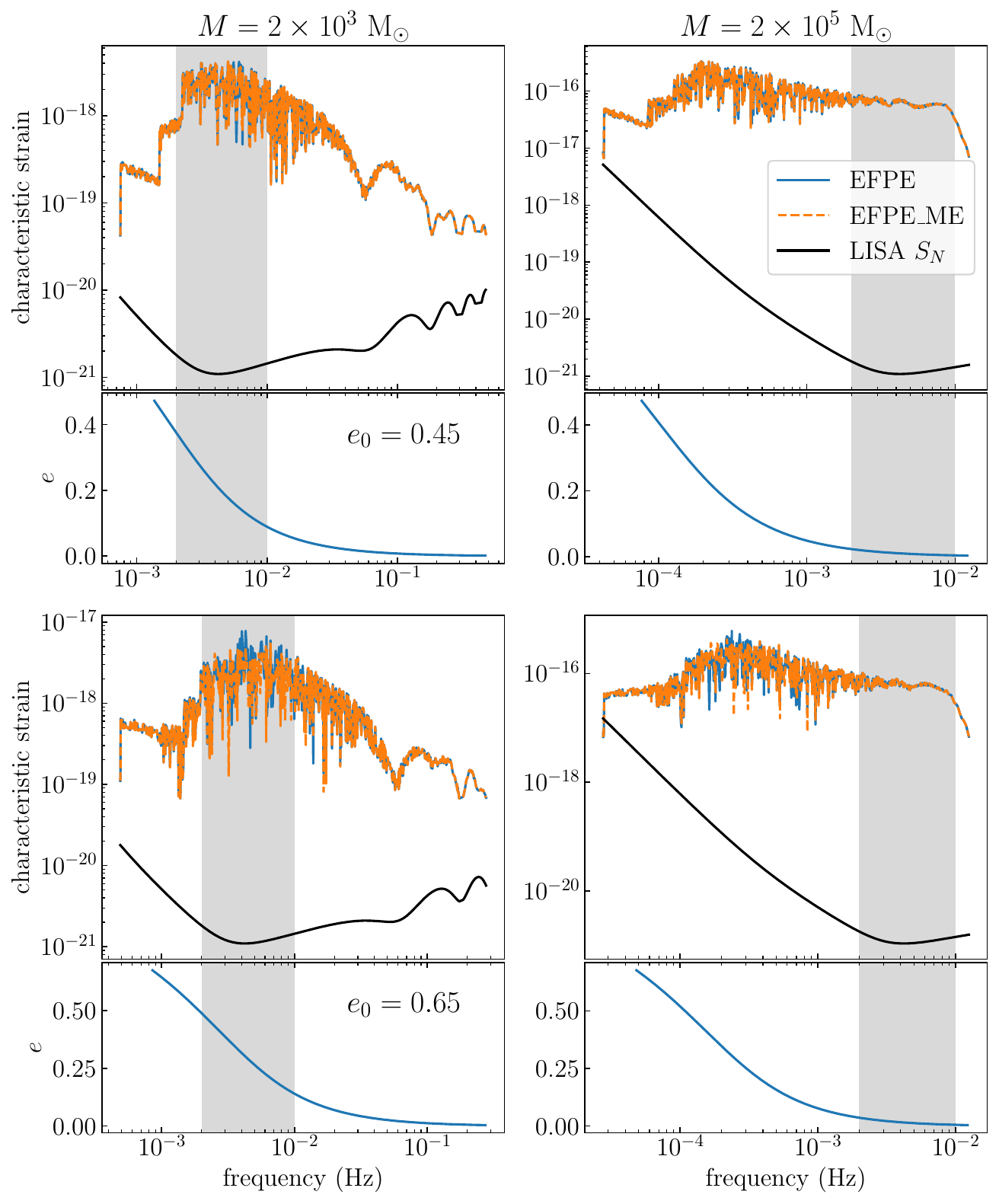}
    \caption{
    Characteristic strains and eccentricity evolution for four of our sample binaries for the EFPE (blue line) and the EFPE\_ME models (dashed orange line).
    In the left column are binaries of total mass $M = 2\times 10^3$ M$_{\odot}$, while in the right column they are of $M = 2\times 10^5$ M$_{\odot}$. The top row are binaries with $e_0=0.45$, while the bottom row have $e_0=0.65$.
    The LISA spectral noise density of the A and E modes is illustrated with the black line.
    Observe that, by the time the higher-mass systems reach the most sensitive part of the LISA band (as illustrated by the shaded regions starting at $2 \times 10^{-3}$ Hz), the eccentricity of these binaries has dropped almost to zero. This explains why the overlaps observed in Fig.~\ref{fig:matchecc} do not drop as rapidly for the high-mass binaries as for the low-mass ones.
    }
    \label{fig:hc_e}
\end{figure*}

With the parameters of our model determined, we proceed to generate the EFPE and EFPE\_ME waveforms for our systems and compute their overlap.
In Fig.~\ref{fig:matchecc}, we show our results, focusing on the range of eccentricities where the overlap goes from being close to 1 at lower eccentricities to dropping as eccentricity increases.
Clearly, this behavior is more pronounced for the lower mass systems. The reason for this is that the higher-mass systems have effectively circularized by the time they reach LISA's most sensitive frequency range in the millihertz, while this is not the case for the lower-mass systems. Let us demonstrate this behavior through the following two arguments.

First, consider removing the frequency-dependence of LISA's sensitivity by setting $S_{N}$ to a constant, i.e.~by considering (flat) white noise. If, indeed, the argument we presented above is true, we would expect that the overlaps would now decrease for the lower-mass and the higher-mass systems in roughly the same way. 
Indeed, this is what we see in in Fig.~\ref{fig:matchecc}.

Let us now consider how rapidly the low- and high-mass systems circularize in the LISA band for the test cases we have studied above. In particular, let us investigate the GW amplitude evolution and the eccentricity evolution, as a function of frequency, for a low-mass ($M=2 \times 10^3\,\Msun$) system and a high-mass ($M=2 \times 10^5\, \Msun$) system, as shown in Fig.~\ref{fig:hc_e}, fixing the initial eccentricities to $e_0= 0.45$ and $e_0= 0.65$.
The GW amplitudes are here computed directly from the EFPE and EFPE\_ME models, where we focus on the total \textit{characteristic strain} of the A and E modes of each model, $\tilde{h}_{i, \mathrm{c}} = 2f \sqrt{|\tilde{h}_{i, \mathrm{A}}|^2 + |\tilde{h}_{i, \mathrm{E}}|^2}$, as these modes dominate over the T mode (at least at low frequencies), and their \textit{characteristic noise strain}, $S_{\rm c} = \sqrt{f S_{\rm A,E}}$, which coincides for these two modes. The square of their ratio, $\left(\tilde{h}_{i, \mathrm{c}}/S_{\rm c}\right)^2$, integrated in logarithmic frequency, is then the signal-to-noise ratio (SNR), and thus, roughly speaking, the area between these two curves is a measure of the detectability of the signal.
The frequency evolution of the eccentricity is obtained by first integrating the radiation-reaction equations (Eq.~\eqref{eq:rr}) to obtain $e^2 = e^2(\dot{\lambda})$, and then mapping $\dot{\lambda}$ to frequency with Eq.~\eqref{eq:tn}, using the quadrupole harmonic $n=2$.

This figure reveals some important information. First, observe from the strain panels, for our choice of initial orbital separation, the higher-mass systems in the right column appear in the LISA band at lower initial frequencies. Thus, even if a high-mass binary starts off at a large $e_0$, causing large discrepancies between the models at early times, the detector will be less sensitive to them because the detector's noise is higher at its low-frequency end.
Second, observe that, when the higher-mass system is inside the most sensitive range of the LISA band (roughly at $4 \times 10^{-3}$ Hz), its eccentricity will be very small (and the system will be effectively circular), as one can see in the eccentricity evolution panels.
This is in contrast to lower-mass binaries on the left column, whose eccentricities remain moderate once the binary enters the most sensitive range of the LISA band. Thus, the effect of eccentricity is larger in lower-mass binaries than in higher-mass binaries, which explains why the overlaps drop for the former much more rapidly than for the latter.

To understand how significant such a loss in overlap can be for interpreting the GWs emitted by eccentric systems, we can use that the expectation value of the match $\mathcal{M}$, which is the overlap maximized over the extrinsic time and phase shifts between two waveforms, at high SNRs is
\begin{equation}
    E[\mathcal{M}] = 1 - \frac{D-1}{2~\rm{SNR}^2},
\end{equation}
where $D$ is the number of parameters in the model \cite{Chatziioannou:2017}.
If the systematic error due to not including our eccentric amplitudes are to be smaller than the statistical errors, then $\mathcal{M}$ must be greater than this expectation value.
While our overlaps are not maximized over extrinsic parameters (although the $D-1$ accounts for the cancellation of the overall amplitude of the waveforms), using this expectation value gives a conservative estimate of the threshold for our overlaps, as the factor of $D$ accounts for our model's larger degree of freedom.
In Fig.~\ref{fig:matchecc}, we indicate this threshold at SNRs of 12, 25, and 50 for $D=17$.
We can therefore see that at masses smaller than $\approx 10^5$ M$_{\odot}$, for equal-mass binaries with $e_0$ greater than $\approx 0.5$, eccentric amplitudes are important even at moderate SNRs; that is, incorrectly modeling the eccentricity in the GW amplitudes would induce a systematic bias that is larger than statistical errors, thus biasing all GW inferences. Keeping these amplitudes as accurate as possible is important for the future detection of intermediate-mass BH binaries with moderate eccentricities that merge within LISA's lifetime.

\subsection{Timing comparison}

In the previous sections, we introduced the EFPE\_ME model with amplitudes accurate for moderate eccentricities as it contains the necessary harmonics to represent each Fourier mode.
In contrast to the previous EFPE model, which has these harmonics expanded in powers of eccentricity, our amplitudes contain the full expressions of the series representations, as shown in Sec.~\ref{sec:amplitudes}.
However, this adds computational costs when evaluating the amplitudes at each value of eccentricity throughout a binary's inspiral.
We benchmark these costs by estimating the evaluation time of generating the frequency-domain waveforms presented in the previous section for initial eccentricities $e_0 \in [ 0.1, 0.8 ]$.

\begin{figure}[!htb]
    \centering
    \includegraphics[width=0.485\textwidth]{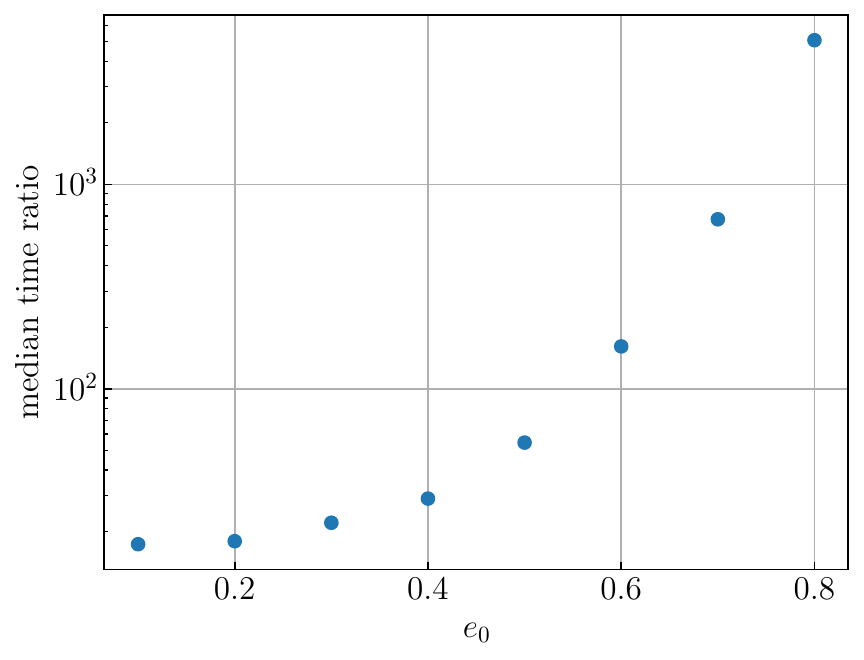}
    \caption{
    The ratios of the median evaluation time of the EFPE\_ME to the EFPE for the binaries described in Sec.~\ref{sub:overlaps}.
    The absolute time for the EFPE model to generate 901 frequency points is approximately 0.2 seconds.
    }
    \label{fig:timing}
\end{figure}

Taking the evaluation time for both models at each mass throughout our set of $e_0$, we find that it is independent of the duration of the signal.
That is because our quadrature scheme, which determines that frequencies at which the waveforms are evaluated, returns the same amount of points, 901; in fact, the EFPE's evaluation time is independent of both $e_0$ and mass, taking approximately 0.2 seconds to evaluate this number of points for all binaries.
What the evaluation time of the EFPE\_ME depends on instead is, as expected, the initial eccentricity.
Therefore, we first fix $e_0$ and a model to  estimate the evaluation times for a set of masses (for $M \in [ 200, 2\times10^5 ]\, \textrm{M}_\odot$) and compute the median; we then repeat this sequence of steps for various choices of $e_0$ so that we can plot the ratio of these medians when computed with the slower model (EFPE\_ME) to the medians computed with the faster model (EFPE), as shown in Fig.~\ref{fig:timing}.
Since the EFPE\_ME adds more harmonics at higher eccentricities, its evaluation time increases with $e_0$ drastically.
At $e_0=0.8$ the EFPE\_ME model takes a thousand times longer to evaluate than the EFPE model (around 22 minutes!), although the latter's accuracy cannot be trusted at this eccentricity.
Nevertheless, a more balanced compromise between efficiency and accuracy has to be reached in order for the EFPE\_ME model to be useful for parameter estimation applications.
A straightforward optimization could be to expand the amplitudes to an appropriate power of eccentricity that makes it fast to evaluate as a polynomial, while keeping enough terms to be as accurate as desired.
The underlying model can also be optimized, such as by computing the waveform at the requested frequencies in parallel.
In this study, however, we have focused on accurately modeling binaries at moderate eccentricities, rather than on reducing the model's computation time.
Once these options for optimization are explored in future work, combining them with the accuracy achieved by the EFPE\_ME would render the EFPE family an excellent tool to analyze GW data.

\section{Conclusions}
\label{sec:conclusion}
As current GW detectors continue to increase their sensitivity and millihertz detectors are poised to come online within the next decade, our scope into the universe will expand, reaching the remnants of stars unlike those that we have seen so far.
If we are to be prepared to characterize new systems properly, we will need models accurate enough to capture the relevant physics.
The extension to the EFPE developed in this work is a step forward in this direction, which allows us to model binaries in a large regime of their parameter space.

Our new EFPE\_ME model extends the EFPE inspiral waveform of~\cite{Klein:2021} for eccentric and spin-precessing binaries from $e\leq0.3$ (for the EFPE model) to $e\leq0.8$ (for the EFPE\_ME model).
We have developed a procedure for determining the harmonics necessary in the Fourier amplitudes of the GW polarizations to ensure that they are convergent and accurate, while retaining computational efficiency.
These harmonics are tabulated and included in our supplementary material \cite{EccentricAmplitudes}.
Unlike the previous implementation of amplitudes expanded in small eccentricity, our amplitudes maintain high accuracy across eccentricities up to $e = 0.8$.
Other waveform models that use the spherical harmonic decomposition can adopt our procedure to more accurately represent moderately-eccentric binaries.
Placing these eccentric amplitudes within the EFPE framework, we now have a new EFPE\_ME model.

One may ask if the inspiral model presented here can be considered at such high eccentricities.
After all, the validity of the SUA and the orbit-averaged equations of motion are based on the assumption that the orbital timescale be the fastest and the radiation-reaction timescale be the longest, with the precession timescale between them (see Sec.~\ref{sec:waveform}), an assumption that could break down at high eccentricities \cite{Loutrel:2018,Fumagalli:2023}.
We find that of the binaries considered in Sec.~\ref{sub:overlaps}, the one that is most likely to violate this hierarchy is the heaviest ($M = 2\times 10^5$ M$_{\odot}$) and most eccentric ($e_0 =0.8$).
Evaluating the leading-order expressions of these timescales for this binary \cite{Gerosa:2023}, we find the ratios of timescales at the beginning of the simulations to be $T_{\rm rr}/T_{\rm pr} \approx 180$ and $T_{\rm rr}/T_{\rm orb} \approx 26000$, large enough to ensure the validity of the assumption.
Of course, this hierarchy breaks down as the orbit decays and the binary approaches merger, at which point the PN approximation also breaks down.
The EFPE\_ME is thus valid for the inspiral.

We computed matches in the LISA band between the EFPE and EFPE\_ME models across eccentricity for different equal-mass binaries merging within the observatory's lifetime and found that the systematic errors of using the small-eccentricity expansions can surpass statistical errors for binaries with masses below $\approx 10^5$ M$_{\odot}$ when $e_0 \geq 0.5$.
Therefore, while binaries with $e_0 < 0.5$ can be modeled with the EFPE, intermediate-mass BH binaries with moderate eccentricities will need accurate amplitudes like ours to be properly interpreted.

The EFPE\_ME, however, is only one of the first steps towards a truly efficient and complete model of a spin-precessing and eccentric binary coalescence.
The inclusion of so many harmonics for large eccentricities increases the computation time significantly, and must be reduced in order to be useful in the inference of a detected binary's parameters. We have not attempted to computationally optimize our implementation of the EFPE\_ME model, but this is indeed possible, and we leave it to future work. 
The EFPE\_ME model is also limited to $e\leq0.8$ by the slow convergence of the Fourier series at higher $e$; a more careful analysis will have to be done to extract a convergent series if one wishes to model binaries with higher eccentricity. Future work could explore various resummations of the Fourier series, perhaps through spectral methods, to improve the convergence of the approximation, which is also an avenue for future work.  
Moreover, the EFPE\_ME model only includes the dominant Newtonian amplitudes of the GW. 
Higher PN-order modes for aligned-spins with eccentricity were recently found up to 2PN in~\cite{Paul:2022}, and could be straightforwardly (though tediously) used to extend the EFPE\_ME to higher PN order. Whether this is necessary or not for parameter estimation remains to be seen and can be the topic of future research. 
While higher PN-order amplitudes are suppressed, they may be significant for stitching the inspiral model to the merger and ringdown model, a critical step toward extending the EFPE family into an inspiral-merger-ringdown model. Currently, this extension is hindered by the lack of accurate, spin-precessing and moderately eccentric numerical relativity simulations of merging BHs.
Once these become available, a phenomenological inspiral-merger-ringdown model based on the EFPE scheme will be developed.


\begin{acknowledgments}
We would like to thank Rohit Chandramouli for helpful discussions.
J.N.A.~acknowledges support from the Alfred P.~Sloan Foundation and the UIUC Graduate College Fellowship. N.Y.~acknowledges support from the National Aeronautics and Space Administration through award 80NSSC22K0806, and from the National Science Foundation through award PHY-2207650.
\end{acknowledgments}


\appendix

\section{Simplifying \texorpdfstring{$\mathcal{E}^m_k$}{EmK}}
\label{app:Emk}
To simplify Eq.~(\ref{eq:Emk}) to a manifestly regular expression, we first note the origin of its apparent divergences.
First,
the binomial coefficient is
\begin{equation}
\label{eq:binom}
    \binom{k-1}{k-m} = \frac{(k-1)!}{(k-m)!\, (m-1)!}.
\end{equation}
If a factorial has a negative argument,
it diverges to infinity\footnote{Recall that $n! = \Gamma(n+1)$, where the gamma function $\Gamma(x)$ can be analytically continued to non-positive $x$ and has simple poles at integers $x = 0, -1, -2, ...$},
which occurs in the denominator for $m>k$,
making the binomial coefficient vanish.
At the same time,
the hypergeometric function can be written as
\begin{equation}
\label{eq:hyper series}
    {}_2F_1(-m,b,c;z) = \sum_{n=0}^m \frac{(-m)_n\, (b)_n}{(c)_n\, n!} z^n
\end{equation}
for $m$ a positive integer \cite{DLMF}.
Here, we have used Pochhammer's symbol $(a)_n = a (a+1) \cdots (a+n-1)$.
For negative $a = -m$,
this symbol can be written as
\begin{equation}
    (-m)_n = (-1)^n \frac{m!}{(m-n)!}
\end{equation}
for $0 \leq n \leq m$ \cite{DLMF},
which is always the case in Eq.~(\ref{eq:hyper series}).
Given this, we can expand the hypergeometric function by substituting in our arguments $b = k$ and $c = k-m+1$ and simplifying, to obtain
\begin{equation}
    \frac{(-m)_n (b)_n}{(c)_n} = \frac{m! (k-m)!}{(k-1)!} \frac{(-1)^n (k+n-1)!}{(k-m+n)! (m-n)!}.
\end{equation}\\
Here, we see that the hypergeometric function can diverge due to the factorial of $k-m$.
However, the whole first fraction can be factored out of the sum and the divergence is then canceled out precisely by the denominator of Eq.~(\ref{eq:binom}).
Together, the binomial coefficient and the hypergeometric function yield the regular polynomial
\begin{multline}
    \binom{k-1}{k-m}\, {}_2F_1(-m,k,k-m+1;z) = \\
    m \sum_{n=0}^m \frac{(-1)^n\, (k+n-1)!}{(k-m+n)!\, (m-n)!\, n!} z^{n}.
\end{multline}
The denominator will still diverge when $m > k$,
but this will occur only for terms in the series where $n < m-k$,
simply making them vanish.
Inserting $z = \beta^2$ into the above equation yields Eq.~(\ref{eq:Emk polynomial}).

\section{Obtaining the initial PN parameter from the time of merger}
\label{app:y0}
To determine the initial value of the PN parameter $y_0 = y(t_0)$ (akin to determining the initial orbital frequency) given the total time of inspiral $t_{\rm f}$, one can consider the leading-order evolution of $y$ from the first term $a_0$ of Eq.~\eqref{eq:dydt}.
This reveals that its evolution is coupled to that of $e$ and depends on the value of $\eta$.
Thus, obtaining $y_0$ from $t_{\rm f}$ is non-trivial if the evolution of $e$ is not ignored.
Obtaining $t_{\rm f}$ given $y_0$ and $e_0$, however, was explored in Ref.~\cite{Tucker:2021}.
In that study, using the leading-order evolution equations with no spin effects, a fitting function $T_{\rm fit}$ for the time of inspiral was obtained as a function of the initial eccentricity and the non-dimensional semi-latus rectum $x = 1/y^2$.
This fit was found to be accurate within 2\% of the true time of inspiral for $e \leq 0.999$. The fitting function
$T_{\rm fit}$ is 
\begin{equation}
    T_{\rm fit}/M = \frac{5}{256\eta} \left( x_0^4 - x(t_{\rm f})^4 \right) \left( 1 + \frac{8(1-e_0^2)}{x_0} \right) \mathcal{G}(e_0),
\end{equation}
where $\mathcal{G}(e_0)$ is a function of the initial eccentricity $e_0$.
Thus, we rewrite $T_{\rm fit}$ as a quintic polynomial in $x_0$ and define $t_{\rm f} = T_{\rm fit}$.
The end of the inspiral is characterized by the innermost stable circular orbit when $x(t_{\rm f}) = 6$.
Taking our initial conditions listed in Sec.~\ref{sub:overlaps}, we solve the polynomial for $x_0$ using the GSL function \verb|gsl_poly_complex_solve| and thus obtain $y_0 = y_0(t_{\rm f}, e_0, \eta)$ for our systems of interest.


\bibliography{main}

\begin{thebibliography}{71}%
\makeatletter
\providecommand \@ifxundefined [1]{%
 \@ifx{#1\undefined}
}%
\providecommand \@ifnum [1]{%
 \ifnum #1\expandafter \@firstoftwo
 \else \expandafter \@secondoftwo
 \fi
}%
\providecommand \@ifx [1]{%
 \ifx #1\expandafter \@firstoftwo
 \else \expandafter \@secondoftwo
 \fi
}%
\providecommand \natexlab [1]{#1}%
\providecommand \enquote  [1]{``#1''}%
\providecommand \bibnamefont  [1]{#1}%
\providecommand \bibfnamefont [1]{#1}%
\providecommand \citenamefont [1]{#1}%
\providecommand \href@noop [0]{\@secondoftwo}%
\providecommand \href [0]{\begingroup \@sanitize@url \@href}%
\providecommand \@href[1]{\@@startlink{#1}\@@href}%
\providecommand \@@href[1]{\endgroup#1\@@endlink}%
\providecommand \@sanitize@url [0]{\catcode `\\12\catcode `\$12\catcode
  `\&12\catcode `\#12\catcode `\^12\catcode `\_12\catcode `\%12\relax}%
\providecommand \@@startlink[1]{}%
\providecommand \@@endlink[0]{}%
\providecommand \url  [0]{\begingroup\@sanitize@url \@url }%
\providecommand \@url [1]{\endgroup\@href {#1}{\urlprefix }}%
\providecommand \urlprefix  [0]{URL }%
\providecommand \Eprint [0]{\href }%
\providecommand \doibase [0]{https://doi.org/}%
\providecommand \selectlanguage [0]{\@gobble}%
\providecommand \bibinfo  [0]{\@secondoftwo}%
\providecommand \bibfield  [0]{\@secondoftwo}%
\providecommand \translation [1]{[#1]}%
\providecommand \BibitemOpen [0]{}%
\providecommand \bibitemStop [0]{}%
\providecommand \bibitemNoStop [0]{.\EOS\space}%
\providecommand \EOS [0]{\spacefactor3000\relax}%
\providecommand \BibitemShut  [1]{\csname bibitem#1\endcsname}%
\let\auto@bib@innerbib\@empty
\bibitem [{\citenamefont {Abbott}\ \emph {et~al.}(2023)\citenamefont {Abbott}
  \emph {et~al.}}]{KAGRA:2021}%
  \BibitemOpen
  \bibfield  {author} {\bibinfo {author} {\bibfnamefont {R.}~\bibnamefont
  {Abbott}} \emph {et~al.} (\bibinfo {collaboration} {KAGRA, VIRGO, LIGO
  Scientific}),\ }\bibfield  {title} {\bibinfo {title} {{GWTC-3: Compact Binary
  Coalescences Observed by LIGO and Virgo during the Second Part of the Third
  Observing Run}},\ }\href {https://doi.org/10.1103/PhysRevX.13.041039}
  {\bibfield  {journal} {\bibinfo  {journal} {Phys. Rev. X}\ }\textbf {\bibinfo
  {volume} {13}},\ \bibinfo {pages} {041039} (\bibinfo {year} {2023})},\
  \Eprint {https://arxiv.org/abs/2111.03606} {arXiv:2111.03606 [gr-qc]}
  \BibitemShut {NoStop}%
\bibitem [{\citenamefont {{Iglesias}}\ \emph {et~al.}(2022)\citenamefont
  {{Iglesias}}, \citenamefont {{Lange}}, \citenamefont {{Bartos}},
  \citenamefont {{Bhaumik}}, \citenamefont {{Gamba}}, \citenamefont
  {{Gayathri}}, \citenamefont {{Jan}}, \citenamefont {{Nowicki}}, \citenamefont
  {{O'Shaughnessy}}, \citenamefont {{Shoemaker}}, \citenamefont
  {{Venkataramanan}},\ and\ \citenamefont {{Wagner}}}]{Iglesias:2022}%
  \BibitemOpen
  \bibfield  {author} {\bibinfo {author} {\bibfnamefont {H.~L.}\ \bibnamefont
  {{Iglesias}}}, \bibinfo {author} {\bibfnamefont {J.}~\bibnamefont {{Lange}}},
  \bibinfo {author} {\bibfnamefont {I.}~\bibnamefont {{Bartos}}}, \bibinfo
  {author} {\bibfnamefont {S.}~\bibnamefont {{Bhaumik}}}, \bibinfo {author}
  {\bibfnamefont {R.}~\bibnamefont {{Gamba}}}, \bibinfo {author} {\bibfnamefont
  {V.}~\bibnamefont {{Gayathri}}}, \bibinfo {author} {\bibfnamefont
  {A.}~\bibnamefont {{Jan}}}, \bibinfo {author} {\bibfnamefont
  {R.}~\bibnamefont {{Nowicki}}}, \bibinfo {author} {\bibfnamefont
  {R.}~\bibnamefont {{O'Shaughnessy}}}, \bibinfo {author} {\bibfnamefont
  {D.}~\bibnamefont {{Shoemaker}}}, \bibinfo {author} {\bibfnamefont
  {R.}~\bibnamefont {{Venkataramanan}}},\ and\ \bibinfo {author} {\bibfnamefont
  {K.}~\bibnamefont {{Wagner}}},\ }\bibfield  {title} {\bibinfo {title}
  {{Eccentricity estimation for five binary black hole mergers with
  higher-order gravitational wave modes}},\ }\href
  {https://doi.org/10.48550/arXiv.2208.01766} {\bibfield  {journal} {\bibinfo
  {journal} {arXiv e-prints}\ ,\ \bibinfo {eid} {arXiv:2208.01766}} (\bibinfo
  {year} {2022})},\ \Eprint {https://arxiv.org/abs/2208.01766}
  {arXiv:2208.01766 [gr-qc]} \BibitemShut {NoStop}%
\bibitem [{\citenamefont {Ramos-Buades}\ \emph
  {et~al.}(2023{\natexlab{a}})\citenamefont {Ramos-Buades}, \citenamefont
  {Buonanno},\ and\ \citenamefont {Gair}}]{Ramos-Buades:2023}%
  \BibitemOpen
  \bibfield  {author} {\bibinfo {author} {\bibfnamefont {A.}~\bibnamefont
  {Ramos-Buades}}, \bibinfo {author} {\bibfnamefont {A.}~\bibnamefont
  {Buonanno}},\ and\ \bibinfo {author} {\bibfnamefont {J.}~\bibnamefont
  {Gair}},\ }\bibfield  {title} {\bibinfo {title} {{Bayesian inference of
  binary black holes with inspiral-merger-ringdown waveforms using two
  eccentric parameters}},\ }\href {https://doi.org/10.1103/PhysRevD.108.124063}
  {\bibfield  {journal} {\bibinfo  {journal} {Phys. Rev. D}\ }\textbf {\bibinfo
  {volume} {108}},\ \bibinfo {pages} {124063} (\bibinfo {year}
  {2023}{\natexlab{a}})},\ \Eprint {https://arxiv.org/abs/2309.15528}
  {arXiv:2309.15528 [gr-qc]} \BibitemShut {NoStop}%
\bibitem [{\citenamefont {Romero-Shaw}\ \emph {et~al.}(2021)\citenamefont
  {Romero-Shaw}, \citenamefont {Lasky},\ and\ \citenamefont
  {Thrane}}]{Romero-Shaw:2021}%
  \BibitemOpen
  \bibfield  {author} {\bibinfo {author} {\bibfnamefont {I.~M.}\ \bibnamefont
  {Romero-Shaw}}, \bibinfo {author} {\bibfnamefont {P.~D.}\ \bibnamefont
  {Lasky}},\ and\ \bibinfo {author} {\bibfnamefont {E.}~\bibnamefont
  {Thrane}},\ }\bibfield  {title} {\bibinfo {title} {{Signs of Eccentricity in
  Two Gravitational-wave Signals May Indicate a Subpopulation of Dynamically
  Assembled Binary Black Holes}},\ }\href
  {https://doi.org/10.3847/2041-8213/ac3138} {\bibfield  {journal} {\bibinfo
  {journal} {Astrophys. J. Lett.}\ }\textbf {\bibinfo {volume} {921}},\
  \bibinfo {pages} {L31} (\bibinfo {year} {2021})},\ \Eprint
  {https://arxiv.org/abs/2108.01284} {arXiv:2108.01284 [astro-ph.HE]}
  \BibitemShut {NoStop}%
\bibitem [{\citenamefont {Romero-Shaw}\ \emph {et~al.}(2022)\citenamefont
  {Romero-Shaw}, \citenamefont {Lasky},\ and\ \citenamefont
  {Thrane}}]{Romero-Shaw:2022xko}%
  \BibitemOpen
  \bibfield  {author} {\bibinfo {author} {\bibfnamefont {I.~M.}\ \bibnamefont
  {Romero-Shaw}}, \bibinfo {author} {\bibfnamefont {P.~D.}\ \bibnamefont
  {Lasky}},\ and\ \bibinfo {author} {\bibfnamefont {E.}~\bibnamefont
  {Thrane}},\ }\bibfield  {title} {\bibinfo {title} {{Four Eccentric Mergers
  Increase the Evidence that
  LIGO\textendash{}Virgo\textendash{}KAGRA\textquoteright{}s Binary Black Holes
  Form Dynamically}},\ }\href {https://doi.org/10.3847/1538-4357/ac9798}
  {\bibfield  {journal} {\bibinfo  {journal} {Astrophys. J.}\ }\textbf
  {\bibinfo {volume} {940}},\ \bibinfo {pages} {171} (\bibinfo {year}
  {2022})},\ \Eprint {https://arxiv.org/abs/2206.14695} {arXiv:2206.14695
  [astro-ph.HE]} \BibitemShut {NoStop}%
\bibitem [{\citenamefont {Lower}\ \emph {et~al.}(2018)\citenamefont {Lower},
  \citenamefont {Thrane}, \citenamefont {Lasky},\ and\ \citenamefont
  {Smith}}]{Lower:2018}%
  \BibitemOpen
  \bibfield  {author} {\bibinfo {author} {\bibfnamefont {M.~E.}\ \bibnamefont
  {Lower}}, \bibinfo {author} {\bibfnamefont {E.}~\bibnamefont {Thrane}},
  \bibinfo {author} {\bibfnamefont {P.~D.}\ \bibnamefont {Lasky}},\ and\
  \bibinfo {author} {\bibfnamefont {R.}~\bibnamefont {Smith}},\ }\bibfield
  {title} {\bibinfo {title} {{Measuring eccentricity in binary black hole
  inspirals with gravitational waves}},\ }\href
  {https://doi.org/10.1103/PhysRevD.98.083028} {\bibfield  {journal} {\bibinfo
  {journal} {Phys. Rev. D}\ }\textbf {\bibinfo {volume} {98}},\ \bibinfo
  {pages} {083028} (\bibinfo {year} {2018})},\ \Eprint
  {https://arxiv.org/abs/1806.05350} {arXiv:1806.05350 [astro-ph.HE]}
  \BibitemShut {NoStop}%
\bibitem [{\citenamefont {Rodriguez}\ \emph {et~al.}(2018)\citenamefont
  {Rodriguez}, \citenamefont {Amaro-Seoane}, \citenamefont {Chatterjee},
  \citenamefont {Kremer}, \citenamefont {Rasio}, \citenamefont {Samsing},
  \citenamefont {Ye},\ and\ \citenamefont {Zevin}}]{Rodriguez:2018}%
  \BibitemOpen
  \bibfield  {author} {\bibinfo {author} {\bibfnamefont {C.~L.}\ \bibnamefont
  {Rodriguez}}, \bibinfo {author} {\bibfnamefont {P.}~\bibnamefont
  {Amaro-Seoane}}, \bibinfo {author} {\bibfnamefont {S.}~\bibnamefont
  {Chatterjee}}, \bibinfo {author} {\bibfnamefont {K.}~\bibnamefont {Kremer}},
  \bibinfo {author} {\bibfnamefont {F.~A.}\ \bibnamefont {Rasio}}, \bibinfo
  {author} {\bibfnamefont {J.}~\bibnamefont {Samsing}}, \bibinfo {author}
  {\bibfnamefont {C.~S.}\ \bibnamefont {Ye}},\ and\ \bibinfo {author}
  {\bibfnamefont {M.}~\bibnamefont {Zevin}},\ }\bibfield  {title} {\bibinfo
  {title} {{Post-Newtonian Dynamics in Dense Star Clusters: Formation, Masses,
  and Merger Rates of Highly-Eccentric Black Hole Binaries}},\ }\href
  {https://doi.org/10.1103/PhysRevD.98.123005} {\bibfield  {journal} {\bibinfo
  {journal} {Phys. Rev. D}\ }\textbf {\bibinfo {volume} {98}},\ \bibinfo
  {pages} {123005} (\bibinfo {year} {2018})},\ \Eprint
  {https://arxiv.org/abs/1811.04926} {arXiv:1811.04926 [astro-ph.HE]}
  \BibitemShut {NoStop}%
\bibitem [{\citenamefont {Zevin}\ \emph {et~al.}(2019)\citenamefont {Zevin},
  \citenamefont {Samsing}, \citenamefont {Rodriguez}, \citenamefont {Haster},\
  and\ \citenamefont {Ramirez-Ruiz}}]{Zevin:2018}%
  \BibitemOpen
  \bibfield  {author} {\bibinfo {author} {\bibfnamefont {M.}~\bibnamefont
  {Zevin}}, \bibinfo {author} {\bibfnamefont {J.}~\bibnamefont {Samsing}},
  \bibinfo {author} {\bibfnamefont {C.}~\bibnamefont {Rodriguez}}, \bibinfo
  {author} {\bibfnamefont {C.-J.}\ \bibnamefont {Haster}},\ and\ \bibinfo
  {author} {\bibfnamefont {E.}~\bibnamefont {Ramirez-Ruiz}},\ }\bibfield
  {title} {\bibinfo {title} {{Eccentric Black Hole Mergers in Dense Star
  Clusters: The Role of Binary\textendash{}Binary Encounters}},\ }\href
  {https://doi.org/10.3847/1538-4357/aaf6ec} {\bibfield  {journal} {\bibinfo
  {journal} {Astrophys. J.}\ }\textbf {\bibinfo {volume} {871}},\ \bibinfo
  {pages} {91} (\bibinfo {year} {2019})},\ \Eprint
  {https://arxiv.org/abs/1810.00901} {arXiv:1810.00901 [astro-ph.HE]}
  \BibitemShut {NoStop}%
\bibitem [{\citenamefont {{Amaro-Seoane}}\ \emph {et~al.}(2017)\citenamefont
  {{Amaro-Seoane}} \emph {et~al.}}]{LISA:2017}%
  \BibitemOpen
  \bibfield  {author} {\bibinfo {author} {\bibfnamefont {P.}~\bibnamefont
  {{Amaro-Seoane}}} \emph {et~al.},\ }\bibfield  {title} {\bibinfo {title}
  {{Laser Interferometer Space Antenna}},\ }\href
  {https://doi.org/10.48550/arXiv.1702.00786} {\bibfield  {journal} {\bibinfo
  {journal} {arXiv e-prints}\ ,\ \bibinfo {eid} {arXiv:1702.00786}} (\bibinfo
  {year} {2017})},\ \Eprint {https://arxiv.org/abs/1702.00786}
  {arXiv:1702.00786 [astro-ph.IM]} \BibitemShut {NoStop}%
\bibitem [{\citenamefont {Luo}\ \emph {et~al.}(2016)\citenamefont {Luo} \emph
  {et~al.}}]{TianQin:2015}%
  \BibitemOpen
  \bibfield  {author} {\bibinfo {author} {\bibfnamefont {J.}~\bibnamefont
  {Luo}} \emph {et~al.} (\bibinfo {collaboration} {TianQin}),\ }\bibfield
  {title} {\bibinfo {title} {{TianQin: a space-borne gravitational wave
  detector}},\ }\href {https://doi.org/10.1088/0264-9381/33/3/035010}
  {\bibfield  {journal} {\bibinfo  {journal} {Class. Quant. Grav.}\ }\textbf
  {\bibinfo {volume} {33}},\ \bibinfo {pages} {035010} (\bibinfo {year}
  {2016})},\ \Eprint {https://arxiv.org/abs/1512.02076} {arXiv:1512.02076
  [astro-ph.IM]} \BibitemShut {NoStop}%
\bibitem [{\citenamefont {Sesana}(2016)}]{Sesana:2016}%
  \BibitemOpen
  \bibfield  {author} {\bibinfo {author} {\bibfnamefont {A.}~\bibnamefont
  {Sesana}},\ }\bibfield  {title} {\bibinfo {title} {{Prospects for Multiband
  Gravitational-Wave Astronomy after GW150914}},\ }\href
  {https://doi.org/10.1103/PhysRevLett.116.231102} {\bibfield  {journal}
  {\bibinfo  {journal} {Phys. Rev. Lett.}\ }\textbf {\bibinfo {volume} {116}},\
  \bibinfo {pages} {231102} (\bibinfo {year} {2016})},\ \Eprint
  {https://arxiv.org/abs/1602.06951} {arXiv:1602.06951 [gr-qc]} \BibitemShut
  {NoStop}%
\bibitem [{\citenamefont {Apostolatos}\ \emph {et~al.}(1994)\citenamefont
  {Apostolatos}, \citenamefont {Cutler}, \citenamefont {Sussman},\ and\
  \citenamefont {Thorne}}]{Apostolatos:1994}%
  \BibitemOpen
  \bibfield  {author} {\bibinfo {author} {\bibfnamefont {T.~A.}\ \bibnamefont
  {Apostolatos}}, \bibinfo {author} {\bibfnamefont {C.}~\bibnamefont {Cutler}},
  \bibinfo {author} {\bibfnamefont {G.~J.}\ \bibnamefont {Sussman}},\ and\
  \bibinfo {author} {\bibfnamefont {K.~S.}\ \bibnamefont {Thorne}},\ }\bibfield
   {title} {\bibinfo {title} {{Spin induced orbital precession and its
  modulation of the gravitational wave forms from merging binaries}},\ }\href
  {https://doi.org/10.1103/PhysRevD.49.6274} {\bibfield  {journal} {\bibinfo
  {journal} {Phys. Rev. D}\ }\textbf {\bibinfo {volume} {49}},\ \bibinfo
  {pages} {6274} (\bibinfo {year} {1994})}\BibitemShut {NoStop}%
\bibitem [{\citenamefont {Seoane}\ \emph {et~al.}(2023)\citenamefont {Seoane}
  \emph {et~al.}}]{LISA:2022}%
  \BibitemOpen
  \bibfield  {author} {\bibinfo {author} {\bibfnamefont {P.~A.}\ \bibnamefont
  {Seoane}} \emph {et~al.} (\bibinfo {collaboration} {LISA}),\ }\bibfield
  {title} {\bibinfo {title} {{Astrophysics with the Laser Interferometer Space
  Antenna}},\ }\href {https://doi.org/10.1007/s41114-022-00041-y} {\bibfield
  {journal} {\bibinfo  {journal} {Living Rev. Rel.}\ }\textbf {\bibinfo
  {volume} {26}},\ \bibinfo {pages} {2} (\bibinfo {year} {2023})},\ \Eprint
  {https://arxiv.org/abs/2203.06016} {arXiv:2203.06016 [gr-qc]} \BibitemShut
  {NoStop}%
\bibitem [{\citenamefont {Antonini}\ and\ \citenamefont
  {Perets}(2012)}]{Antonini:2012}%
  \BibitemOpen
  \bibfield  {author} {\bibinfo {author} {\bibfnamefont {F.}~\bibnamefont
  {Antonini}}\ and\ \bibinfo {author} {\bibfnamefont {H.~B.}\ \bibnamefont
  {Perets}},\ }\bibfield  {title} {\bibinfo {title} {{Secular evolution of
  compact binaries near massive black holes: Gravitational wave sources and
  other exotica}},\ }\href {https://doi.org/10.1088/0004-637X/757/1/27}
  {\bibfield  {journal} {\bibinfo  {journal} {Astrophys. J.}\ }\textbf
  {\bibinfo {volume} {757}},\ \bibinfo {pages} {27} (\bibinfo {year} {2012})},\
  \Eprint {https://arxiv.org/abs/1203.2938} {arXiv:1203.2938 [astro-ph.GA]}
  \BibitemShut {NoStop}%
\bibitem [{\citenamefont {Antonini}\ \emph {et~al.}(2016)\citenamefont
  {Antonini}, \citenamefont {Chatterjee}, \citenamefont {Rodriguez},
  \citenamefont {Morscher}, \citenamefont {Pattabiraman}, \citenamefont
  {Kalogera},\ and\ \citenamefont {Rasio}}]{Antonini:2015}%
  \BibitemOpen
  \bibfield  {author} {\bibinfo {author} {\bibfnamefont {F.}~\bibnamefont
  {Antonini}}, \bibinfo {author} {\bibfnamefont {S.}~\bibnamefont
  {Chatterjee}}, \bibinfo {author} {\bibfnamefont {C.~L.}\ \bibnamefont
  {Rodriguez}}, \bibinfo {author} {\bibfnamefont {M.}~\bibnamefont {Morscher}},
  \bibinfo {author} {\bibfnamefont {B.}~\bibnamefont {Pattabiraman}}, \bibinfo
  {author} {\bibfnamefont {V.}~\bibnamefont {Kalogera}},\ and\ \bibinfo
  {author} {\bibfnamefont {F.~A.}\ \bibnamefont {Rasio}},\ }\bibfield  {title}
  {\bibinfo {title} {{Black hole mergers and blue stragglers from hierarchical
  triples formed in globular clusters}},\ }\href
  {https://doi.org/10.3847/0004-637X/816/2/65} {\bibfield  {journal} {\bibinfo
  {journal} {Astrophys. J.}\ }\textbf {\bibinfo {volume} {816}},\ \bibinfo
  {pages} {65} (\bibinfo {year} {2016})},\ \Eprint
  {https://arxiv.org/abs/1509.05080} {arXiv:1509.05080 [astro-ph.GA]}
  \BibitemShut {NoStop}%
\bibitem [{\citenamefont {Samsing}(2018)}]{Samsing:2017}%
  \BibitemOpen
  \bibfield  {author} {\bibinfo {author} {\bibfnamefont {J.}~\bibnamefont
  {Samsing}},\ }\bibfield  {title} {\bibinfo {title} {{Eccentric Black Hole
  Mergers Forming in Globular Clusters}},\ }\href
  {https://doi.org/10.1103/PhysRevD.97.103014} {\bibfield  {journal} {\bibinfo
  {journal} {Phys. Rev. D}\ }\textbf {\bibinfo {volume} {97}},\ \bibinfo
  {pages} {103014} (\bibinfo {year} {2018})},\ \Eprint
  {https://arxiv.org/abs/1711.07452} {arXiv:1711.07452 [astro-ph.HE]}
  \BibitemShut {NoStop}%
\bibitem [{\citenamefont {Samsing}\ \emph {et~al.}(2014)\citenamefont
  {Samsing}, \citenamefont {MacLeod},\ and\ \citenamefont
  {Ramirez-Ruiz}}]{Samsing:2013}%
  \BibitemOpen
  \bibfield  {author} {\bibinfo {author} {\bibfnamefont {J.}~\bibnamefont
  {Samsing}}, \bibinfo {author} {\bibfnamefont {M.}~\bibnamefont {MacLeod}},\
  and\ \bibinfo {author} {\bibfnamefont {E.}~\bibnamefont {Ramirez-Ruiz}},\
  }\bibfield  {title} {\bibinfo {title} {{The Formation of Eccentric Compact
  Binary Inspirals and the Role of Gravitational Wave Emission in Binary-Single
  Stellar Encounters}},\ }\href {https://doi.org/10.1088/0004-637X/784/1/71}
  {\bibfield  {journal} {\bibinfo  {journal} {Astrophys. J.}\ }\textbf
  {\bibinfo {volume} {784}},\ \bibinfo {pages} {71} (\bibinfo {year} {2014})},\
  \Eprint {https://arxiv.org/abs/1308.2964} {arXiv:1308.2964 [astro-ph.HE]}
  \BibitemShut {NoStop}%
\bibitem [{\citenamefont {Liu}\ and\ \citenamefont {Lai}(2021)}]{Liu:2020}%
  \BibitemOpen
  \bibfield  {author} {\bibinfo {author} {\bibfnamefont {B.}~\bibnamefont
  {Liu}}\ and\ \bibinfo {author} {\bibfnamefont {D.}~\bibnamefont {Lai}},\
  }\bibfield  {title} {\bibinfo {title} {{Hierarchical Black-Hole Mergers in
  Multiple Systems: Constrain the Formation of GW190412, GW190814 and
  GW190521-like events}},\ }\href {https://doi.org/10.1093/mnras/stab178}
  {\bibfield  {journal} {\bibinfo  {journal} {Mon. Not. Roy. Astron. Soc.}\
  }\textbf {\bibinfo {volume} {502}},\ \bibinfo {pages} {2049} (\bibinfo {year}
  {2021})},\ \Eprint {https://arxiv.org/abs/2009.10068} {arXiv:2009.10068
  [astro-ph.HE]} \BibitemShut {NoStop}%
\bibitem [{\citenamefont {Rodriguez}\ \emph {et~al.}(2016)\citenamefont
  {Rodriguez}, \citenamefont {Zevin}, \citenamefont {Pankow}, \citenamefont
  {Kalogera},\ and\ \citenamefont {Rasio}}]{Rodriguez:2016}%
  \BibitemOpen
  \bibfield  {author} {\bibinfo {author} {\bibfnamefont {C.~L.}\ \bibnamefont
  {Rodriguez}}, \bibinfo {author} {\bibfnamefont {M.}~\bibnamefont {Zevin}},
  \bibinfo {author} {\bibfnamefont {C.}~\bibnamefont {Pankow}}, \bibinfo
  {author} {\bibfnamefont {V.}~\bibnamefont {Kalogera}},\ and\ \bibinfo
  {author} {\bibfnamefont {F.~A.}\ \bibnamefont {Rasio}},\ }\bibfield  {title}
  {\bibinfo {title} {{Illuminating Black Hole Binary Formation Channels with
  Spins in Advanced LIGO}},\ }\href
  {https://doi.org/10.3847/2041-8205/832/1/L2} {\bibfield  {journal} {\bibinfo
  {journal} {Astrophys. J. Lett.}\ }\textbf {\bibinfo {volume} {832}},\
  \bibinfo {pages} {L2} (\bibinfo {year} {2016})},\ \Eprint
  {https://arxiv.org/abs/1609.05916} {arXiv:1609.05916 [astro-ph.HE]}
  \BibitemShut {NoStop}%
\bibitem [{\citenamefont {{Afshordi}}\ \emph {et~al.}(2023)\citenamefont
  {{Afshordi}} \emph {et~al.}}]{LISA:2023}%
  \BibitemOpen
  \bibfield  {author} {\bibinfo {author} {\bibfnamefont {N.}~\bibnamefont
  {{Afshordi}}} \emph {et~al.} (\bibinfo {collaboration} {LISA Consortium
  Waveform Working Group}),\ }\href {https://doi.org/10.48550/arXiv.2311.01300}
  {\bibfield  {journal} {\bibinfo  {journal} {arXiv e-prints}\ ,\ \bibinfo
  {eid} {arXiv:2311.01300}} (\bibinfo {year} {2023})},\ \Eprint
  {https://arxiv.org/abs/2311.01300} {arXiv:2311.01300 [gr-qc]} \BibitemShut
  {NoStop}%
\bibitem [{\citenamefont {Amaro-Seoane}\ and\ \citenamefont
  {Chen}(2016)}]{Amaro-Seoane:2015}%
  \BibitemOpen
  \bibfield  {author} {\bibinfo {author} {\bibfnamefont {P.}~\bibnamefont
  {Amaro-Seoane}}\ and\ \bibinfo {author} {\bibfnamefont {X.}~\bibnamefont
  {Chen}},\ }\bibfield  {title} {\bibinfo {title} {{Relativistic mergers of
  black hole binaries have large, similar masses, low spins and are
  circular}},\ }\href {https://doi.org/10.1093/mnras/stw503} {\bibfield
  {journal} {\bibinfo  {journal} {Mon. Not. Roy. Astron. Soc.}\ }\textbf
  {\bibinfo {volume} {458}},\ \bibinfo {pages} {3075} (\bibinfo {year}
  {2016})},\ \Eprint {https://arxiv.org/abs/1512.04897} {arXiv:1512.04897
  [astro-ph.CO]} \BibitemShut {NoStop}%
\bibitem [{\citenamefont {Khan}\ \emph {et~al.}(2019)\citenamefont {Khan},
  \citenamefont {Chatziioannou}, \citenamefont {Hannam},\ and\ \citenamefont
  {Ohme}}]{Khan:2018}%
  \BibitemOpen
  \bibfield  {author} {\bibinfo {author} {\bibfnamefont {S.}~\bibnamefont
  {Khan}}, \bibinfo {author} {\bibfnamefont {K.}~\bibnamefont {Chatziioannou}},
  \bibinfo {author} {\bibfnamefont {M.}~\bibnamefont {Hannam}},\ and\ \bibinfo
  {author} {\bibfnamefont {F.}~\bibnamefont {Ohme}},\ }\bibfield  {title}
  {\bibinfo {title} {{Phenomenological model for the gravitational-wave signal
  from precessing binary black holes with two-spin effects}},\ }\href
  {https://doi.org/10.1103/PhysRevD.100.024059} {\bibfield  {journal} {\bibinfo
   {journal} {Phys. Rev. D}\ }\textbf {\bibinfo {volume} {100}},\ \bibinfo
  {pages} {024059} (\bibinfo {year} {2019})},\ \Eprint
  {https://arxiv.org/abs/1809.10113} {arXiv:1809.10113 [gr-qc]} \BibitemShut
  {NoStop}%
\bibitem [{\citenamefont {Garc\'\i{}a-Quir\'os}\ \emph
  {et~al.}(2020)\citenamefont {Garc\'\i{}a-Quir\'os}, \citenamefont {Colleoni},
  \citenamefont {Husa}, \citenamefont {Estell\'es}, \citenamefont {Pratten},
  \citenamefont {Ramos-Buades}, \citenamefont {Mateu-Lucena},\ and\
  \citenamefont {Jaume}}]{IMRPhenomXHM}%
  \BibitemOpen
  \bibfield  {author} {\bibinfo {author} {\bibfnamefont {C.}~\bibnamefont
  {Garc\'\i{}a-Quir\'os}}, \bibinfo {author} {\bibfnamefont {M.}~\bibnamefont
  {Colleoni}}, \bibinfo {author} {\bibfnamefont {S.}~\bibnamefont {Husa}},
  \bibinfo {author} {\bibfnamefont {H.}~\bibnamefont {Estell\'es}}, \bibinfo
  {author} {\bibfnamefont {G.}~\bibnamefont {Pratten}}, \bibinfo {author}
  {\bibfnamefont {A.}~\bibnamefont {Ramos-Buades}}, \bibinfo {author}
  {\bibfnamefont {M.}~\bibnamefont {Mateu-Lucena}},\ and\ \bibinfo {author}
  {\bibfnamefont {R.}~\bibnamefont {Jaume}},\ }\bibfield  {title} {\bibinfo
  {title} {{Multimode frequency-domain model for the gravitational wave signal
  from nonprecessing black-hole binaries}},\ }\href
  {https://doi.org/10.1103/PhysRevD.102.064002} {\bibfield  {journal} {\bibinfo
   {journal} {Phys. Rev. D}\ }\textbf {\bibinfo {volume} {102}},\ \bibinfo
  {pages} {064002} (\bibinfo {year} {2020})},\ \Eprint
  {https://arxiv.org/abs/2001.10914} {arXiv:2001.10914 [gr-qc]} \BibitemShut
  {NoStop}%
\bibitem [{\citenamefont {Pratten}\ \emph {et~al.}(2021)\citenamefont {Pratten}
  \emph {et~al.}}]{IMRPhenomXPHM}%
  \BibitemOpen
  \bibfield  {author} {\bibinfo {author} {\bibfnamefont {G.}~\bibnamefont
  {Pratten}} \emph {et~al.},\ }\bibfield  {title} {\bibinfo {title}
  {{Computationally efficient models for the dominant and subdominant harmonic
  modes of precessing binary black holes}},\ }\href
  {https://doi.org/10.1103/PhysRevD.103.104056} {\bibfield  {journal} {\bibinfo
   {journal} {Phys. Rev. D}\ }\textbf {\bibinfo {volume} {103}},\ \bibinfo
  {pages} {104056} (\bibinfo {year} {2021})},\ \Eprint
  {https://arxiv.org/abs/2004.06503} {arXiv:2004.06503 [gr-qc]} \BibitemShut
  {NoStop}%
\bibitem [{\citenamefont {Estell\'es}\ \emph {et~al.}(2021)\citenamefont
  {Estell\'es}, \citenamefont {Ramos-Buades}, \citenamefont {Husa},
  \citenamefont {Garc\'\i{}a-Quir\'os}, \citenamefont {Colleoni}, \citenamefont
  {Haegel},\ and\ \citenamefont {Jaume}}]{IMRPhenomT}%
  \BibitemOpen
  \bibfield  {author} {\bibinfo {author} {\bibfnamefont {H.}~\bibnamefont
  {Estell\'es}}, \bibinfo {author} {\bibfnamefont {A.}~\bibnamefont
  {Ramos-Buades}}, \bibinfo {author} {\bibfnamefont {S.}~\bibnamefont {Husa}},
  \bibinfo {author} {\bibfnamefont {C.}~\bibnamefont {Garc\'\i{}a-Quir\'os}},
  \bibinfo {author} {\bibfnamefont {M.}~\bibnamefont {Colleoni}}, \bibinfo
  {author} {\bibfnamefont {L.}~\bibnamefont {Haegel}},\ and\ \bibinfo {author}
  {\bibfnamefont {R.}~\bibnamefont {Jaume}},\ }\bibfield  {title} {\bibinfo
  {title} {{Phenomenological time domain model for dominant quadrupole
  gravitational wave signal of coalescing binary black holes}},\ }\href
  {https://doi.org/10.1103/PhysRevD.103.124060} {\bibfield  {journal} {\bibinfo
   {journal} {Phys. Rev. D}\ }\textbf {\bibinfo {volume} {103}},\ \bibinfo
  {pages} {124060} (\bibinfo {year} {2021})},\ \Eprint
  {https://arxiv.org/abs/2004.08302} {arXiv:2004.08302 [gr-qc]} \BibitemShut
  {NoStop}%
\bibitem [{\citenamefont {Estell\'es}\ \emph {et~al.}(2022)\citenamefont
  {Estell\'es}, \citenamefont {Colleoni}, \citenamefont {Garc\'\i{}a-Quir\'os},
  \citenamefont {Husa}, \citenamefont {Keitel}, \citenamefont {Mateu-Lucena},
  \citenamefont {Planas},\ and\ \citenamefont {Ramos-Buades}}]{IMRPhenomTPHM}%
  \BibitemOpen
  \bibfield  {author} {\bibinfo {author} {\bibfnamefont {H.}~\bibnamefont
  {Estell\'es}}, \bibinfo {author} {\bibfnamefont {M.}~\bibnamefont
  {Colleoni}}, \bibinfo {author} {\bibfnamefont {C.}~\bibnamefont
  {Garc\'\i{}a-Quir\'os}}, \bibinfo {author} {\bibfnamefont {S.}~\bibnamefont
  {Husa}}, \bibinfo {author} {\bibfnamefont {D.}~\bibnamefont {Keitel}},
  \bibinfo {author} {\bibfnamefont {M.}~\bibnamefont {Mateu-Lucena}}, \bibinfo
  {author} {\bibfnamefont {M.~d.~L.}\ \bibnamefont {Planas}},\ and\ \bibinfo
  {author} {\bibfnamefont {A.}~\bibnamefont {Ramos-Buades}},\ }\bibfield
  {title} {\bibinfo {title} {{New twists in compact binary waveform modeling: A
  fast time-domain model for precession}},\ }\href
  {https://doi.org/10.1103/PhysRevD.105.084040} {\bibfield  {journal} {\bibinfo
   {journal} {Phys. Rev. D}\ }\textbf {\bibinfo {volume} {105}},\ \bibinfo
  {pages} {084040} (\bibinfo {year} {2022})},\ \Eprint
  {https://arxiv.org/abs/2105.05872} {arXiv:2105.05872 [gr-qc]} \BibitemShut
  {NoStop}%
\bibitem [{\citenamefont {Ramos-Buades}\ \emph {et~al.}(2022)\citenamefont
  {Ramos-Buades}, \citenamefont {Buonanno}, \citenamefont {Khalil},\ and\
  \citenamefont {Ossokine}}]{SEOBNRv4EHM}%
  \BibitemOpen
  \bibfield  {author} {\bibinfo {author} {\bibfnamefont {A.}~\bibnamefont
  {Ramos-Buades}}, \bibinfo {author} {\bibfnamefont {A.}~\bibnamefont
  {Buonanno}}, \bibinfo {author} {\bibfnamefont {M.}~\bibnamefont {Khalil}},\
  and\ \bibinfo {author} {\bibfnamefont {S.}~\bibnamefont {Ossokine}},\
  }\bibfield  {title} {\bibinfo {title} {{Effective-one-body multipolar
  waveforms for eccentric binary black holes with nonprecessing spins}},\
  }\href {https://doi.org/10.1103/PhysRevD.105.044035} {\bibfield  {journal}
  {\bibinfo  {journal} {Phys. Rev. D}\ }\textbf {\bibinfo {volume} {105}},\
  \bibinfo {pages} {044035} (\bibinfo {year} {2022})},\ \Eprint
  {https://arxiv.org/abs/2112.06952} {arXiv:2112.06952 [gr-qc]} \BibitemShut
  {NoStop}%
\bibitem [{\citenamefont {Pompili}\ \emph {et~al.}(2023)\citenamefont {Pompili}
  \emph {et~al.}}]{SEOBNRv5HM}%
  \BibitemOpen
  \bibfield  {author} {\bibinfo {author} {\bibfnamefont {L.}~\bibnamefont
  {Pompili}} \emph {et~al.},\ }\bibfield  {title} {\bibinfo {title} {{Laying
  the foundation of the effective-one-body waveform models SEOBNRv5: Improved
  accuracy and efficiency for spinning nonprecessing binary black holes}},\
  }\href {https://doi.org/10.1103/PhysRevD.108.124035} {\bibfield  {journal}
  {\bibinfo  {journal} {Phys. Rev. D}\ }\textbf {\bibinfo {volume} {108}},\
  \bibinfo {pages} {124035} (\bibinfo {year} {2023})},\ \Eprint
  {https://arxiv.org/abs/2303.18039} {arXiv:2303.18039 [gr-qc]} \BibitemShut
  {NoStop}%
\bibitem [{\citenamefont {Ramos-Buades}\ \emph
  {et~al.}(2023{\natexlab{b}})\citenamefont {Ramos-Buades}, \citenamefont
  {Buonanno}, \citenamefont {Estell\'es}, \citenamefont {Khalil}, \citenamefont
  {Mihaylov}, \citenamefont {Ossokine}, \citenamefont {Pompili},\ and\
  \citenamefont {Shiferaw}}]{SEOBNRv5PHM}%
  \BibitemOpen
  \bibfield  {author} {\bibinfo {author} {\bibfnamefont {A.}~\bibnamefont
  {Ramos-Buades}}, \bibinfo {author} {\bibfnamefont {A.}~\bibnamefont
  {Buonanno}}, \bibinfo {author} {\bibfnamefont {H.}~\bibnamefont
  {Estell\'es}}, \bibinfo {author} {\bibfnamefont {M.}~\bibnamefont {Khalil}},
  \bibinfo {author} {\bibfnamefont {D.~P.}\ \bibnamefont {Mihaylov}}, \bibinfo
  {author} {\bibfnamefont {S.}~\bibnamefont {Ossokine}}, \bibinfo {author}
  {\bibfnamefont {L.}~\bibnamefont {Pompili}},\ and\ \bibinfo {author}
  {\bibfnamefont {M.}~\bibnamefont {Shiferaw}},\ }\bibfield  {title} {\bibinfo
  {title} {{Next generation of accurate and efficient multipolar
  precessing-spin effective-one-body waveforms for binary black holes}},\
  }\href {https://doi.org/10.1103/PhysRevD.108.124037} {\bibfield  {journal}
  {\bibinfo  {journal} {Phys. Rev. D}\ }\textbf {\bibinfo {volume} {108}},\
  \bibinfo {pages} {124037} (\bibinfo {year} {2023}{\natexlab{b}})},\ \Eprint
  {https://arxiv.org/abs/2303.18046} {arXiv:2303.18046 [gr-qc]} \BibitemShut
  {NoStop}%
\bibitem [{\citenamefont {Nagar}\ and\ \citenamefont
  {Rettegno}(2021)}]{TEOBResumS-Dali}%
  \BibitemOpen
  \bibfield  {author} {\bibinfo {author} {\bibfnamefont {A.}~\bibnamefont
  {Nagar}}\ and\ \bibinfo {author} {\bibfnamefont {P.}~\bibnamefont
  {Rettegno}},\ }\bibfield  {title} {\bibinfo {title} {{Next generation: Impact
  of high-order analytical information on effective one body waveform models
  for noncircularized, spin-aligned black hole binaries}},\ }\href
  {https://doi.org/10.1103/PhysRevD.104.104004} {\bibfield  {journal} {\bibinfo
   {journal} {Phys. Rev. D}\ }\textbf {\bibinfo {volume} {104}},\ \bibinfo
  {pages} {104004} (\bibinfo {year} {2021})},\ \Eprint
  {https://arxiv.org/abs/2108.02043} {arXiv:2108.02043 [gr-qc]} \BibitemShut
  {NoStop}%
\bibitem [{\citenamefont {Gamba}\ \emph {et~al.}(2022)\citenamefont {Gamba},
  \citenamefont {Ak\c{c}ay}, \citenamefont {Bernuzzi},\ and\ \citenamefont
  {Williams}}]{TEOBResumS-Giotto}%
  \BibitemOpen
  \bibfield  {author} {\bibinfo {author} {\bibfnamefont {R.}~\bibnamefont
  {Gamba}}, \bibinfo {author} {\bibfnamefont {S.}~\bibnamefont {Ak\c{c}ay}},
  \bibinfo {author} {\bibfnamefont {S.}~\bibnamefont {Bernuzzi}},\ and\
  \bibinfo {author} {\bibfnamefont {J.}~\bibnamefont {Williams}},\ }\bibfield
  {title} {\bibinfo {title} {{Effective-one-body waveforms for precessing
  coalescing compact binaries with post-Newtonian twist}},\ }\href
  {https://doi.org/10.1103/PhysRevD.106.024020} {\bibfield  {journal} {\bibinfo
   {journal} {Phys. Rev. D}\ }\textbf {\bibinfo {volume} {106}},\ \bibinfo
  {pages} {024020} (\bibinfo {year} {2022})},\ \Eprint
  {https://arxiv.org/abs/2111.03675} {arXiv:2111.03675 [gr-qc]} \BibitemShut
  {NoStop}%
\bibitem [{\citenamefont {{Liu}}\ \emph {et~al.}(2023)\citenamefont {{Liu}},
  \citenamefont {{Cao}},\ and\ \citenamefont {{Zhu}}}]{SEOBNRPE}%
  \BibitemOpen
  \bibfield  {author} {\bibinfo {author} {\bibfnamefont {X.}~\bibnamefont
  {{Liu}}}, \bibinfo {author} {\bibfnamefont {Z.}~\bibnamefont {{Cao}}},\ and\
  \bibinfo {author} {\bibfnamefont {Z.-H.}\ \bibnamefont {{Zhu}}},\ }\bibfield
  {title} {\bibinfo {title} {{Effective-One-Body Numerical-Relativity waveform
  model for Eccentric spin-precessing binary black hole coalescence}},\ }\href
  {https://doi.org/10.48550/arXiv.2310.04552} {\bibfield  {journal} {\bibinfo
  {journal} {arXiv e-prints}\ ,\ \bibinfo {eid} {arXiv:2310.04552}} (\bibinfo
  {year} {2023})},\ \Eprint {https://arxiv.org/abs/2310.04552}
  {arXiv:2310.04552 [gr-qc]} \BibitemShut {NoStop}%
\bibitem [{\citenamefont {{Wang}}\ \emph {et~al.}(2023)\citenamefont {{Wang}},
  \citenamefont {{Zou}}, \citenamefont {{Wu}},\ and\ \citenamefont
  {{Liu}}}]{Wang:2023}%
  \BibitemOpen
  \bibfield  {author} {\bibinfo {author} {\bibfnamefont {H.}~\bibnamefont
  {{Wang}}}, \bibinfo {author} {\bibfnamefont {Y.-C.}\ \bibnamefont {{Zou}}},
  \bibinfo {author} {\bibfnamefont {Q.-W.}\ \bibnamefont {{Wu}}},\ and\
  \bibinfo {author} {\bibfnamefont {Y.}~\bibnamefont {{Liu}}},\ }\bibfield
  {title} {\bibinfo {title} {{Unveiling the Fingerprint of Eccentric Binary
  Black Hole Mergers}},\ }\href {https://doi.org/10.48550/arXiv.2311.08822}
  {\bibfield  {journal} {\bibinfo  {journal} {arXiv e-prints}\ ,\ \bibinfo
  {eid} {arXiv:2311.08822}} (\bibinfo {year} {2023})},\ \Eprint
  {https://arxiv.org/abs/2311.08822} {arXiv:2311.08822 [gr-qc]} \BibitemShut
  {NoStop}%
\bibitem [{\citenamefont {{Carullo}}\ \emph {et~al.}(2023)\citenamefont
  {{Carullo}}, \citenamefont {{Albanesi}}, \citenamefont {{Nagar}},
  \citenamefont {{Gamba}}, \citenamefont {{Bernuzzi}}, \citenamefont
  {{Andrade}},\ and\ \citenamefont {{Trenado}}}]{Carullo:2023}%
  \BibitemOpen
  \bibfield  {author} {\bibinfo {author} {\bibfnamefont {G.}~\bibnamefont
  {{Carullo}}}, \bibinfo {author} {\bibfnamefont {S.}~\bibnamefont
  {{Albanesi}}}, \bibinfo {author} {\bibfnamefont {A.}~\bibnamefont {{Nagar}}},
  \bibinfo {author} {\bibfnamefont {R.}~\bibnamefont {{Gamba}}}, \bibinfo
  {author} {\bibfnamefont {S.}~\bibnamefont {{Bernuzzi}}}, \bibinfo {author}
  {\bibfnamefont {T.}~\bibnamefont {{Andrade}}},\ and\ \bibinfo {author}
  {\bibfnamefont {J.}~\bibnamefont {{Trenado}}},\ }\bibfield  {title} {\bibinfo
  {title} {{Unveiling the merger structure of black hole binaries in generic
  planar orbits}},\ }\href {https://doi.org/10.48550/arXiv.2309.07228}
  {\bibfield  {journal} {\bibinfo  {journal} {arXiv e-prints}\ ,\ \bibinfo
  {eid} {arXiv:2309.07228}} (\bibinfo {year} {2023})},\ \Eprint
  {https://arxiv.org/abs/2309.07228} {arXiv:2309.07228 [gr-qc]} \BibitemShut
  {NoStop}%
\bibitem [{\citenamefont {Hannam}\ \emph {et~al.}(2014)\citenamefont {Hannam},
  \citenamefont {Schmidt}, \citenamefont {Boh\'e}, \citenamefont {Haegel},
  \citenamefont {Husa}, \citenamefont {Ohme}, \citenamefont {Pratten},\ and\
  \citenamefont {P\"urrer}}]{Hannam:2013}%
  \BibitemOpen
  \bibfield  {author} {\bibinfo {author} {\bibfnamefont {M.}~\bibnamefont
  {Hannam}}, \bibinfo {author} {\bibfnamefont {P.}~\bibnamefont {Schmidt}},
  \bibinfo {author} {\bibfnamefont {A.}~\bibnamefont {Boh\'e}}, \bibinfo
  {author} {\bibfnamefont {L.}~\bibnamefont {Haegel}}, \bibinfo {author}
  {\bibfnamefont {S.}~\bibnamefont {Husa}}, \bibinfo {author} {\bibfnamefont
  {F.}~\bibnamefont {Ohme}}, \bibinfo {author} {\bibfnamefont {G.}~\bibnamefont
  {Pratten}},\ and\ \bibinfo {author} {\bibfnamefont {M.}~\bibnamefont
  {P\"urrer}},\ }\bibfield  {title} {\bibinfo {title} {{Simple Model of
  Complete Precessing Black-Hole-Binary Gravitational Waveforms}},\ }\href
  {https://doi.org/10.1103/PhysRevLett.113.151101} {\bibfield  {journal}
  {\bibinfo  {journal} {Phys. Rev. Lett.}\ }\textbf {\bibinfo {volume} {113}},\
  \bibinfo {pages} {151101} (\bibinfo {year} {2014})},\ \Eprint
  {https://arxiv.org/abs/1308.3271} {arXiv:1308.3271 [gr-qc]} \BibitemShut
  {NoStop}%
\bibitem [{\citenamefont {Yunes}\ \emph {et~al.}(2009)\citenamefont {Yunes},
  \citenamefont {Arun}, \citenamefont {Berti},\ and\ \citenamefont
  {Will}}]{Yunes:2009}%
  \BibitemOpen
  \bibfield  {author} {\bibinfo {author} {\bibfnamefont {N.}~\bibnamefont
  {Yunes}}, \bibinfo {author} {\bibfnamefont {K.~G.}\ \bibnamefont {Arun}},
  \bibinfo {author} {\bibfnamefont {E.}~\bibnamefont {Berti}},\ and\ \bibinfo
  {author} {\bibfnamefont {C.~M.}\ \bibnamefont {Will}},\ }\bibfield  {title}
  {\bibinfo {title} {{Post-Circular Expansion of Eccentric Binary Inspirals:
  Fourier-Domain Waveforms in the Stationary Phase Approximation}},\ }\href
  {https://doi.org/10.1103/PhysRevD.80.084001} {\bibfield  {journal} {\bibinfo
  {journal} {Phys. Rev. D}\ }\textbf {\bibinfo {volume} {80}},\ \bibinfo
  {pages} {084001} (\bibinfo {year} {2009})},\ \bibinfo {note} {[Erratum:
  Phys.Rev.D 89, 109901 (2014)]},\ \Eprint {https://arxiv.org/abs/0906.0313}
  {arXiv:0906.0313 [gr-qc]} \BibitemShut {NoStop}%
\bibitem [{\citenamefont {Favata}\ \emph {et~al.}(2022)\citenamefont {Favata},
  \citenamefont {Kim}, \citenamefont {Arun}, \citenamefont {Kim},\ and\
  \citenamefont {Lee}}]{Favata:2021}%
  \BibitemOpen
  \bibfield  {author} {\bibinfo {author} {\bibfnamefont {M.}~\bibnamefont
  {Favata}}, \bibinfo {author} {\bibfnamefont {C.}~\bibnamefont {Kim}},
  \bibinfo {author} {\bibfnamefont {K.~G.}\ \bibnamefont {Arun}}, \bibinfo
  {author} {\bibfnamefont {J.}~\bibnamefont {Kim}},\ and\ \bibinfo {author}
  {\bibfnamefont {H.~W.}\ \bibnamefont {Lee}},\ }\bibfield  {title} {\bibinfo
  {title} {{Constraining the orbital eccentricity of inspiralling compact
  binary systems with Advanced LIGO}},\ }\href
  {https://doi.org/10.1103/PhysRevD.105.023003} {\bibfield  {journal} {\bibinfo
   {journal} {Phys. Rev. D}\ }\textbf {\bibinfo {volume} {105}},\ \bibinfo
  {pages} {023003} (\bibinfo {year} {2022})},\ \Eprint
  {https://arxiv.org/abs/2108.05861} {arXiv:2108.05861 [gr-qc]} \BibitemShut
  {NoStop}%
\bibitem [{\citenamefont {{Divyajyoti}}\ \emph {et~al.}(2023)\citenamefont
  {{Divyajyoti}}, \citenamefont {{Kumar}}, \citenamefont {{Tibrewal}},
  \citenamefont {{Romero-Shaw}},\ and\ \citenamefont {{Kant
  Mishra}}}]{Divyajyoti:2023}%
  \BibitemOpen
  \bibfield  {author} {\bibinfo {author} {\bibnamefont {{Divyajyoti}}},
  \bibinfo {author} {\bibfnamefont {S.}~\bibnamefont {{Kumar}}}, \bibinfo
  {author} {\bibfnamefont {S.}~\bibnamefont {{Tibrewal}}}, \bibinfo {author}
  {\bibfnamefont {I.~M.}\ \bibnamefont {{Romero-Shaw}}},\ and\ \bibinfo
  {author} {\bibfnamefont {C.}~\bibnamefont {{Kant Mishra}}},\ }\bibfield
  {title} {\bibinfo {title} {{Blind spots and biases: the dangers of ignoring
  eccentricity in gravitational-wave signals from binary black holes}},\ }\href
  {https://doi.org/10.48550/arXiv.2309.16638} {\bibfield  {journal} {\bibinfo
  {journal} {arXiv e-prints}\ ,\ \bibinfo {eid} {arXiv:2309.16638}} (\bibinfo
  {year} {2023})},\ \Eprint {https://arxiv.org/abs/2309.16638}
  {arXiv:2309.16638 [gr-qc]} \BibitemShut {NoStop}%
\bibitem [{\citenamefont {Moore}\ \emph {et~al.}(2018)\citenamefont {Moore},
  \citenamefont {Robson}, \citenamefont {Loutrel},\ and\ \citenamefont
  {Yunes}}]{Moore:2018}%
  \BibitemOpen
  \bibfield  {author} {\bibinfo {author} {\bibfnamefont {B.}~\bibnamefont
  {Moore}}, \bibinfo {author} {\bibfnamefont {T.}~\bibnamefont {Robson}},
  \bibinfo {author} {\bibfnamefont {N.}~\bibnamefont {Loutrel}},\ and\ \bibinfo
  {author} {\bibfnamefont {N.}~\bibnamefont {Yunes}},\ }\bibfield  {title}
  {\bibinfo {title} {{Towards a Fourier domain waveform for non-spinning
  binaries with arbitrary eccentricity}},\ }\href
  {https://doi.org/10.1088/1361-6382/aaea00} {\bibfield  {journal} {\bibinfo
  {journal} {Class. Quant. Grav.}\ }\textbf {\bibinfo {volume} {35}},\ \bibinfo
  {pages} {235006} (\bibinfo {year} {2018})},\ \Eprint
  {https://arxiv.org/abs/1807.07163} {arXiv:1807.07163 [gr-qc]} \BibitemShut
  {NoStop}%
\bibitem [{\citenamefont {Moore}\ and\ \citenamefont
  {Yunes}(2019)}]{Moore:2019}%
  \BibitemOpen
  \bibfield  {author} {\bibinfo {author} {\bibfnamefont {B.}~\bibnamefont
  {Moore}}\ and\ \bibinfo {author} {\bibfnamefont {N.}~\bibnamefont {Yunes}},\
  }\bibfield  {title} {\bibinfo {title} {{A 3PN Fourier Domain Waveform for
  Non-Spinning Binaries with Moderate Eccentricity}},\ }\href
  {https://doi.org/10.1088/1361-6382/ab3778} {\bibfield  {journal} {\bibinfo
  {journal} {Class. Quant. Grav.}\ }\textbf {\bibinfo {volume} {36}},\ \bibinfo
  {pages} {185003} (\bibinfo {year} {2019})},\ \Eprint
  {https://arxiv.org/abs/1903.05203} {arXiv:1903.05203 [gr-qc]} \BibitemShut
  {NoStop}%
\bibitem [{\citenamefont {Romero-Shaw}\ \emph {et~al.}(2023)\citenamefont
  {Romero-Shaw}, \citenamefont {Gerosa},\ and\ \citenamefont
  {Loutrel}}]{Romero-Shaw:2022}%
  \BibitemOpen
  \bibfield  {author} {\bibinfo {author} {\bibfnamefont {I.~M.}\ \bibnamefont
  {Romero-Shaw}}, \bibinfo {author} {\bibfnamefont {D.}~\bibnamefont
  {Gerosa}},\ and\ \bibinfo {author} {\bibfnamefont {N.}~\bibnamefont
  {Loutrel}},\ }\bibfield  {title} {\bibinfo {title} {{Eccentricity or spin
  precession? Distinguishing subdominant effects in gravitational-wave data}},\
  }\href {https://doi.org/10.1093/mnras/stad031} {\bibfield  {journal}
  {\bibinfo  {journal} {Mon. Not. Roy. Astron. Soc.}\ }\textbf {\bibinfo
  {volume} {519}},\ \bibinfo {pages} {5352} (\bibinfo {year} {2023})},\ \Eprint
  {https://arxiv.org/abs/2211.07528} {arXiv:2211.07528 [astro-ph.HE]}
  \BibitemShut {NoStop}%
\bibitem [{\citenamefont {{Klein}}(2021)}]{Klein:2021}%
  \BibitemOpen
  \bibfield  {author} {\bibinfo {author} {\bibfnamefont {A.}~\bibnamefont
  {{Klein}}},\ }\bibfield  {title} {\bibinfo {title} {{EFPE: Efficient fully
  precessing eccentric gravitational waveforms for binaries with long
  inspirals}},\ }\href@noop {} {\bibfield  {journal} {\bibinfo  {journal}
  {arXiv e-prints}\ ,\ \bibinfo {eid} {arXiv:2106.10291}} (\bibinfo {year}
  {2021})},\ \Eprint {https://arxiv.org/abs/2106.10291} {arXiv:2106.10291
  [gr-qc]} \BibitemShut {NoStop}%
\bibitem [{\citenamefont {Klein}\ \emph {et~al.}(2018)\citenamefont {Klein},
  \citenamefont {Boetzel}, \citenamefont {Gopakumar}, \citenamefont {Jetzer},\
  and\ \citenamefont {de~Vittori}}]{Klein:2018}%
  \BibitemOpen
  \bibfield  {author} {\bibinfo {author} {\bibfnamefont {A.}~\bibnamefont
  {Klein}}, \bibinfo {author} {\bibfnamefont {Y.}~\bibnamefont {Boetzel}},
  \bibinfo {author} {\bibfnamefont {A.}~\bibnamefont {Gopakumar}}, \bibinfo
  {author} {\bibfnamefont {P.}~\bibnamefont {Jetzer}},\ and\ \bibinfo {author}
  {\bibfnamefont {L.}~\bibnamefont {de~Vittori}},\ }\bibfield  {title}
  {\bibinfo {title} {{Fourier domain gravitational waveforms for precessing
  eccentric binaries}},\ }\href {https://doi.org/10.1103/PhysRevD.98.104043}
  {\bibfield  {journal} {\bibinfo  {journal} {Phys. Rev. D}\ }\textbf {\bibinfo
  {volume} {98}},\ \bibinfo {pages} {104043} (\bibinfo {year} {2018})},\
  \Eprint {https://arxiv.org/abs/1801.08542} {arXiv:1801.08542 [gr-qc]}
  \BibitemShut {NoStop}%
\bibitem [{\citenamefont {P\"urrer}(2016)}]{Purrer:2015}%
  \BibitemOpen
  \bibfield  {author} {\bibinfo {author} {\bibfnamefont {M.}~\bibnamefont
  {P\"urrer}},\ }\bibfield  {title} {\bibinfo {title} {{Frequency domain
  reduced order model of aligned-spin effective-one-body waveforms with generic
  mass-ratios and spins}},\ }\href {https://doi.org/10.1103/PhysRevD.93.064041}
  {\bibfield  {journal} {\bibinfo  {journal} {Phys. Rev. D}\ }\textbf {\bibinfo
  {volume} {93}},\ \bibinfo {pages} {064041} (\bibinfo {year} {2016})},\
  \Eprint {https://arxiv.org/abs/1512.02248} {arXiv:1512.02248 [gr-qc]}
  \BibitemShut {NoStop}%
\bibitem [{\citenamefont {Smith}\ \emph {et~al.}(2016)\citenamefont {Smith},
  \citenamefont {Field}, \citenamefont {Blackburn}, \citenamefont {Haster},
  \citenamefont {P\"urrer}, \citenamefont {Raymond},\ and\ \citenamefont
  {Schmidt}}]{Smith:2016}%
  \BibitemOpen
  \bibfield  {author} {\bibinfo {author} {\bibfnamefont {R.}~\bibnamefont
  {Smith}}, \bibinfo {author} {\bibfnamefont {S.~E.}\ \bibnamefont {Field}},
  \bibinfo {author} {\bibfnamefont {K.}~\bibnamefont {Blackburn}}, \bibinfo
  {author} {\bibfnamefont {C.-J.}\ \bibnamefont {Haster}}, \bibinfo {author}
  {\bibfnamefont {M.}~\bibnamefont {P\"urrer}}, \bibinfo {author}
  {\bibfnamefont {V.}~\bibnamefont {Raymond}},\ and\ \bibinfo {author}
  {\bibfnamefont {P.}~\bibnamefont {Schmidt}},\ }\bibfield  {title} {\bibinfo
  {title} {{Fast and accurate inference on gravitational waves from precessing
  compact binaries}},\ }\href {https://doi.org/10.1103/PhysRevD.94.044031}
  {\bibfield  {journal} {\bibinfo  {journal} {Phys. Rev. D}\ }\textbf {\bibinfo
  {volume} {94}},\ \bibinfo {pages} {044031} (\bibinfo {year} {2016})},\
  \Eprint {https://arxiv.org/abs/1604.08253} {arXiv:1604.08253 [gr-qc]}
  \BibitemShut {NoStop}%
\bibitem [{\citenamefont {Thomas}\ \emph {et~al.}(2022)\citenamefont {Thomas},
  \citenamefont {Pratten},\ and\ \citenamefont {Schmidt}}]{Thomas:2022}%
  \BibitemOpen
  \bibfield  {author} {\bibinfo {author} {\bibfnamefont {L.~M.}\ \bibnamefont
  {Thomas}}, \bibinfo {author} {\bibfnamefont {G.}~\bibnamefont {Pratten}},\
  and\ \bibinfo {author} {\bibfnamefont {P.}~\bibnamefont {Schmidt}},\
  }\bibfield  {title} {\bibinfo {title} {{Accelerating multimodal gravitational
  waveforms from precessing compact binaries with artificial neural
  networks}},\ }\href {https://doi.org/10.1103/PhysRevD.106.104029} {\bibfield
  {journal} {\bibinfo  {journal} {Phys. Rev. D}\ }\textbf {\bibinfo {volume}
  {106}},\ \bibinfo {pages} {104029} (\bibinfo {year} {2022})},\ \Eprint
  {https://arxiv.org/abs/2205.14066} {arXiv:2205.14066 [gr-qc]} \BibitemShut
  {NoStop}%
\bibitem [{\citenamefont {Chatziioannou}\ \emph {et~al.}(2017)\citenamefont
  {Chatziioannou}, \citenamefont {Klein}, \citenamefont {Yunes},\ and\
  \citenamefont {Cornish}}]{Chatziioannou:2017}%
  \BibitemOpen
  \bibfield  {author} {\bibinfo {author} {\bibfnamefont {K.}~\bibnamefont
  {Chatziioannou}}, \bibinfo {author} {\bibfnamefont {A.}~\bibnamefont
  {Klein}}, \bibinfo {author} {\bibfnamefont {N.}~\bibnamefont {Yunes}},\ and\
  \bibinfo {author} {\bibfnamefont {N.}~\bibnamefont {Cornish}},\ }\bibfield
  {title} {\bibinfo {title} {{Constructing Gravitational Waves from Generic
  Spin-Precessing Compact Binary Inspirals}},\ }\href
  {https://doi.org/10.1103/PhysRevD.95.104004} {\bibfield  {journal} {\bibinfo
  {journal} {Phys. Rev. D}\ }\textbf {\bibinfo {volume} {95}},\ \bibinfo
  {pages} {104004} (\bibinfo {year} {2017})},\ \Eprint
  {https://arxiv.org/abs/1703.03967} {arXiv:1703.03967 [gr-qc]} \BibitemShut
  {NoStop}%
\bibitem [{\citenamefont {Schmidt}\ \emph {et~al.}(2012)\citenamefont
  {Schmidt}, \citenamefont {Hannam},\ and\ \citenamefont
  {Husa}}]{Schmidt:2012}%
  \BibitemOpen
  \bibfield  {author} {\bibinfo {author} {\bibfnamefont {P.}~\bibnamefont
  {Schmidt}}, \bibinfo {author} {\bibfnamefont {M.}~\bibnamefont {Hannam}},\
  and\ \bibinfo {author} {\bibfnamefont {S.}~\bibnamefont {Husa}},\ }\bibfield
  {title} {\bibinfo {title} {{Towards models of gravitational waveforms from
  generic binaries: A simple approximate mapping between precessing and
  non-precessing inspiral signals}},\ }\href
  {https://doi.org/10.1103/PhysRevD.86.104063} {\bibfield  {journal} {\bibinfo
  {journal} {Phys. Rev. D}\ }\textbf {\bibinfo {volume} {86}},\ \bibinfo
  {pages} {104063} (\bibinfo {year} {2012})},\ \Eprint
  {https://arxiv.org/abs/1207.3088} {arXiv:1207.3088 [gr-qc]} \BibitemShut
  {NoStop}%
\bibitem [{\citenamefont {Klein}\ and\ \citenamefont
  {Jetzer}(2010)}]{Klein:2010}%
  \BibitemOpen
  \bibfield  {author} {\bibinfo {author} {\bibfnamefont {A.}~\bibnamefont
  {Klein}}\ and\ \bibinfo {author} {\bibfnamefont {P.}~\bibnamefont {Jetzer}},\
  }\bibfield  {title} {\bibinfo {title} {{Spin effects in the phasing of
  gravitational waves from binaries on eccentric orbits}},\ }\href
  {https://doi.org/10.1103/PhysRevD.81.124001} {\bibfield  {journal} {\bibinfo
  {journal} {Phys. Rev. D}\ }\textbf {\bibinfo {volume} {81}},\ \bibinfo
  {pages} {124001} (\bibinfo {year} {2010})},\ \Eprint
  {https://arxiv.org/abs/1005.2046} {arXiv:1005.2046 [gr-qc]} \BibitemShut
  {NoStop}%
\bibitem [{\citenamefont {Racine}(2008)}]{Racine:2008}%
  \BibitemOpen
  \bibfield  {author} {\bibinfo {author} {\bibfnamefont {E.}~\bibnamefont
  {Racine}},\ }\bibfield  {title} {\bibinfo {title} {{Analysis of spin
  precession in binary black hole systems including quadrupole-monopole
  interaction}},\ }\href {https://doi.org/10.1103/PhysRevD.78.044021}
  {\bibfield  {journal} {\bibinfo  {journal} {Phys. Rev. D}\ }\textbf {\bibinfo
  {volume} {78}},\ \bibinfo {pages} {044021} (\bibinfo {year} {2008})},\
  \Eprint {https://arxiv.org/abs/0803.1820} {arXiv:0803.1820 [gr-qc]}
  \BibitemShut {NoStop}%
\bibitem [{\citenamefont {Brizard}(2007)}]{Brizard:2007}%
  \BibitemOpen
  \bibfield  {author} {\bibinfo {author} {\bibfnamefont {A.~J.}\ \bibnamefont
  {Brizard}},\ }\href@noop {} {\bibinfo {title} {A primer on elliptic functions
  with applications in classical mechanics}} (\bibinfo {year} {2007}),\ \Eprint
  {https://arxiv.org/abs/0711.4064} {arXiv:0711.4064 [physics.class-ph]}
  \BibitemShut {NoStop}%
\bibitem [{\citenamefont {Kidder}(2008)}]{Kidder:2007}%
  \BibitemOpen
  \bibfield  {author} {\bibinfo {author} {\bibfnamefont {L.~E.}\ \bibnamefont
  {Kidder}},\ }\bibfield  {title} {\bibinfo {title} {{Using full information
  when computing modes of post-Newtonian waveforms from inspiralling compact
  binaries in circular orbit}},\ }\href
  {https://doi.org/10.1103/PhysRevD.77.044016} {\bibfield  {journal} {\bibinfo
  {journal} {Phys. Rev. D}\ }\textbf {\bibinfo {volume} {77}},\ \bibinfo
  {pages} {044016} (\bibinfo {year} {2008})},\ \Eprint
  {https://arxiv.org/abs/0710.0614} {arXiv:0710.0614 [gr-qc]} \BibitemShut
  {NoStop}%
\bibitem [{\citenamefont {Blanchet}\ \emph {et~al.}(2008)\citenamefont
  {Blanchet}, \citenamefont {Faye}, \citenamefont {Iyer},\ and\ \citenamefont
  {Sinha}}]{Blanchet:2008}%
  \BibitemOpen
  \bibfield  {author} {\bibinfo {author} {\bibfnamefont {L.}~\bibnamefont
  {Blanchet}}, \bibinfo {author} {\bibfnamefont {G.}~\bibnamefont {Faye}},
  \bibinfo {author} {\bibfnamefont {B.~R.}\ \bibnamefont {Iyer}},\ and\
  \bibinfo {author} {\bibfnamefont {S.}~\bibnamefont {Sinha}},\ }\bibfield
  {title} {\bibinfo {title} {{The Third post-Newtonian gravitational wave
  polarisations and associated spherical harmonic modes for inspiralling
  compact binaries in quasi-circular orbits}},\ }\href
  {https://doi.org/10.1088/0264-9381/25/16/165003} {\bibfield  {journal}
  {\bibinfo  {journal} {Class. Quant. Grav.}\ }\textbf {\bibinfo {volume}
  {25}},\ \bibinfo {pages} {165003} (\bibinfo {year} {2008})},\ \bibinfo {note}
  {[Erratum: Class.Quant.Grav. 29, 239501 (2012)]},\ \Eprint
  {https://arxiv.org/abs/0802.1249} {arXiv:0802.1249 [gr-qc]} \BibitemShut
  {NoStop}%
\bibitem [{\citenamefont {Schmidt}\ \emph {et~al.}(2011)\citenamefont
  {Schmidt}, \citenamefont {Hannam}, \citenamefont {Husa},\ and\ \citenamefont
  {Ajith}}]{Schmidt:2010}%
  \BibitemOpen
  \bibfield  {author} {\bibinfo {author} {\bibfnamefont {P.}~\bibnamefont
  {Schmidt}}, \bibinfo {author} {\bibfnamefont {M.}~\bibnamefont {Hannam}},
  \bibinfo {author} {\bibfnamefont {S.}~\bibnamefont {Husa}},\ and\ \bibinfo
  {author} {\bibfnamefont {P.}~\bibnamefont {Ajith}},\ }\bibfield  {title}
  {\bibinfo {title} {{Tracking the precession of compact binaries from their
  gravitational-wave signal}},\ }\href
  {https://doi.org/10.1103/PhysRevD.84.024046} {\bibfield  {journal} {\bibinfo
  {journal} {Phys. Rev. D}\ }\textbf {\bibinfo {volume} {84}},\ \bibinfo
  {pages} {024046} (\bibinfo {year} {2011})},\ \Eprint
  {https://arxiv.org/abs/1012.2879} {arXiv:1012.2879 [gr-qc]} \BibitemShut
  {NoStop}%
\bibitem [{\citenamefont {O'Shaughnessy}\ \emph {et~al.}(2011)\citenamefont
  {O'Shaughnessy}, \citenamefont {Vaishnav}, \citenamefont {Healy},
  \citenamefont {Meeks},\ and\ \citenamefont {Shoemaker}}]{OShaughnessy:2011}%
  \BibitemOpen
  \bibfield  {author} {\bibinfo {author} {\bibfnamefont {R.}~\bibnamefont
  {O'Shaughnessy}}, \bibinfo {author} {\bibfnamefont {B.}~\bibnamefont
  {Vaishnav}}, \bibinfo {author} {\bibfnamefont {J.}~\bibnamefont {Healy}},
  \bibinfo {author} {\bibfnamefont {Z.}~\bibnamefont {Meeks}},\ and\ \bibinfo
  {author} {\bibfnamefont {D.}~\bibnamefont {Shoemaker}},\ }\bibfield  {title}
  {\bibinfo {title} {{Efficient asymptotic frame selection for binary black
  hole spacetimes using asymptotic radiation}},\ }\href
  {https://doi.org/10.1103/PhysRevD.84.124002} {\bibfield  {journal} {\bibinfo
  {journal} {Phys. Rev. D}\ }\textbf {\bibinfo {volume} {84}},\ \bibinfo
  {pages} {124002} (\bibinfo {year} {2011})},\ \Eprint
  {https://arxiv.org/abs/1109.5224} {arXiv:1109.5224 [gr-qc]} \BibitemShut
  {NoStop}%
\bibitem [{\citenamefont {Boyle}\ \emph {et~al.}(2011)\citenamefont {Boyle},
  \citenamefont {Owen},\ and\ \citenamefont {Pfeiffer}}]{Boyle:2011}%
  \BibitemOpen
  \bibfield  {author} {\bibinfo {author} {\bibfnamefont {M.}~\bibnamefont
  {Boyle}}, \bibinfo {author} {\bibfnamefont {R.}~\bibnamefont {Owen}},\ and\
  \bibinfo {author} {\bibfnamefont {H.~P.}\ \bibnamefont {Pfeiffer}},\
  }\bibfield  {title} {\bibinfo {title} {{A geometric approach to the
  precession of compact binaries}},\ }\href
  {https://doi.org/10.1103/PhysRevD.84.124011} {\bibfield  {journal} {\bibinfo
  {journal} {Phys. Rev. D}\ }\textbf {\bibinfo {volume} {84}},\ \bibinfo
  {pages} {124011} (\bibinfo {year} {2011})},\ \Eprint
  {https://arxiv.org/abs/1110.2965} {arXiv:1110.2965 [gr-qc]} \BibitemShut
  {NoStop}%
\bibitem [{\citenamefont {Mishra}\ \emph {et~al.}(2015)\citenamefont {Mishra},
  \citenamefont {Arun},\ and\ \citenamefont {Iyer}}]{Mishra:2015}%
  \BibitemOpen
  \bibfield  {author} {\bibinfo {author} {\bibfnamefont {C.~K.}\ \bibnamefont
  {Mishra}}, \bibinfo {author} {\bibfnamefont {K.~G.}\ \bibnamefont {Arun}},\
  and\ \bibinfo {author} {\bibfnamefont {B.~R.}\ \bibnamefont {Iyer}},\
  }\bibfield  {title} {\bibinfo {title} {{Third post-Newtonian gravitational
  waveforms for compact binary systems in general orbits: Instantaneous
  terms}},\ }\href {https://doi.org/10.1103/PhysRevD.91.084040} {\bibfield
  {journal} {\bibinfo  {journal} {Phys. Rev. D}\ }\textbf {\bibinfo {volume}
  {91}},\ \bibinfo {pages} {084040} (\bibinfo {year} {2015})},\ \Eprint
  {https://arxiv.org/abs/1501.07096} {arXiv:1501.07096 [gr-qc]} \BibitemShut
  {NoStop}%
\bibitem [{\citenamefont {Klein}\ \emph {et~al.}(2014)\citenamefont {Klein},
  \citenamefont {Cornish},\ and\ \citenamefont {Yunes}}]{Klein:2014}%
  \BibitemOpen
  \bibfield  {author} {\bibinfo {author} {\bibfnamefont {A.}~\bibnamefont
  {Klein}}, \bibinfo {author} {\bibfnamefont {N.}~\bibnamefont {Cornish}},\
  and\ \bibinfo {author} {\bibfnamefont {N.}~\bibnamefont {Yunes}},\ }\bibfield
   {title} {\bibinfo {title} {{Fast Frequency-domain Waveforms for
  Spin-Precessing Binary Inspirals}},\ }\href
  {https://doi.org/10.1103/PhysRevD.90.124029} {\bibfield  {journal} {\bibinfo
  {journal} {Phys. Rev. D}\ }\textbf {\bibinfo {volume} {90}},\ \bibinfo
  {pages} {124029} (\bibinfo {year} {2014})},\ \Eprint
  {https://arxiv.org/abs/1408.5158} {arXiv:1408.5158 [gr-qc]} \BibitemShut
  {NoStop}%
\bibitem [{\citenamefont {Boetzel}\ \emph {et~al.}(2017)\citenamefont
  {Boetzel}, \citenamefont {Susobhanan}, \citenamefont {Gopakumar},
  \citenamefont {Klein},\ and\ \citenamefont {Jetzer}}]{Boetzel:2017}%
  \BibitemOpen
  \bibfield  {author} {\bibinfo {author} {\bibfnamefont {Y.}~\bibnamefont
  {Boetzel}}, \bibinfo {author} {\bibfnamefont {A.}~\bibnamefont {Susobhanan}},
  \bibinfo {author} {\bibfnamefont {A.}~\bibnamefont {Gopakumar}}, \bibinfo
  {author} {\bibfnamefont {A.}~\bibnamefont {Klein}},\ and\ \bibinfo {author}
  {\bibfnamefont {P.}~\bibnamefont {Jetzer}},\ }\bibfield  {title} {\bibinfo
  {title} {{Solving post-Newtonian accurate Kepler Equation}},\ }\href
  {https://doi.org/10.1103/PhysRevD.96.044011} {\bibfield  {journal} {\bibinfo
  {journal} {Phys. Rev. D}\ }\textbf {\bibinfo {volume} {96}},\ \bibinfo
  {pages} {044011} (\bibinfo {year} {2017})},\ \Eprint
  {https://arxiv.org/abs/1707.02088} {arXiv:1707.02088 [gr-qc]} \BibitemShut
  {NoStop}%
\bibitem [{\citenamefont {Pierro}\ \emph {et~al.}(2001)\citenamefont {Pierro},
  \citenamefont {Pinto}, \citenamefont {Spallicci}, \citenamefont {Laserra},\
  and\ \citenamefont {Recano}}]{Pierro:2000}%
  \BibitemOpen
  \bibfield  {author} {\bibinfo {author} {\bibfnamefont {V.}~\bibnamefont
  {Pierro}}, \bibinfo {author} {\bibfnamefont {I.~M.}\ \bibnamefont {Pinto}},
  \bibinfo {author} {\bibfnamefont {A.~D.}\ \bibnamefont {Spallicci}}, \bibinfo
  {author} {\bibfnamefont {E.}~\bibnamefont {Laserra}},\ and\ \bibinfo {author}
  {\bibfnamefont {F.}~\bibnamefont {Recano}},\ }\bibfield  {title} {\bibinfo
  {title} {{Fast and accurate computation tools for gravitational wave forms
  from binary stars with any orbital eccentricity}},\ }\href
  {https://doi.org/10.1046/j.1365-8711.2001.04442.x} {\bibfield  {journal}
  {\bibinfo  {journal} {Mon. Not. Roy. Astron. Soc.}\ }\textbf {\bibinfo
  {volume} {325}},\ \bibinfo {pages} {358} (\bibinfo {year} {2001})},\ \Eprint
  {https://arxiv.org/abs/gr-qc/0005044} {arXiv:gr-qc/0005044} \BibitemShut
  {NoStop}%
\bibitem [{\citenamefont {Tessmer}\ and\ \citenamefont
  {Schaefer}(2010)}]{Tessmer:2010}%
  \BibitemOpen
  \bibfield  {author} {\bibinfo {author} {\bibfnamefont {M.}~\bibnamefont
  {Tessmer}}\ and\ \bibinfo {author} {\bibfnamefont {G.}~\bibnamefont
  {Schaefer}},\ }\bibfield  {title} {\bibinfo {title} {{Full-analytic
  frequency-domain 1pN-accurate gravitational wave forms from eccentric compact
  binaries}},\ }\href {https://doi.org/10.1103/PhysRevD.82.124064} {\bibfield
  {journal} {\bibinfo  {journal} {Phys. Rev. D}\ }\textbf {\bibinfo {volume}
  {82}},\ \bibinfo {pages} {124064} (\bibinfo {year} {2010})},\ \Eprint
  {https://arxiv.org/abs/1006.3714} {arXiv:1006.3714 [gr-qc]} \BibitemShut
  {NoStop}%
\bibitem [{\citenamefont {Arredondo}(2024)}]{EccentricAmplitudes}%
  \BibitemOpen
  \bibfield  {author} {\bibinfo {author} {\bibfnamefont {J.~N.}\ \bibnamefont
  {Arredondo}},\ }\href {https://doi.org/10.5281/zenodo.10574033} {\bibinfo
  {title} {{Nijaid/EccentricAmplitudes: v1-beta}}} (\bibinfo {year}
  {2024})\BibitemShut {NoStop}%
\bibitem [{\citenamefont {Tinto}\ and\ \citenamefont
  {Dhurandhar}(2005)}]{Tinto:2004}%
  \BibitemOpen
  \bibfield  {author} {\bibinfo {author} {\bibfnamefont {M.}~\bibnamefont
  {Tinto}}\ and\ \bibinfo {author} {\bibfnamefont {S.~V.}\ \bibnamefont
  {Dhurandhar}},\ }\bibfield  {title} {\bibinfo {title} {{Time Delay
  Interferometry}},\ }\href {https://doi.org/10.12942/lrr-2005-4} {\bibfield
  {journal} {\bibinfo  {journal} {Living Rev. Rel.}\ }\textbf {\bibinfo
  {volume} {8}},\ \bibinfo {pages} {4} (\bibinfo {year} {2005})},\ \Eprint
  {https://arxiv.org/abs/gr-qc/0409034} {arXiv:gr-qc/0409034} \BibitemShut
  {NoStop}%
\bibitem [{\citenamefont {Schmidt}\ \emph {et~al.}(2015)\citenamefont
  {Schmidt}, \citenamefont {Ohme},\ and\ \citenamefont
  {Hannam}}]{Schmidt:2014}%
  \BibitemOpen
  \bibfield  {author} {\bibinfo {author} {\bibfnamefont {P.}~\bibnamefont
  {Schmidt}}, \bibinfo {author} {\bibfnamefont {F.}~\bibnamefont {Ohme}},\ and\
  \bibinfo {author} {\bibfnamefont {M.}~\bibnamefont {Hannam}},\ }\bibfield
  {title} {\bibinfo {title} {{Towards models of gravitational waveforms from
  generic binaries II: Modelling precession effects with a single effective
  precession parameter}},\ }\href {https://doi.org/10.1103/PhysRevD.91.024043}
  {\bibfield  {journal} {\bibinfo  {journal} {Phys. Rev. D}\ }\textbf {\bibinfo
  {volume} {91}},\ \bibinfo {pages} {024043} (\bibinfo {year} {2015})},\
  \Eprint {https://arxiv.org/abs/1408.1810} {arXiv:1408.1810 [gr-qc]}
  \BibitemShut {NoStop}%
\bibitem [{\citenamefont {{Gil Choi}}\ \emph {et~al.}(2022)\citenamefont {{Gil
  Choi}}, \citenamefont {{Yang}},\ and\ \citenamefont {{Lee}}}]{GilChoi:2022}%
  \BibitemOpen
  \bibfield  {author} {\bibinfo {author} {\bibfnamefont {H.}~\bibnamefont {{Gil
  Choi}}}, \bibinfo {author} {\bibfnamefont {T.}~\bibnamefont {{Yang}}},\ and\
  \bibinfo {author} {\bibfnamefont {H.~M.}\ \bibnamefont {{Lee}}},\ }\bibfield
  {title} {\bibinfo {title} {{Importance of Eccentricities in Parameter
  Estimation of Compact Binary Inspirals with Decihertz Gravitational-Wave
  Detectors}},\ }\href {https://doi.org/10.48550/arXiv.2210.09541} {\bibfield
  {journal} {\bibinfo  {journal} {arXiv e-prints}\ ,\ \bibinfo {eid}
  {arXiv:2210.09541}} (\bibinfo {year} {2022})},\ \Eprint
  {https://arxiv.org/abs/2210.09541} {arXiv:2210.09541 [gr-qc]} \BibitemShut
  {NoStop}%
\bibitem [{\citenamefont {Loutrel}\ \emph {et~al.}(2019)\citenamefont
  {Loutrel}, \citenamefont {Liebersbach}, \citenamefont {Yunes},\ and\
  \citenamefont {Cornish}}]{Loutrel:2018}%
  \BibitemOpen
  \bibfield  {author} {\bibinfo {author} {\bibfnamefont {N.}~\bibnamefont
  {Loutrel}}, \bibinfo {author} {\bibfnamefont {S.}~\bibnamefont
  {Liebersbach}}, \bibinfo {author} {\bibfnamefont {N.}~\bibnamefont {Yunes}},\
  and\ \bibinfo {author} {\bibfnamefont {N.}~\bibnamefont {Cornish}},\
  }\bibfield  {title} {\bibinfo {title} {{The eccentric behavior of
  inspiralling compact binaries}},\ }\href
  {https://doi.org/10.1088/1361-6382/aaf2a9} {\bibfield  {journal} {\bibinfo
  {journal} {Class. Quant. Grav.}\ }\textbf {\bibinfo {volume} {36}},\ \bibinfo
  {pages} {025004} (\bibinfo {year} {2019})},\ \Eprint
  {https://arxiv.org/abs/1810.03521} {arXiv:1810.03521 [gr-qc]} \BibitemShut
  {NoStop}%
\bibitem [{\citenamefont {Fumagalli}\ and\ \citenamefont
  {Gerosa}(2023)}]{Fumagalli:2023}%
  \BibitemOpen
  \bibfield  {author} {\bibinfo {author} {\bibfnamefont {G.}~\bibnamefont
  {Fumagalli}}\ and\ \bibinfo {author} {\bibfnamefont {D.}~\bibnamefont
  {Gerosa}},\ }\bibfield  {title} {\bibinfo {title} {{Spin-eccentricity
  interplay in merging binary black holes}},\ }\href
  {https://doi.org/10.1103/PhysRevD.108.124055} {\bibfield  {journal} {\bibinfo
   {journal} {Phys. Rev. D}\ }\textbf {\bibinfo {volume} {108}},\ \bibinfo
  {pages} {124055} (\bibinfo {year} {2023})},\ \Eprint
  {https://arxiv.org/abs/2310.16893} {arXiv:2310.16893 [gr-qc]} \BibitemShut
  {NoStop}%
\bibitem [{\citenamefont {Gerosa}\ \emph {et~al.}(2023)\citenamefont {Gerosa},
  \citenamefont {Fumagalli}, \citenamefont {Mould}, \citenamefont {Cavallotto},
  \citenamefont {Monroy}, \citenamefont {Gangardt},\ and\ \citenamefont
  {De~Renzis}}]{Gerosa:2023}%
  \BibitemOpen
  \bibfield  {author} {\bibinfo {author} {\bibfnamefont {D.}~\bibnamefont
  {Gerosa}}, \bibinfo {author} {\bibfnamefont {G.}~\bibnamefont {Fumagalli}},
  \bibinfo {author} {\bibfnamefont {M.}~\bibnamefont {Mould}}, \bibinfo
  {author} {\bibfnamefont {G.}~\bibnamefont {Cavallotto}}, \bibinfo {author}
  {\bibfnamefont {D.~P.}\ \bibnamefont {Monroy}}, \bibinfo {author}
  {\bibfnamefont {D.}~\bibnamefont {Gangardt}},\ and\ \bibinfo {author}
  {\bibfnamefont {V.}~\bibnamefont {De~Renzis}},\ }\bibfield  {title} {\bibinfo
  {title} {{Efficient multi-timescale dynamics of precessing black-hole
  binaries}},\ }\href {https://doi.org/10.1103/PhysRevD.108.024042} {\bibfield
  {journal} {\bibinfo  {journal} {Phys. Rev. D}\ }\textbf {\bibinfo {volume}
  {108}},\ \bibinfo {pages} {024042} (\bibinfo {year} {2023})},\ \Eprint
  {https://arxiv.org/abs/2304.04801} {arXiv:2304.04801 [gr-qc]} \BibitemShut
  {NoStop}%
\bibitem [{\citenamefont {Paul}\ and\ \citenamefont
  {Mishra}(2023)}]{Paul:2022}%
  \BibitemOpen
  \bibfield  {author} {\bibinfo {author} {\bibfnamefont {K.}~\bibnamefont
  {Paul}}\ and\ \bibinfo {author} {\bibfnamefont {C.~K.}\ \bibnamefont
  {Mishra}},\ }\bibfield  {title} {\bibinfo {title} {{Spin effects in spherical
  harmonic modes of gravitational waves from eccentric compact binary
  inspirals}},\ }\href {https://doi.org/10.1103/PhysRevD.108.024023} {\bibfield
   {journal} {\bibinfo  {journal} {Phys. Rev. D}\ }\textbf {\bibinfo {volume}
  {108}},\ \bibinfo {pages} {024023} (\bibinfo {year} {2023})},\ \Eprint
  {https://arxiv.org/abs/2211.04155} {arXiv:2211.04155 [gr-qc]} \BibitemShut
  {NoStop}%
\bibitem [{{\relax DLMF}()}]{DLMF}%
  \BibitemOpen
  {\relax DLMF},\ \href {https://dlmf.nist.gov/} {\bibinfo {title} {{\it NIST
  Digital Library of Mathematical Functions}}},\ \bibinfo {howpublished}
  {\url{https://dlmf.nist.gov/}, Release 1.1.9 of 2023-03-15},\ \bibinfo {note}
  {f.~W.~J. Olver, A.~B. {Olde Daalhuis}, D.~W. Lozier, B.~I. Schneider, R.~F.
  Boisvert, C.~W. Clark, B.~R. Miller, B.~V. Saunders, H.~S. Cohl, and M.~A.
  McClain, eds.}\BibitemShut {Stop}%
\bibitem [{\citenamefont {Tucker}\ and\ \citenamefont
  {Will}(2021)}]{Tucker:2021}%
  \BibitemOpen
  \bibfield  {author} {\bibinfo {author} {\bibfnamefont {A.}~\bibnamefont
  {Tucker}}\ and\ \bibinfo {author} {\bibfnamefont {C.~M.}\ \bibnamefont
  {Will}},\ }\bibfield  {title} {\bibinfo {title} {{Residual eccentricity of
  inspiralling orbits at the gravitational-wave detection threshold: Accurate
  estimates using post-Newtonian theory}},\ }\href
  {https://doi.org/10.1103/PhysRevD.104.104023} {\bibfield  {journal} {\bibinfo
   {journal} {Phys. Rev. D}\ }\textbf {\bibinfo {volume} {104}},\ \bibinfo
  {pages} {104023} (\bibinfo {year} {2021})},\ \Eprint
  {https://arxiv.org/abs/2108.12210} {arXiv:2108.12210 [gr-qc]} \BibitemShut
  {NoStop}%
\end{thebibliography}%

\end{document}